%% file: main.tex
\documentclass[trackchanges,twocolumn]{aastex701}
\usepackage{amsmath}
\usepackage{booktabs}

\begin{document}

\title{Follow-up of SN\,2025wny II: Superluminous Supernova Physics at Cosmic Noon}

\include{affiliations.tex}
\input{authors.tex}

\begin{abstract}
SN\,2025wny is a gravitationally lensed, hydrogen-poor superluminous supernova (SLSN-I) at $z=2.015$. To date, it is the most extensively observed high-redshift core-collapse SN and has the most detailed rest-frame UV observations of any SLSN. We present densely sampled rest-frame UV-to-optical photometry and spectroscopy out to $+80$\,d post-peak (rest frame) from several facilities, including \textit{JWST}, Keck, \textit{VLT}, Gemini, the Palomar 200-inch, the Fraunhofer Telescope at Wendelstein, and the Liverpool Telescope. Correcting for lensing magnification, SN\,2025wny reaches a peak pseudo-bolometric luminosity of $L_{\rm peak}\gtrsim4\times10^{44}$\,erg\,s$^{-1}$ over rest-frame 1500--4230\,{\AA}, placing it within the luminosity range of typical SLSNe-I. SN\,2025wny exhibits several unusual features, including a continuum excess and sharp spectral features in the FUV from $+20$--60\,d that coincide with an FUV light-curve plateau and higher inferred blackbody temperatures. SN\,2025wny's spectra also show little to no UV line blanketing, no obvious \ion{O}{2} absorption despite high temperatures, and evidence for \ion{C}{2}, H$\alpha$, and possible \ion{He}{1}. 
Light-curve modeling suggests that SN\,2025wny may require a hybrid or non-standard power source. This work provides some of the first detailed constraints on high-redshift SLSNe and establishes SN\,2025wny as an essential spectral and photometric reference for identifying and interpreting high-redshift SLSNe discovered by \textit{Rubin} and \textit{Roman}.
\end{abstract}

\section{Introduction}
At high redshift, stellar evolution proceeds in environments that differ substantially from the local Universe, notably at lower metallicities \citep{2006ApJ...644..813E, 2014ApJ...795..165S, 2021ApJ...914...19S, 2026arXiv260530513L}. Lower metallicities should weaken metal line-driven winds, decreasing mass loss and allowing stars to retain more mass \citep{2001A&A...369..574V} and greater angular momenta \citep{2001A&A...373..555M}, producing more massive stars and increasing the core-collapse supernova rate \citep{2013ApJ...765L..43I}.

Core-collapse supernovae (CCSNe) are one of the few probes of massive stars at high redshift, where they trace recent star formation and reveal the end-of-life properties of massive stars. Superluminous supernovae (SLSNe), which reach peak absolute magnitudes of $\mathrm{M}_{\mathrm{AB}}\lesssim -21$ and outshine normal CCSNe by factors of 10--100 \citep{2008ApJ...673.1014U, 2007ApJ...668L..99Q, Quimby2011_SLSNeI, GalYam2012_SLSN_Review}, are especially powerful in this context. Their extreme luminosities allow them to be detected at much larger distances than ordinary CCSNe, while their long light-curve durations enable extended follow-up, especially when combined with cosmological time dilation. At high redshift, SLSNe also provide efficient access to the rest-frame UV through observer-frame optical observations, probing a spectral region where metal-line blanketing, ionization structure, and signatures of the underlying power source are imprinted. Typical power sources invoked for hydrogen-poor SLSNe (SLSNe-I), which is the focus of this work, include radioactive
$^{56}$Ni decay, often in the context of pair instability explosions
\citep{Barkat1967, Kasen2011},
magnetar spin-down \citep{Kasen2010_MagnetarModel, Woosley2010_MagnetarModel}, interaction between the ejecta and H-poor circumstellar
material \citep{Chevalier2011_Shell, Chatzopoulos2012, Moriya2013,
Chatzopoulos2013}, and fallback accretion \citep{Dexter2013_FallbackAccretion}. Only a small number of slowly evolving SLSNe-I have been discussed as
possible pair-instability supernovae (PISNe) or extreme $^{56}$Ni-powered candidates
(e.g., SN\,2007bi, SN\,2018ibb; \citealt{2009Natur.462..624G, Schulze2024_18ibb}), although
simple PISN models predict spectra that are redder and more
strongly line-blanketed than observed.

However, SLSNe remain sparsely sampled both in the local Universe and at high redshift. Observationally, they are rare, comprising $\approx1.8\%$ of all SNe reported to the Transient Name Server (TNS\footnote{https://www.wis-tns.org/}), with the overwhelming majority identified at low redshift. The existing sample of 21 SLSNe at $z\gtrsim1$ \citep{2012Natur.491..228C, 2013ApJ...779...98H, 2015MNRAS.449.1215P, Smith2018_DES16C2nm, 2018ApJ...852...81L, Angus2019_DES_SLSNe} remains extremely sparsely sampled, limited to near-peak photometry and typically one epoch of rest-frame UV spectroscopy, leaving their photometric and spectroscopic evolution, as well as the rest-frame optical properties, poorly constrained. Furthermore, the powering mechanisms of SLSNe remain poorly constrained even in the local Universe, in large part because the strongest diagnostics require late-time observations. These observations are already challenging for nearby events and become increasingly difficult for SLSNe at high redshift.

So far, the most practical way to discover and monitor high-$z$ SNe is for them to be extremely intrinsically luminous or ``boosted" in brightness by gravitational lensing. Strong gravitational lensing in particular can amplify the observed flux of high-$z$ SNe by factors of up to $\sim100$ \citep{2026ApJ..1003L..47L}, enabling high-SNR photometric and spectroscopic observations that would otherwise be inaccessible. This approach has been demonstrated to be successful for Type Ia SNe, including SN H0pe \citep{2025ApJ...979...13P}, SN Requiem \citep{2021NatAs...5.1118R}, SN Encore \citep{2024ApJ...967L..37P}, PS1-10afx \citep{2013ApJ...767..162C, Quimby2013_PS1-10afx}, iPTF16geu \citep{2017Sci...356..291G}, and SN Zwicky \citep{2023NatAs...7.1098G}. On the other hand, only four high-$z$ CCSNe have been discovered through gravitational lensing, including SN Refsdal \citep{2016ApJ...831..205K}, a SN\,1987-like Type~II SN at $z=1.49$, SN Eos \citep{Coulter2026_z5}, a normal Type~II CCSN at $z=5.13$, SN\,2025mkn \citep{2026ApJ..1003L..47L}, a Type~II SN at $z=1.37$, and SN\,2025wny \citep{Taubenberger2025_SN2025wny, Johansson2025_SN2025wny}, a SLSN-I at $z=2.015$ which is the focus of this work. Of all gravitationally-lensed high-$z$ CCSNe, SN\,2025wny has by far the most well-sampled dataset.

In this paper, we present follow-up observations and a detailed analysis of SN\,2025wny.
In Section~\ref{sec:observations}, we describe the photometric and spectroscopic observations. In Section~\ref{sec:light-curve-analysis}, we present the observed light curves, analyze their temporal evolution, construct the pseudo-bolometric light curve, and compare our results to those of other SLSNe-I. In Section~\ref{sec:light-curve-modeling}, we present light curve fitting results. In Section~\ref{sec:spectra}, we present line identifications and the spectroscopic evolution. In Section~\ref{sec:discussion}, we discuss the implications of these results, and in Section~\ref{sec:conclusions}, we summarize our conclusions. In this work, we adopt the $\Lambda$CDM cosmology of \citet{Planck_2020} and AB magnitudes unless otherwise specified.

This work, which focuses on SN science, is part of a series of papers on SN\,2025wny. \citet{Johansson2025_SN2025wny} originally presented the discovery and classification of SN\,2025wny as a SLSN-I at $z=2.015$. \citet{Goobar2026} present additional Hubble and James Webb Space Telescope observations, which are used in this work. \citet{Johansson2026_inprep} present spatially-resolved spectroscopy of all supernova images and calculate spectroscopic time delays. \citet{Townsend2026} present scene-modelling photometry of ground-based imaging in the \textit{grizJ} filters, from which they measure the time delays and relative magnification ratios between the multiple images. Lens modeling, magnification estimates, and $H_0$ inference are presented in \citet{Mortsell2026}. Analysis of the host galaxy of SN\,2025wny is presented in \citet{Qin2026}. \citet{Hjortlund2026} simulate seven years of ZTF operations to estimate the expected rate of lensed SLSNe-I and quantify selection biases in the inferred magnification and intrinsic luminosity of SN 2025wny.

\section{Observations}
\label{sec:observations}

\begin{figure*}[t]
    \centering
    \includegraphics[width=\linewidth]{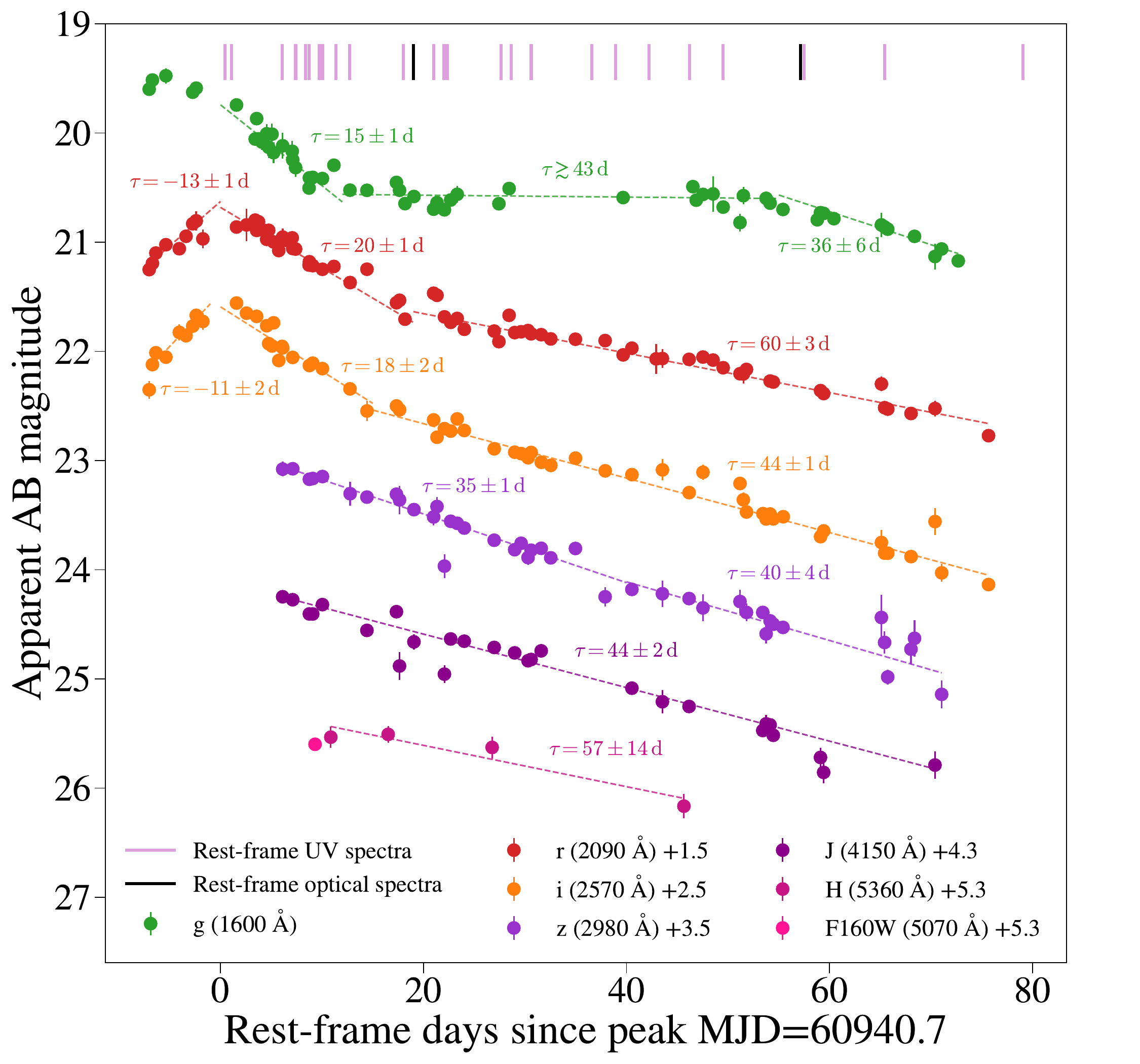}
    \caption{S-corrected observed light curves of SN\,2025wny (image A). Phases where spectra were taken are marked as thick lines at the top of the figure. Dashed lines indicate exponential fits to various parts of each light curve, with corresponding $e$-folding times labeled. The magnitudes are corrected for Galactic extinction.  The zero value on the x-axis is with respect to the pseudo-bolometric maximum ($\mathrm{MJD}=60940.7$). We do not correct for lensing magnification. In the legend, we label the observed band and approximate corresponding rest-frame effective wavelength in parentheses.}.
    \label{fig:light-curve}
\end{figure*}

\subsection{Discovery of SN\,2025wny}
SN\,2025wny was initially reported to TNS by the Gravitational-wave Optical Transient Observer (GOTO; \citealt{2022MNRAS.511.2405S}) on 2025 September 1. The event was also detected by the Zwicky Transient Facility (ZTF; \citealt{Graham2019, Bellm2019_ztf, Bellm2019_surveys, Dekany2020, Masci2019}) with the internal name ZTF25abnjznp, with its first detection on 2025 August 23. SN\,2025wny was later classified as a SLSN-I at $z=2.01$ \citep{Johansson2025_WinnyClassification}, gravitationally lensed by two foreground galaxies and producing at least five resolved images \citep{2025TNSAN.330....1A}. Photometric and spectroscopic observations of SN\,2025wny spanning  $\sim2$ months after the first ZTF detection are presented in \citet{Taubenberger2025_SN2025wny} and \citet{Johansson2025_SN2025wny}. In this work, we focus on image A of SN\,2025wny, which is the brightest of the observed lensed images.

The redshift initially reported for SN\,2025wny, $z=2.011$, was inferred from narrow absorption features in ground-based spectra. In this work, we adopt an updated redshift of $z=2.0151\pm0.0001$, determined from H$_\beta$ and the \ion{O}{3} $\lambda4959$ and $\lambda5007$ emission lines; further details are presented in \citet{Qin2026}.

\subsection{Imaging}
\subsubsection{Resolved photometry}
Due to the lensing geometry, light from the exploding source will traverse different optical paths, resulting in different Images with different photon travel times. For SN\,2025wny ($z=2.015$), gravitational lensing by two foreground galaxies at $z\approx0.375$ produce at least five Images (A, B, C, D, E) of the SN, the first four being resolvable from ground. Since the underlying emission originates from a single SN event, each Image traces the same intrinsic rise and decline of the explosion, albeit offset from each other in time.

Optical and IR imaging from the companion paper \citet{Townsend2026} have sufficient spatial resolution to isolate the flux from the brightest image of SN\,2025wny (image A) with scene modeling using the \textsc{lightcurver} code \citep{2024JOSS....9.6775D}, an end-to-end photometry pipeline that performs scene modeling using the ``starlet" functionality from the \textsc{STARRED} code \citep{2023JOSS....8.5340M}. Detailed explanations of the imaging observations, image reductions, and scene modeling to extract the image A component of SN\,2025wny are presented 
by \citet{Townsend2026}. 

We also include an epoch of resolved \textit{HST}/WFC3 F160W imaging from \citet{Goobar2026}, for which the image A flux was measured with custom scene modeling code using \textsc{lmfit} \citep{newville_2025_16175987}, \textsc{astropy} \citep{astropy:2013, astropy:2018, astropy:2022}, and \textsc{scipy} \citep{2020SciPy-NMeth}. We further include P200/WIRC $H$-band photometry, with the image A component extracted using the \textsc{galfit} code \citep{2002AJ....124..266P}. Because the current IR dataset is limited, we defer a more detailed and homogeneous IR photometric analysis to future work, when additional \textit{JWST} imaging epochs will be available. We do not use upper limits in this work because their primary purpose is to constrain the explosion time, and SN\,2025wny was behind the Sun before the first detection. 

The imaging observations used in this work come from the Liverpool Telescope (LT) with the IO:O optical imager and LOCI instrument in $griz$; Gemini North telescope with the GMOS instrument in $gr$; Large Binocular Telescope (LBT) with the Large Binocular Camera (LBC) in $griz$; the Palomar 60-inch telescope (P60) with the Spectral Energy Distribution Machine (SEDM) in $gri$ \citep{2018PASP..130c5003B, Rigault_2019}; with the Three Channel Imager (3KK; \citealt{2016SPIE.9908E..44L}) on the 2.1-m Fraunhofer Telescope at Wendelstein Observatory (FTW; \citealt{2014SPIE.9145E..2DH}) in $grizJ$; the Wide-field Infrared Camera (WIRC) on the Palomar 200-inch telescope (P200) in $H$; and Wide Field Camera 3 (WFC3) on the Hubble Space Telescope (\textit{HST}) in F160W. Please see \citet{Townsend2026} for reduction details.

\subsubsection{Photometric corrections}
\label{subsec:photometric-corrections}
We present the corrected multi-wavelength resolved light curves of SN\,2025wny (image A) in Figure~\ref{fig:light-curve} and Appendix~\ref{app:photometry-table}. The photometry is shown in the observer frame and has not been converted to rest-frame flux densities. The following corrections were applied.

Because the photometry combines data from multiple instruments, we apply S-corrections to transform all measurements onto a common photometric system. We adopt the  Fraunhofer Telescope at Wendelstein Observatory (FTW) $grizJ$ bandpasses as the reference system, since they provide the densest temporal coverage. Following \citet{2002AJ....124.2100S}, 
the correction from an instrument's native bandpass to the corresponding FTW bandpass is computed as

\begin{equation*}
    S(t) = -2.5 \log_{10}( \frac{F_{\lambda,\mathrm{ref}}(t)}{F_{\lambda}
    (t)}),
\end{equation*}

with

\begin{equation*}
    F_{\lambda,\mathrm{X}}(t)=\frac{\int f_{\lambda}(t) T_\mathrm{X}(\lambda) \lambda d\lambda}{\int T_\mathrm{X}(\lambda) \lambda d\lambda}
\end{equation*}

and where $t$ is the epoch, $f_\lambda(t)$ is the spectrum of an object at time $t$, $F_{\lambda,\mathrm{X}}(t)$ is the synthetic flux of an object at epoch $t$ and through bandpass $X$, and $T_\mathrm{X}$ is the total system throughput of bandpass $X$, accounting for filter transmission, detector response, telescope optics, and atmospheric transmission.

\begin{figure}
    \centering
    \includegraphics[width=\linewidth]{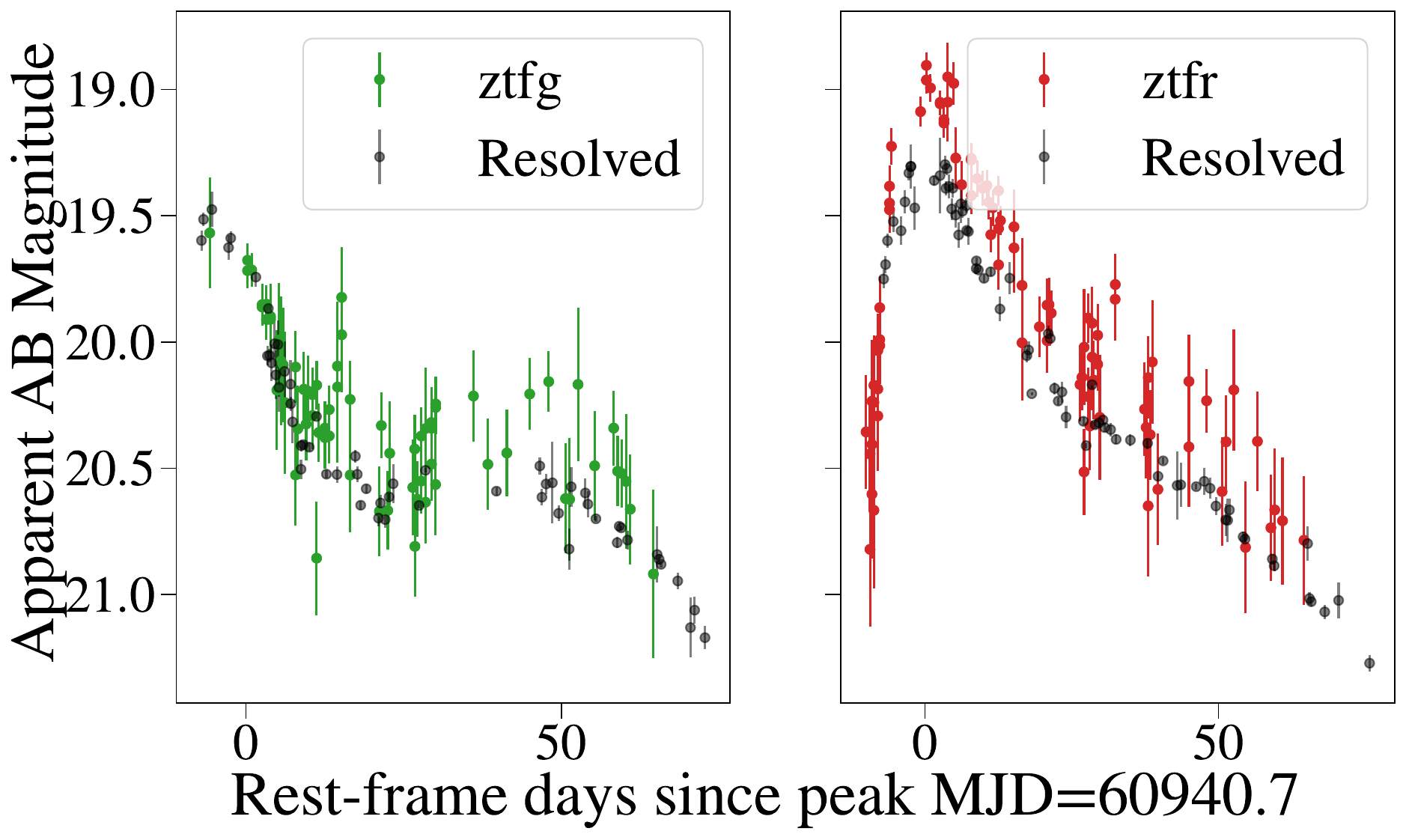}
    \caption{Unresolved ``GOOD" and ``OK" ZTF photometry in the $g$ (left) and $r$ (right) filters, with resolved and resolved S-corrected photometry plotted in black for comparison. All photometry shown is corrected for Milky Way extinction.}
    \label{fig:ztf_unresolved_lightcurve}
\end{figure}

However, the bandpasses available for many instruments are limited to filter transmission curves, and only in some cases include the detector response or telescope optics. None include atmospheric transmission or account for the observing conditions of individual nights. Therefore, the S-corrections can only partially correct for calibration systematics, especially for data obtained under poor conditions. In this work, we exclude photometry taken in very poor weather, where scene modeling is unreliable; the full dataset is presented in \citet{Townsend2026}.

Because S-corrections depend on the source spectrum at each epoch, we use the available spectra of SN\,2025wny to compute them. The spectra are first absolute flux-calibrated to the FTW light curves using photometry interpolated from a Gaussian Process (GP). We perform GP fitting using the Python package \textsc{george} \citep{2015ITPAM..38..252A}. We then use the calibrated spectra to measure synthetic photometry through native instrument and FTW bandpasses. We use the resulting synthetic flux ratios to compute and apply the S-corrections.

We do not apply S-corrections for the P200/WIRC $H$-band and \textit{HST}/WFC3 F160W photometry, since few spectra overlap with their photometric epochs. However, we do use the P200/WIRC $H$-band photometry, together with the FTW/WWFI photometry, to place the \textit{JWST} spectra on an absolute flux scale.

Following the S-corrections, we correct the observer-frame photometry for Milky Way extinction. We adopt a Galactic reddening of $E(B-V)=0.0529$ mag from \citet{2011ApJ...737..103S} and assume the extinction law of \citet{2007ApJ...663..320F} with $R_V=3.1$. We do not correct for extinction from the host galaxy or foreground lensing galaxies since the spectra do not exhibit strong narrow Na I D absorption features, and the light curves of SN\,2025wny are not significantly red at pre-peak phases. The differential extinction between SN images introduced by the foreground lensing galaxies is also found to be small ($\Delta E(B-V)\sim0.07-0.09\:\mathrm{mag}$ between image A and images B/C; \citealt{Goobar2026, Townsend2026}).

\subsubsection{Unresolved photometry}

We present unresolved photometry obtained by the Zwicky Transient Facility (ZTF) \citep{Bellm2019_ztf, 2019PASP..131g8001G, Masci2019, 2020PASP..132c8001D} on the Palomar 48-inch telescope (P48) in Figure~\ref{fig:ztf_unresolved_lightcurve} and Appendix~\ref{app:photometry-table}. We present data retrieved from ZTF Forced Photometry service \citep{2023arXiv230516279M} on Fritz \citep{2019JOSS....4.1247V, 2023ApJS..267...31C}, measured through PSF fitting on reference-subtracted images. Due to the large ZTF plate scale (1\farcs01 pixel$^{-1}$) and typical seeing conditions ($\gtrsim1\farcs5$), scene modeling cannot reliably isolate the image A flux, so we exclude these data from our analysis. Still, we present these data as a reference, since future discoveries of gravitationally lensed high-$z$ CCSNe will likely have unresolved light curves.

The unresolved measurements contain a seeing-dependent mixture of flux from image A and the other lensed images of SN\,2025wny. The unresolved ZTF light curves are therefore generally brighter than the resolved image A light curves. They also show substantially larger scatter, likely due to variable observing conditions and possible time-delay effects among the lensed images.

We discard measurements with $\mathrm{S/N}<3$ or seeing $>3.0\arcsec$, and classify the remainder as ``OK'' for $2.0\arcsec<$ seeing $<3.0\arcsec$ or $3\leq\mathrm{S/N}<5$, and ``GOOD'' for seeing $<2.0\arcsec$ and $\mathrm{S/N}\geq5$. Difference-image cutouts are visually inspected, and measurements without a clear residual or with obvious artifacts are flagged as ``BAD.''

\subsection{Spectroscopy}
We obtained an extensive spectroscopic sequence of SN\,2025wny spanning rest-frame phases of $\approx1$--$80$\,d relative to bolometric maximum, for which we adopt $\mathrm{MJD}_{\rm peak}=60940.7$ (Section~\ref{subsec:bol-lc}). Most epochs cover the rest-frame UV, from $\approx1200$--3300\,{\AA}\, using ground-based optical spectroscopy. In addition, two \textit{JWST}/NIRSpec spectra at $+20$\,d and $+57$\,d extend the wavelength coverage to longer wavelengths. One covers the rest-frame optical, $\approx3300$--6500\,{\AA}, while the other covers the rest-frame optical to near-infrared, $\approx3300$--17500\,{\AA}. A description of the spectroscopic observations is provided in Appendix~\ref{app:spectra-log}.

As described in Section~\ref{subsec:photometric-corrections}, spectra are placed on an absolute flux scale using contemporaneous FTW/WWFI $grizJ$ and P200/WIRC $H$-band photometry, which provide the densest temporal coverage in their respective bands. We do not absolute flux-calibrate the +57\,d \textit{JWST}/NIRSpec spectrum, as it lies beyond the $H$-band light curve. For each band, we interpolate the light curve with a GP and use the resulting photometric estimates to flux calibrate the spectra. 

\section{Light curve analysis}
\label{sec:light-curve-analysis}

\begin{figure}
    \centering
    \includegraphics[width=\linewidth]{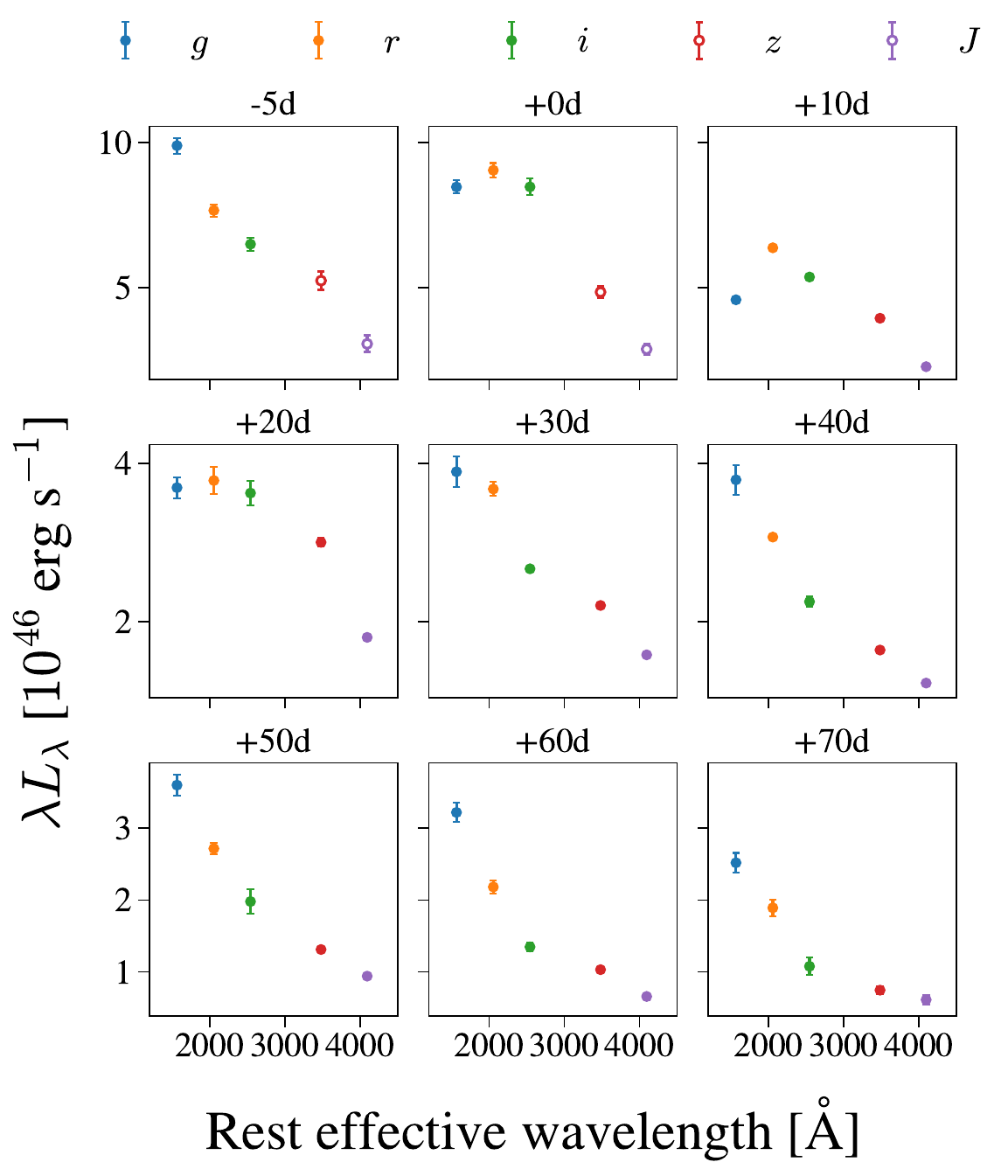}
    \caption{Rest-frame SED evolution of SN\,2025wny, constructed from a GP interpolation on the S-corrected $grizJ$ photometry. Open symbols denote $z$- and $J$-band points extrapolated beyond the observed phase range. No magnification correction is applied.}
    \label{fig:sed}
\end{figure}

\subsection{Magnification from gravitational lensing}
\label{sec:magnification-correction}

Lensing by foreground gravitational potentials can magnify and time-delay the light from background transients. The magnification, which is in princple achromatic, can increase the observed flux, while differences in the light-travel paths through the lensing potentials can produce multiple ``images" of the background transient that appear at different times. Based on the lens modeling presented in \citet{Mortsell2026}, the observed light from image A is magnified by a factor of $\mu=-4.7^{+3.4}_{-37.5}$ (95\% confidence interval). This factor includes effects of microlensing by stars in the lensing galaxies. Based on convention, a negative sign indicates that lensing boosts the observed flux.

Recovering intrinsic luminosities requires marginalizing over the magnification posterior. For a measured luminosity $L_{\rm obs}$ with uncertainty $\sigma_{\rm L}$, the posterior on the intrinsic luminosity is

\begin{equation}
\begin{split}
    p(L_{\rm int} | L_{{\rm obs}, \mu}) \propto
    \pi(L_{\rm int}) p(L_{\rm obs} | \mu L_{\rm int}) p(\mu),
\end{split}
\end{equation}
\label{eq:magnification-correction}

where $p(\mu)$ is the magnification posterior from \citet{Mortsell2026}, $\pi(L_{\rm int})$ is the prior on the intrinsic luminosity (assumed flat), and $p(L_{\rm obs} | \mu L_{\rm int})$ is the Gaussian likelihood between the observed luminosity and predicted intrinsic luminosity.

Because the magnification posterior for image A is extremely broad due to microlensing uncertainties, we instead use image D as a proxy for the intrinsic luminosity. image D has the narrowest magnification posterior of the four images, with $\mu_{\rm D}=+3.2^{+8.8}_{-0.7}$ at 95\% confidence. This choice requires the observed luminosity of image D, which we obtain with
\begin{align}
L_{\rm obs,D} = F_{\rm D/A} L_{\rm obs,A},
\end{align}
where $F_{\rm D/A}$ is the flux ratio between Images A and D and $L_{\rm obs,A}$ is the observed luminosity of image A. To determine $F_{\rm D/A}$, we GP interpolate the ground-based $gri$ light curves of image A to the phases corresponding to the resolved \textit{HST} F475W, F625W, and F814W observations of image D \citep{Goobar2026}, accounting for the measured time delay between Images A and D ($\Delta t_{\rm AD}=-65.7\pm3.5$\,d;  \citealt{Johansson2026_inprep}). We display the GP parameters and fits in Appendix~\ref{app:gp-for-magnification}. Propagating the GP, time-delay, and photometric uncertainties yields mutually consistent ratios across the three bands, which combine to give $F_{\rm D/A}=0.160\pm0.008$. Finally, we use Equation~\ref{eq:magnification-correction} and present the intrinsic pseudo-bolometric luminosities of SN\,2025wny corrected using this scheme in Section~\ref{subsec:bol-lc}.

We caution that this correction remains model dependent, as the microlensing posterior depends on assumptions about the stellar mass distributions of the lensing galaxies. While the image D-based correction is more stable than directly correcting image A, the inferred intrinsic luminosities remain subject to systematic uncertainties from the microlensing model.


\subsection{Light curve evolution}
\label{subsec:lc_evolution}

\begin{figure}
    \centering
    \includegraphics[width=\linewidth]{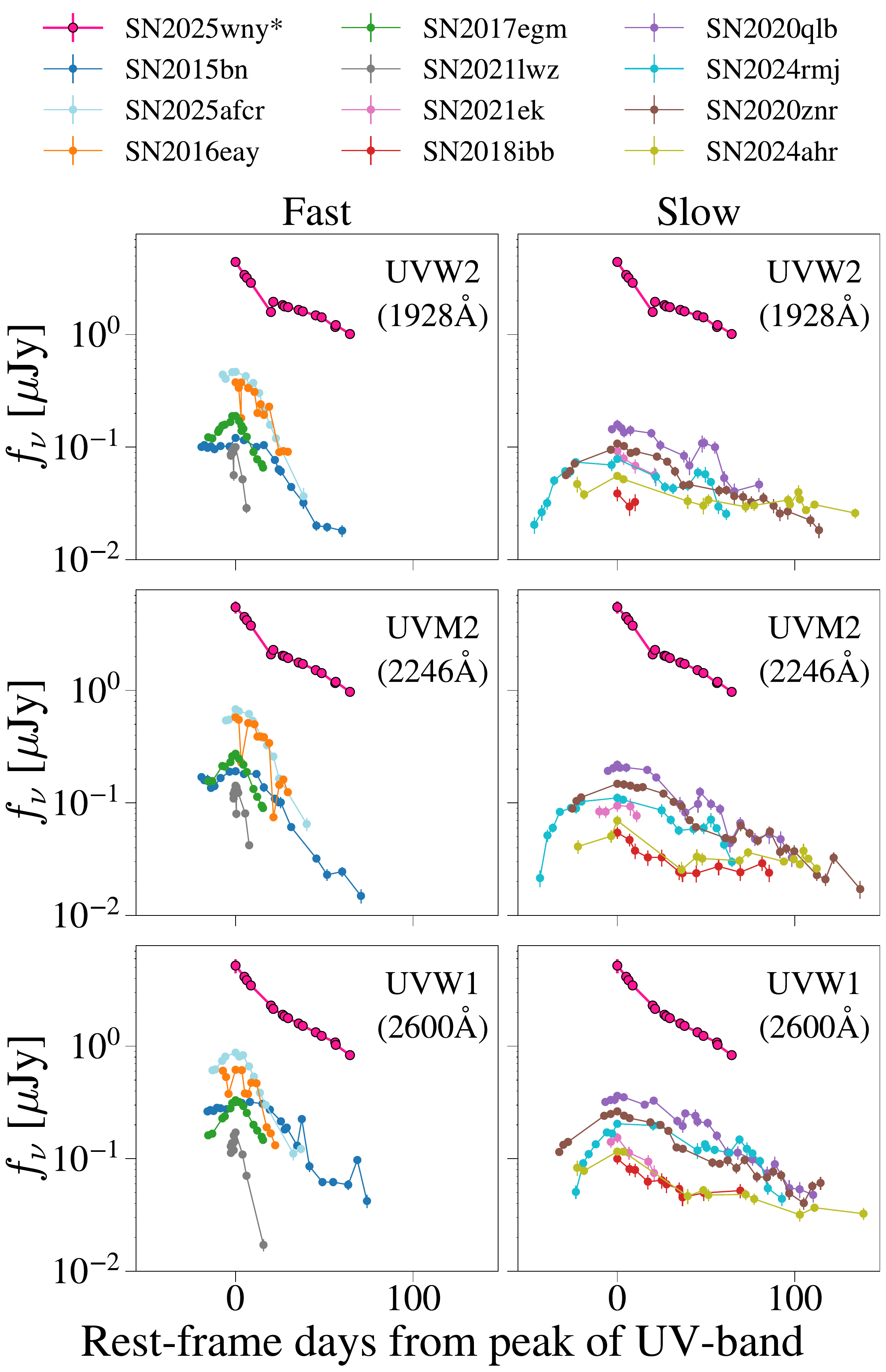}
    \caption{Comparison of well-sampled \textit{Swift}/UVOT SLSN-I light curves with synthetic SN\,2025wny photometry from rest-frame UV spectra. We scale SN\,2025wny down by a factor of 4.7 (from the median magnification of image A). Events are separated into fast (left) and slow-evolving events (right), based on their $e$-folding decline times in Appendix~\ref{app:uv-slsn-timescales}). Light curves are converted to rest-frame phases and flux densities by dividing by $(1+z)$, and then rescaled to the luminosity distance of SN,2025wny. Light curves are plotted with respect to the per-band peak. Sources for the photometric data used for the comparison objects are listed in Appendix~\ref{app:comparison-lc-refs}.}
    \label{fig:uv-lc-comparison}
\end{figure}

As shown in Figure~\ref{fig:light-curve}, 
at approximately +20\,d post-peak\footnote{Throughout this paper, we define the peak phase as the epoch of maximum bolometric luminosity $\mathrm{MJD}=60940.7$, derived from the empirical bolometric light curve (Section~\ref{subsec:bol-lc}).}, the observed $g$-band light curve, which probes the rest-frame FUV, transitions from a rapid decline to an extended plateau that persists until around $\sim$60\,d. In contrast, the $rizJH$ bands continue to fade over the same interval. The observed $g$ band also reaches peak brightness earlier than the $r$ and $i$ bands. This behavior suggests the presence of an additional source of FUV flux at around $+20$--60\,d post-peak.

We support this interpretation by fitting exponential declines to the rest-frame flux densities. The resulting $e$-folding timescales (Appendix~\ref{app:25wny_timescales}) show the $g$-band plateau at +20--60\,d. Photometric SEDs constructed from a GP interpolation of the light curves also show the same FUV excess (Figure~\ref{fig:sed}).

\begin{figure}
    \centering
    \includegraphics[width=\linewidth]{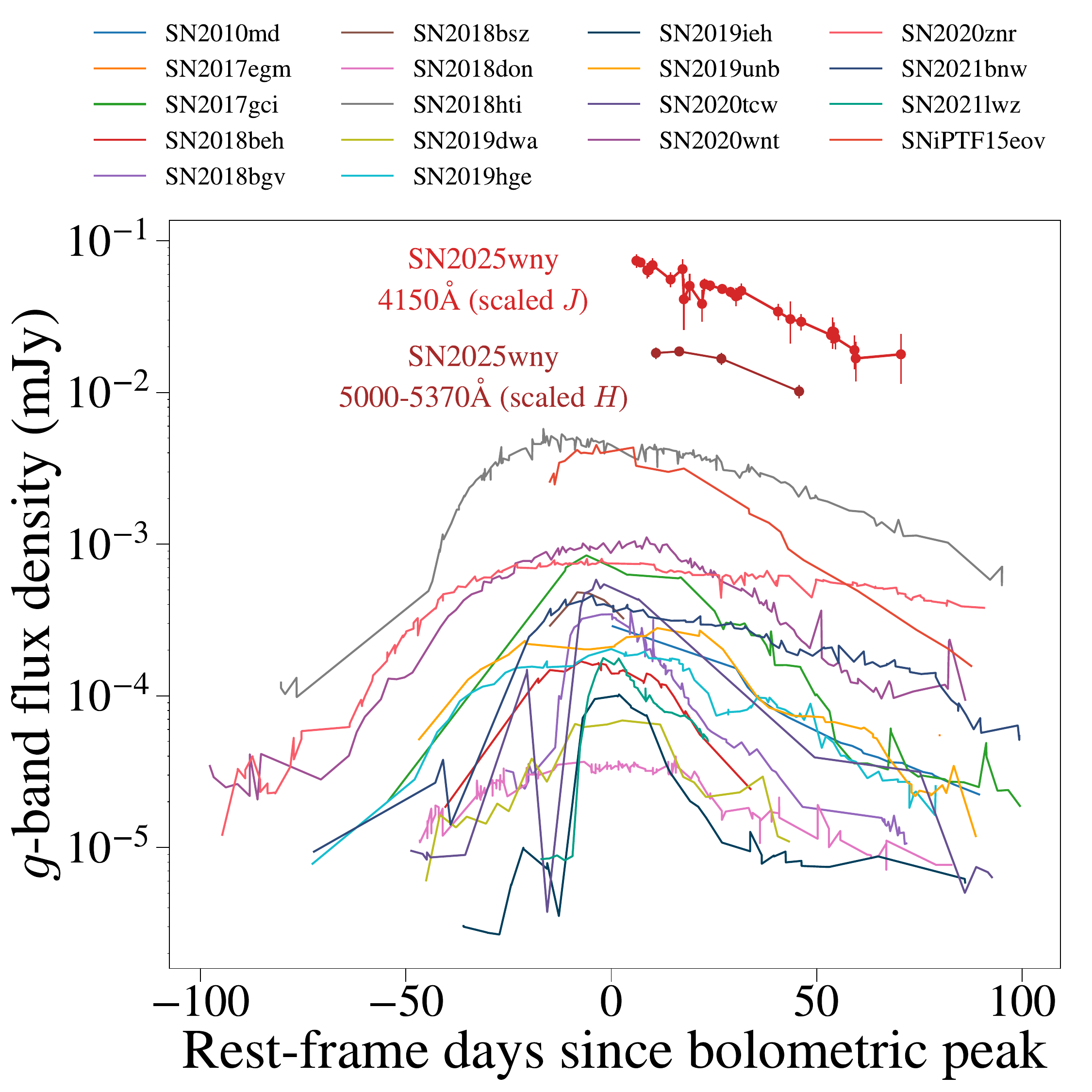}
    \caption{Comparison of SN\,2025wny's observed $J$- and $H$-band photometry with observed $g$-band light curves of low-$z$ ($z<0.1$) SLSNe-I from \citet{Gomez2024_SLSNeI}. The $J$ and $H$-bands of SN\,2025wny approximately probe the rest-frame $g$ band. We restrict the comparison to low-$z$ events to minimize the difference between observed and rest-frame $g$-band photometry. We list sources for the photometric data used for the comparison objects in Appendix~\ref{app:comparison-lc-refs}.}
    \label{fig:gband-compare}
\end{figure}

To determine whether the rest-frame FUV plateau is unusual, we compare SN\,2025wny to SLSNe-I with well-sampled \textit{Swift}/UVOT light curves (Figure~\ref{fig:uv-lc-comparison}). All comparison light curves are corrected for Milky Way extinction, converted to rest-frame flux densities, and scaled to the luminosity distance of SN\,2025wny. For SN\,2025wny, we construct synthetic \textit{Swift}/UVOT photometry by integrating the absolute flux-calibrated, extinction-corrected spectra through the UVW1, UVM2, and UVW2 filter transmission curves. We do not use the observed $g$ and $r$ light curves directly because their bandpasses differ substantially from the UVOT filters. We note that the synthetic photometry is approximate since we do not account for effects like detector response, telescope optics, and atmospheric transmission. However, the evolution of the synthetic light curves (plateau beginning at +20\,d) are independently supported by the observed $g$-band light curve.

SN\,2025wny does not follow a simple fast- or slow-evolving UV light curve. Its early decline is comparable to fast SLSNe-I, but after $\sim20$\,d post-peak, the UV evolution flattens and resembles the slower events. The measured $e$-folding timescales for the comparison sample (Table~\ref{tab:uv-slsn-timescales}; Appendix~\ref{app:uv-slsn-timescales}) and synthetic \textit{Swift/UVOT} photometry of SN\,2025wny (Table~\ref{tab:efolding-times}) confirms this.

\begin{figure}
    \centering
    \includegraphics[width=\linewidth]{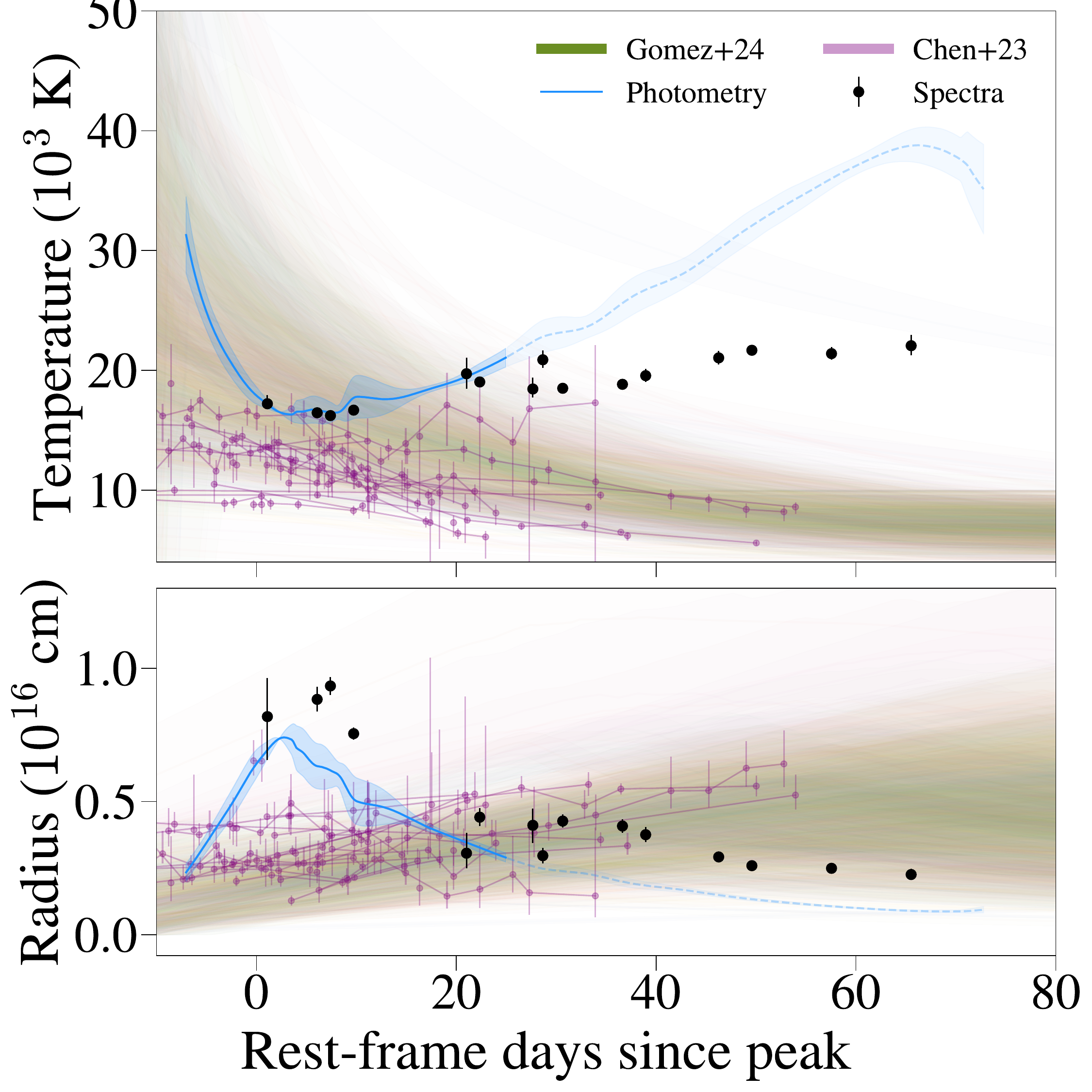}
    \caption{Temperature (top) and radius (bottom) evolution of SN\,2025wny derived from blackbody fits to its spectra (black) and its photometric SED (blue) compared to SLSNe-I from the \citet{Gomez2024_SLSNeI} sample (faded lines) and SLSNe-I with \textit{Swift/UVOT} data from \citet{Chen2023_PopulationStudy} (faded purple). Beyond +20\,d, the temperature inferred from the photometric SED fit is affected by the emergence of strong FUV emission lines and is therefore unreliable, shown with dashed lines. We scale down the radius evolution of SN\,2025wny by a factor of $\sqrt{4.7}$.}
    \label{fig:temperature-and-radius}
\end{figure}

As a check on the rest-frame optical evolution, we compare the observed $J$- and $H$-band photometry of SN\,2025wny, which probe rest-frame blue/green optical wavelengths, to observed $g$-band light curves of low-redshift ($z<0.1$) SLSNe-I (Figure~\ref{fig:gband-compare}). We do not compare absolute rest-frame $g$-band magnitudes because of the current sparse time sampling in SN\,2025wny's NIR data. The purpose of Figure~\ref{fig:gband-compare} is to illustrate that the rest-frame optical light curve is broadly similar to that of some local SLSN-I.

\subsection{Evidence for photospheric reheating}
\label{subsec:bb-lc}

We present the blackbody radius and temperature evolution in Figure~\ref{fig:temperature-and-radius}. We estimate these quantities using two independent methods. First, we fit blackbody functions to selected absolute flux-calibrated spectra with substantial wavelength coverage, masking out non-continuum features. However, as the spectra do not sample pre-peak phases, we also fit blackbody functions to the photometric SEDs constructed in Section~\ref{subsec:lc_evolution}. We perform all blackbody fits using the Markov Chain Monte Carlo (MCMC) methods implemented in \textsc{emcee}.

The two methods agree on the temperature and radius evolution at early phases but diverge after $\sim+20$\,d post-peak, when excess continuum and potential broad emission features emerge in the UV (Section~\ref{subsubsec:uv-excess}). This excess falls in the observed $g$ band and is therefore included in the photometric blackbody fits, while the spectral fits exclude them. As a result, the photometry-derived temperatures become systematically higher once the FUV excess strengthens.

We compare the blackbody temperature and radius evolution of SN\,2025wny to the SLSN-I sample in \citet{Gomez2024_SLSNeI} and \citet{Chen2023_PopulationStudy}(Figure~\ref{fig:temperature-and-radius}). \citet{Gomez2024_SLSNeI} measures luminosities and blackbody properties using the \textsc{extrabol} package. \citet{Chen2023_PopulationStudy} only reports temperatures for SLSNe-I with \textit{Swift/UVOT} data, for which the broader UV--optical coverage provides more reliable blackbody constraints. While the radius evolution is broadly consistent with that of other SLSNe-I, SN\,2025wny exhibits a notable temperature reheating beginning at $\sim$20\,d post-peak, in both the photometric and spectral fits. With the exception of SN\,2019szu \citep{2024MNRAS.52711970A} this behavior is not seen in the comparison sample. However, we note that UV observations for most published SLSNe-I are often sparse and typically do not extend to similarly late phases. Consequently, comparable late-time reheating episodes may have gone undetected in previous events.

SN\,2019szu has been interpreted as a possible pulsational pair-instability event, with late-time reheating attributed to interaction with recently ejected material. In SN\,2025wny, we do not detect the forbidden emission lines expected for a PPISN-like interpretation; however, the qualitative similarity may still point to an interaction-related origin for the reheating episode. We discuss an interaction powering source for SN\,2025wny in Section~\ref{sec:discussion}. 

\subsection{Bolometric light curve}
\label{subsec:bol-lc}

\begin{figure}
    \centering
    \includegraphics[width=\linewidth]{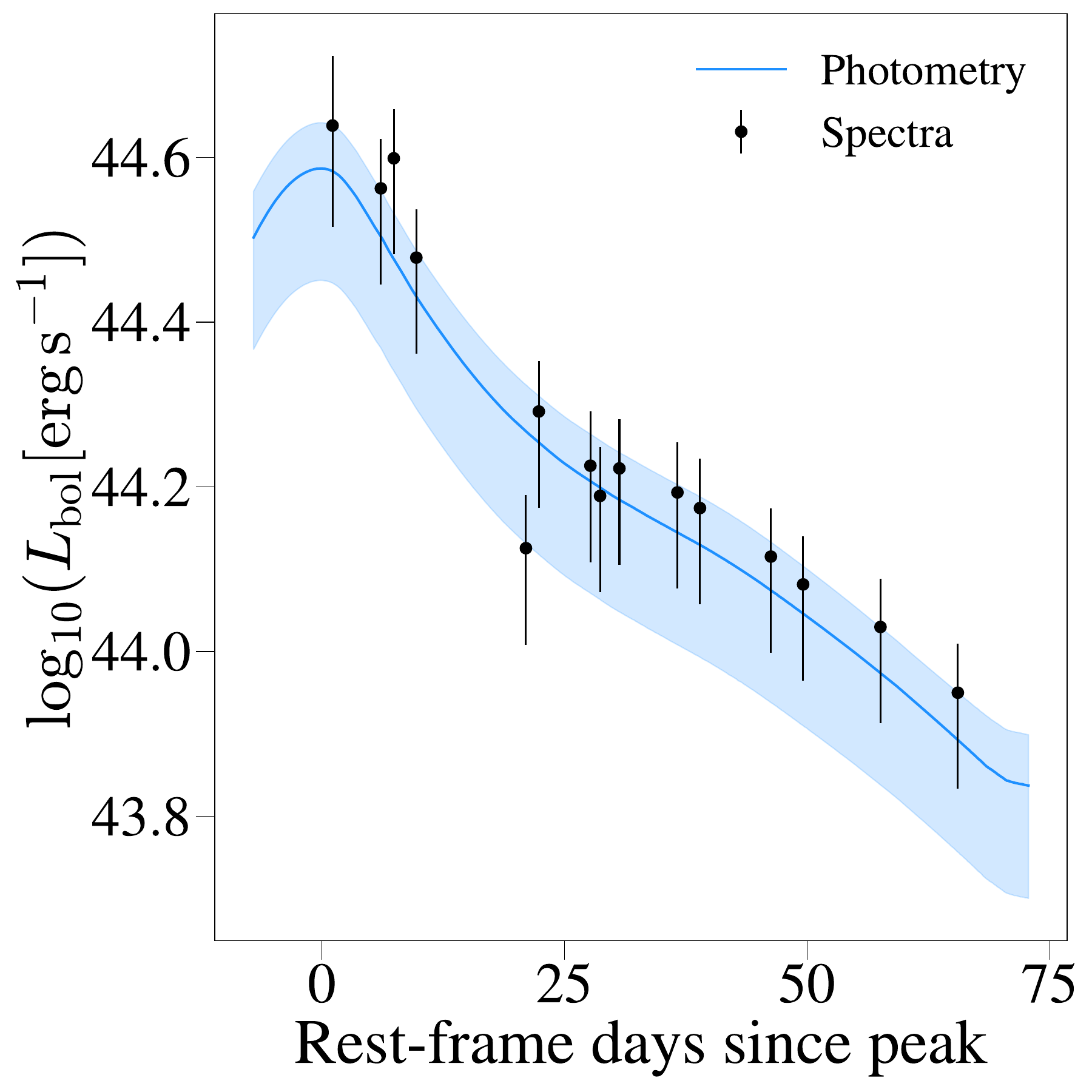}
    \caption{Empirical pseudo-bolometric light curves derived from photometric $grizJ$ SEDs (blue) and from the spectra (black). They are obtained by integrating over 1500--4230\,{\AA}, so the pseudo-bolometric light curves represent lower limits. Shaded regions and error bars indicate the 68\% credible region of the luminosity posteriors after demagnifying following the methodology in Section~\ref{sec:magnification-correction}.}
    \label{fig:bolometric-lc}
\end{figure}

\input{tables/bolometric_lc_table.tex}

We construct a spectra-derived pseudo-bolometric light curve by integrating select rest-frame UV spectra over their observed wavelength ranges. We fit blackbodies to the spectra to estimate flux in instrumental gaps and beyond the spectral coverage. The integration is restricted to 1500--4230\,{\AA}, matching the rest-frame wavelength range of the observed $grizJ$ photometry. We do not extrapolate beyond this range because most spectra extend only to rest-frame $\sim$3300\,{\AA}, causing blackbody fits to be UV-biased and overestimate the unobserved red-tail flux. Uncertainties in the observed regions were propagated in quadrature from the spectral flux errors; for unobserved regions, uncertainties were estimated from MCMC blackbody posterior samples. The uncertainties from the observed and extrapolated components were then combined in quadrature.

We also construct a photometry-derived pseudo-bolometric light curve by integrating the observed SEDs from Section~\ref{subsec:lc_evolution} over the observed $grizJ$ effective wavelength range (rest-frame 1500--4230\,{\AA}). For epochs without $z$- and $J$-band data (e.g. during the rise), we fit blackbody models to the photometric SEDs and integrate the best-fit blackbodies over the wavelength range spanned by the observed $i$ and $J$ bands to estimate the missing red flux. The pseudo-bolometric luminosity at these epochs is taken as the sum of the observed SED integral and this blackbody-derived correction.

\begin{figure}
    \centering
    \includegraphics[width=\linewidth]{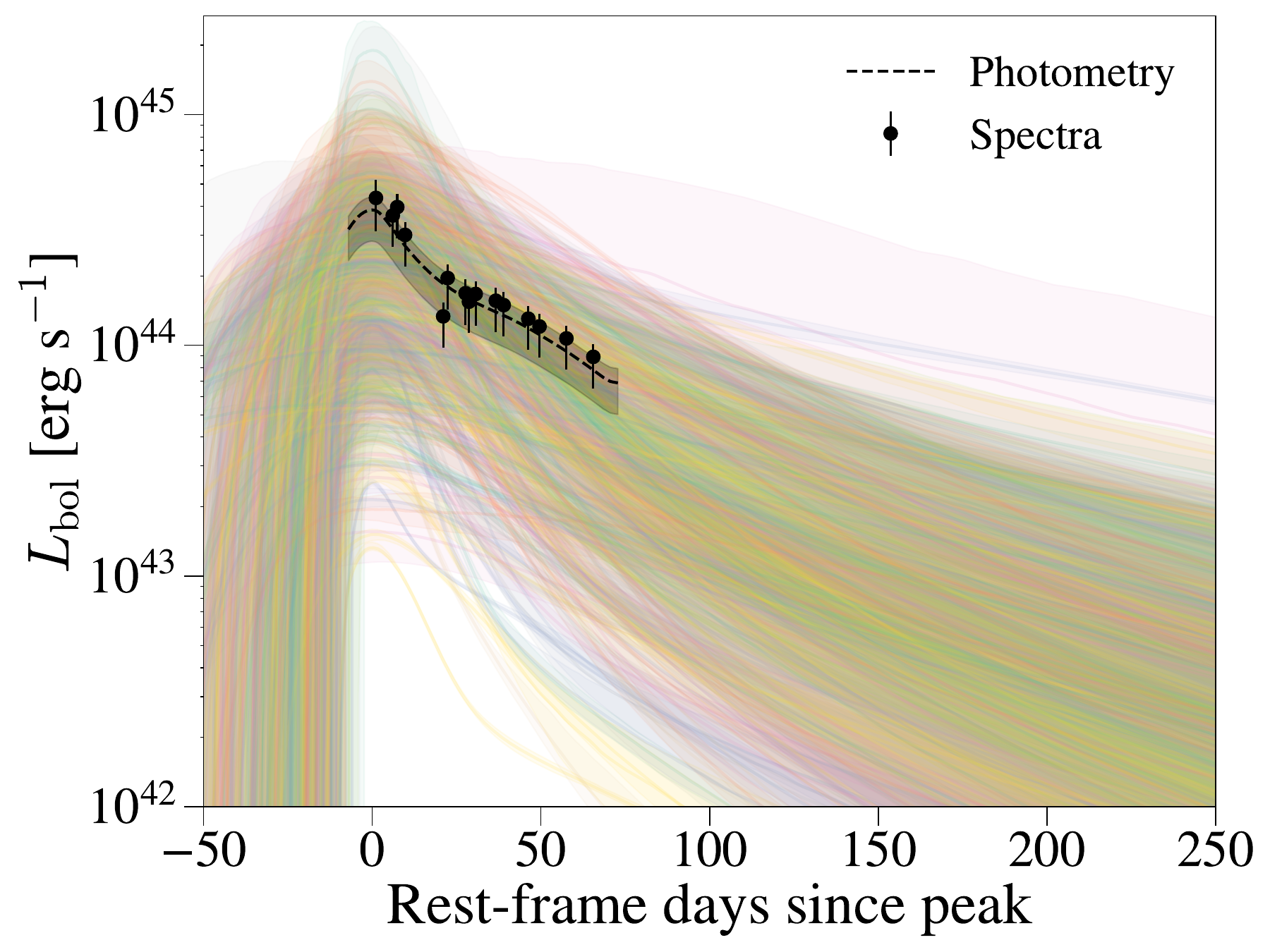}
    \includegraphics[width=\linewidth]{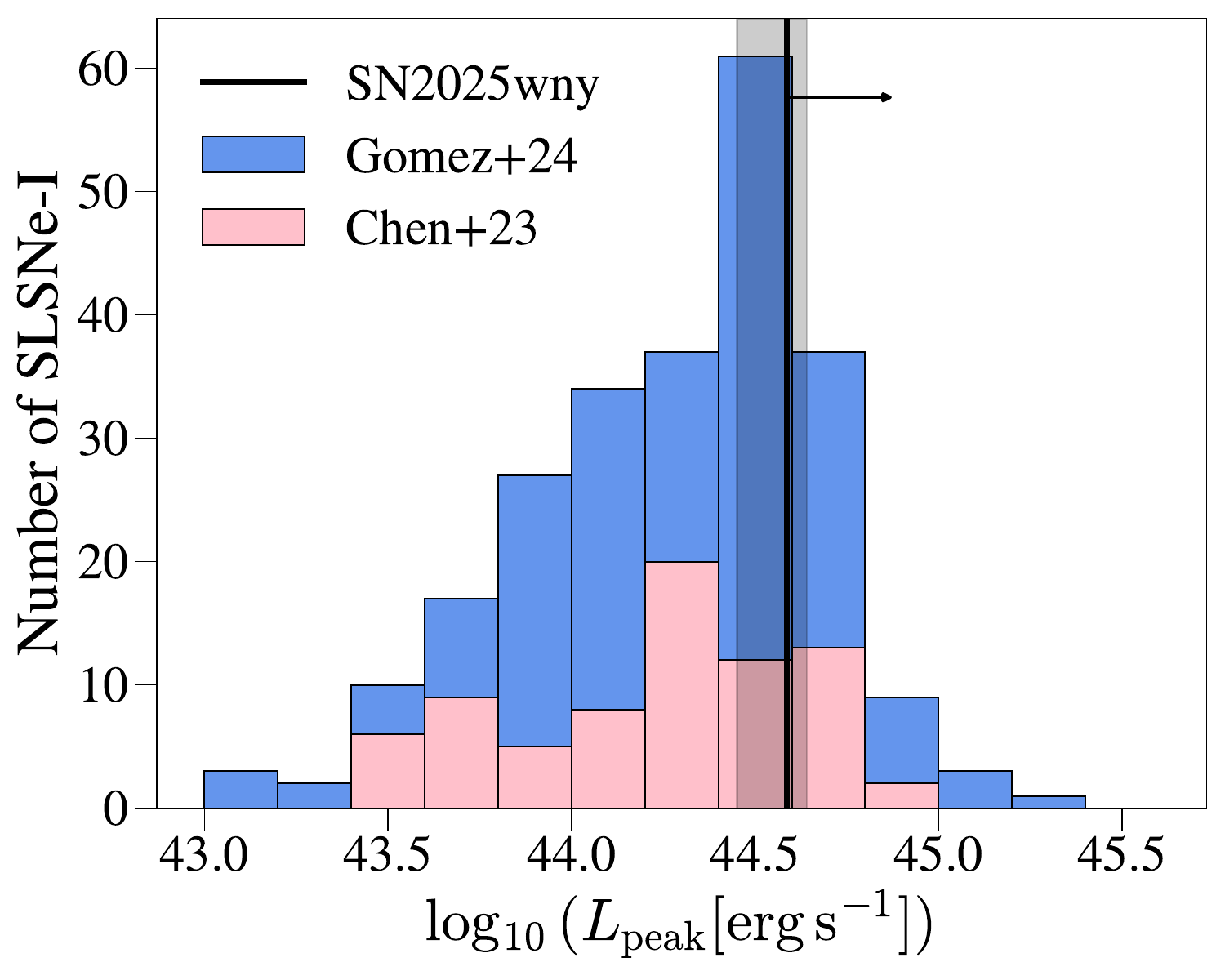}
    \caption{\textit{Top:} Comparison of the photometrically derived and spectroscopically derived pseudo-bolometric light curves of SN\,2025wny (black) with the Gold and Silver SLSN-I samples from \citet{Gomez2024_SLSNeI}. \textit{Bottom:} Histogram of peak pseudo-bolometric luminosities of SLSNe-I from \citet{Chen2023_PopulationStudy} (pink) and from the Gold and Silver samples from \citet{Gomez2024_SLSNeI} (blue). The inferred peak pseudo-bolometric luminosity of SN\,2025wny is shown in black. The most luminous events ($\gtrsim10^{45}$\,erg\,s$^{-1}$) as reported in \citet{Gomez2024_SLSNeI} are likely unreliable. Since the luminosity is calculated by integrating only the observed wavelength range, the value for SN\,2025wny represents a lower limit. Peak luminosities are reported as the 68\% credible interval. We demagnify SN\,2025wny using the methodology in Section~\ref{sec:magnification-correction}.}
    \label{fig:bol-lc-compare}
\end{figure}

We emphasize that these pseudo-bolometric luminosities likely underestimate the total bolometric output of the event, since the integration is restricted to the rest-frame wavelength range 1500--4230\,{\AA}.

\begin{figure}[t]
    \centering
    \includegraphics[width=\linewidth]{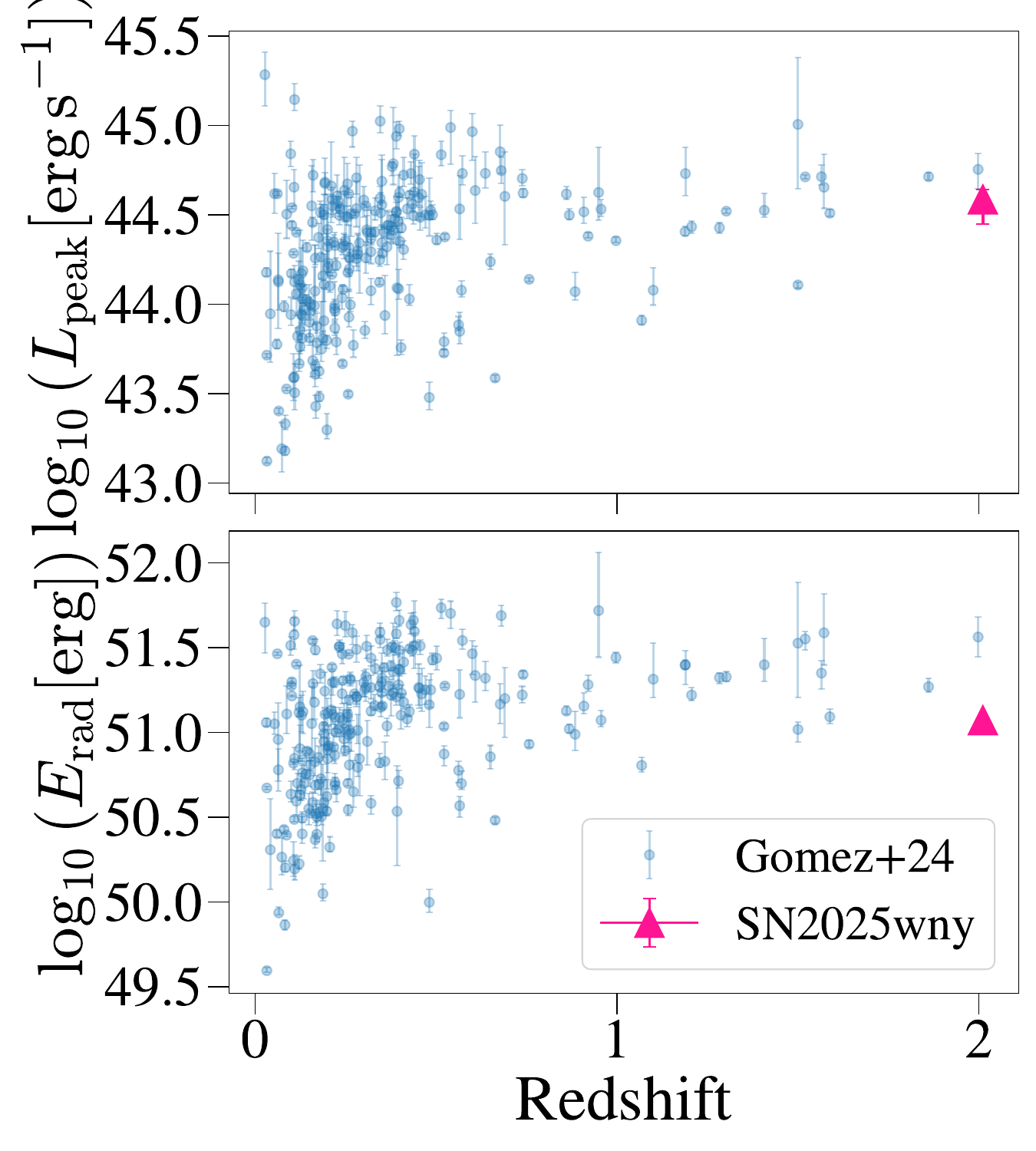}
    \caption{Peak pseudo-bolometric luminosity (top) and integrated radiated energy (bottom) of the Gold and Silver SLSN-I samples from \citet{Gomez2024_SLSNeI} as a function of redshift. Overplotted are the lower limits inferred from the photometric pseudo-bolometric light curve of SN\,2025wny (pink); the spectra-derived estimate is almost identical. The most luminous events ($\gtrsim10^{45}$\,erg\,s$^{-1}$) as reported in \citet{Gomez2024_SLSNeI} are likely unreliable. Peak luminosities are reported as the 68\% credible interval. We demagnify SN\,2025wny using the methodology in Section~\ref{sec:magnification-correction}.}
    \label{fig:Lbol_Erad_vs_redshift}
\end{figure}

\begin{figure}
    \centering
    \includegraphics[width=\linewidth]{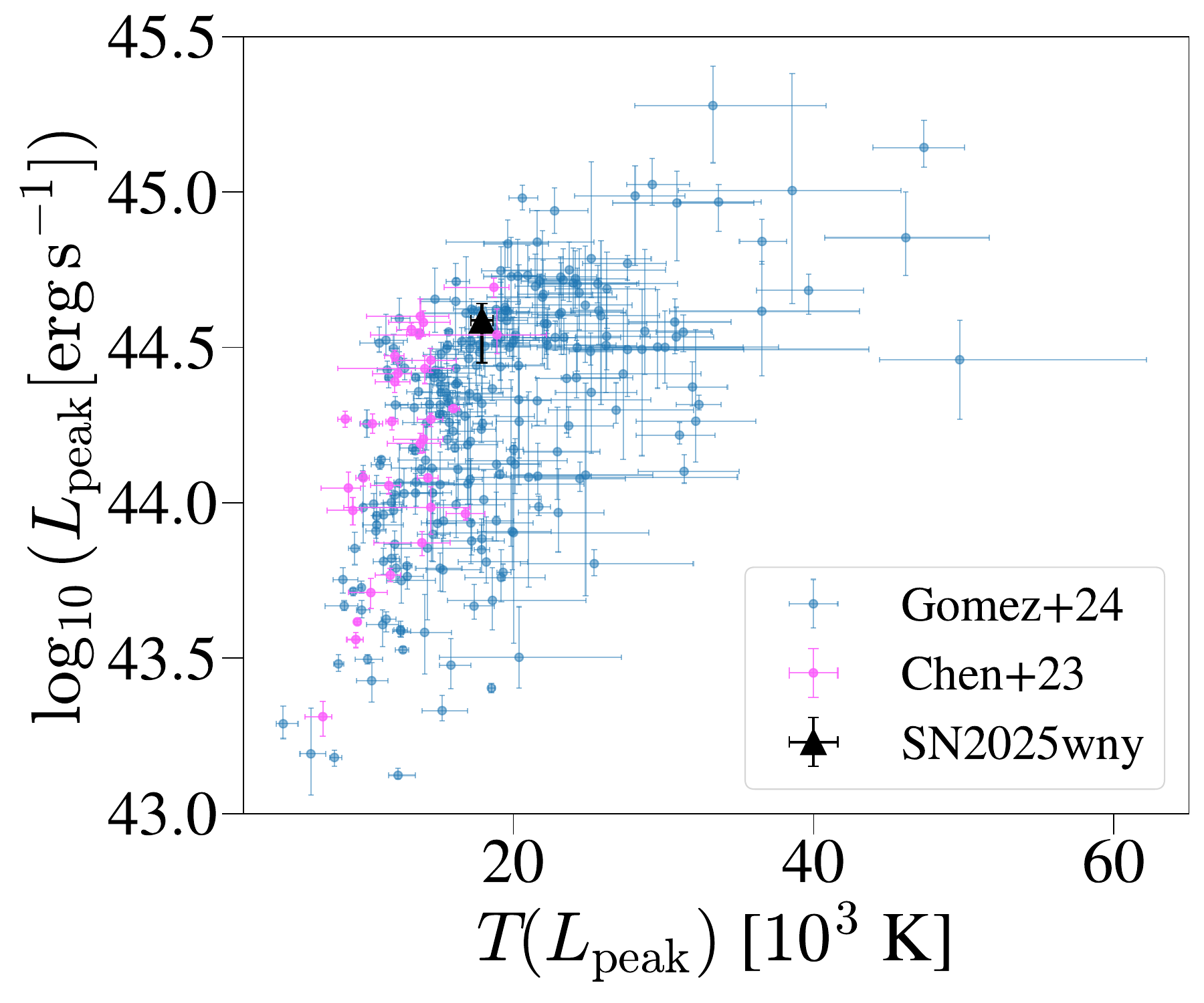}
    \caption{Pseudo-bolometric luminosities and blackbody temperatures of the SLSNe-I from the \citet{Chen2023_PopulationStudy} (purple) and Gold and Silver  samples from \citet{Gomez2024_SLSNeI} (blue). Overplotted are the lower limits inferred from the photometric pseudo-bolometric light curve of SN\,2025wny (pink); the spectra-derived estimate is almost identical. The most luminous events ($\gtrsim10^{45}$\,erg\,s$^{-1}$) as reported in \citet{Gomez2024_SLSNeI} are likely unreliable. Peak luminosities are reported as the 68\% credible interval. We demagnify SN\,2025wny using the methodology in Section~\ref{sec:magnification-correction}.}
    \label{fig:lum-vs-temp-dist}
\end{figure}

Finally, we correct for magnification following the methodology in Section~\ref{sec:magnification-correction} and report the 16th-50th-85th percentiles on the pseudo-bolometric luminosities, shown in Figure~\ref{fig:bolometric-lc}. Both methods yield broadly consistent results, declining with an $e$-folding time of $\tau\sim30$\,d until $+20$\,d post-peak and then flatten to $\tau\sim50$\,d. Peak luminosities and lower limits on the radiated energy, measured by integrating the photometry-derived pseudo-bolometric light curve over 0--70\,d, are listed in Table~\ref{tab:bol-lc-results}.

The spectra-derived luminosities are somewhat systematically higher, which likely reflects biased spectral blackbody extrapolations. Unlike the photometry, the spectra extend only to rest-frame $\sim3300$\,{\AA} and lack the rest-frame optical anchor provided by the observed $J$ band. As a result, the spectral blackbody fits are biased to the observed $gri$ bands (rest-frame FUV), likely overestimating the red tail of the SED. For this reason, we do not show luminosities computed directly from the spectral blackbody parameters ($L_\mathrm{BB}=4\pi\sigma_{\mathrm{SB}}R^2T^4$).

We compare the pseudo-bolometric light curve results to the \citet{Gomez2024_SLSNeI} and \citet{Chen2023_PopulationStudy} SLSN-I samples, which also report 16th-50th-84th percentile results. We restrict the comparison to the Gold and Silver samples from \citet{Gomez2024_SLSNeI}, excluding events that lack a spectroscopic SLSN classification. We caution that the most luminous events reported in \citet{Gomez2024_SLSNeI} ($\gtrsim10^{45}$\,erg\,s$^{-1}$) appear unreliable, as their reported peak luminosities are inconsistent with measurements made in the corresponding discovery papers. We also note that most comparison events lack well-sampled rest-frame UV observations, especially at late phases, which may lead to underestimated pseudo-bolometric luminosities and radiated energies.

As shown in Figures~\ref{fig:bol-lc-compare} and \ref{fig:Lbol_Erad_vs_redshift}, after demagnification, SN\,2025wny falls within the typical luminosity and radiated energy ranges of local SLSNe-I. SN\,2025wny also remains consistent with the trend in Figure~\ref{fig:lum-vs-temp-dist}, which shows a broad correlation between peak luminosity and blackbody temperature, with hotter events being more luminous. 


\input{tables/lbol-priors.tex}

\section{Light curve modeling}
\label{sec:light-curve-modeling}
\subsection{Semi-analytic bolometric fits}

We fit the demagnified photometric pseudo-bolometric light curve with simple semi-analytic luminosity models, following the methods of \citet{2018ApJ...867L..31C}. The model time is measured as rest-frame days since explosion, as follows:

\begin{equation}
t = t' + t_{\rm rise}.
\end{equation}

where $t'$ are the rest-frame days since the pseudo-bolometric peak $\mathrm{MJD}_{\rm peak}=60940.7$. We set $t_{\rm rise}$ as a free parameter. We consider magnetar spin-down, radioactive
$^{56}$Ni decay, and ejecta--circumstellar-material (CSM) interaction. The magnetar input goes as \citep{1998A&A...333L..87D, 2001ApJ...552L..35Z}

\begin{equation}
L_{\rm mag}(t) =
\frac{E_{\rm p}/t_{\rm p}}{(1+t/t_{\rm p})^2},
\end{equation}
where $E_{\rm p}$ is the initial rotational energy and $t_{\rm p}$ is the
spin-down timescale. These quantities are related to the magnetar initial spin
period, $P$, and the dipole magnetic field strength, $B$, by

\begin{equation}
\begin{split}
E_{\rm p} &= \frac{1}{2} I \Omega^2 \\
&\approx 2.0\times10^{52}
\left(\frac{P}{1\,{\rm ms}}\right)^{-2}
\left(\frac{I}{10^{45}\,{\rm g\,cm^2}}\right)\,{\rm erg},
\end{split}
\end{equation}

and

\begin{equation}
\begin{split}
t_{\rm p} \approx
1.3\times10^{5}
\left(\frac{B}{10^{14}\,{\rm G}}\right)^{-2}
\left(\frac{P}{1\,{\rm ms}}\right)^2 \\
\times \left(\frac{R_{\rm NS}}{10^6\,{\rm cm}}\right)^{-6}
\left(\frac{I}{10^{45}\,{\rm g\,cm^2}}\right)
\,{\rm sec}.
\end{split}
\end{equation}

\begin{figure}
    \centering
    \includegraphics[width=\linewidth]{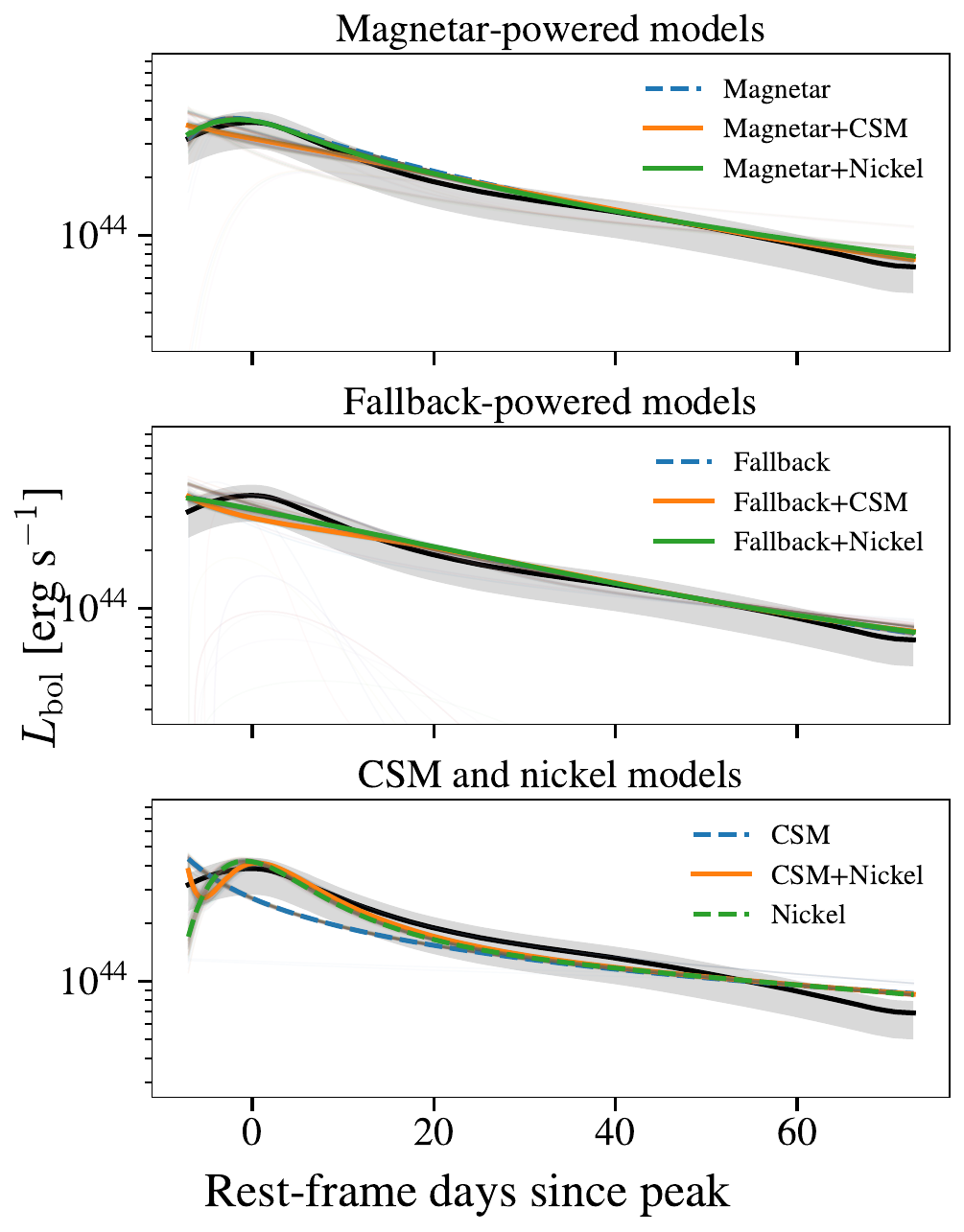}
    \caption{Posterior draws from various model fits to the demagnified pseudo-bolometric light curve (black). The shaded regions indicate the 68\% credible interval of the pseudo-bolometric light curve.}
    \label{fig:bolometric_model_fits}
\end{figure}

The radioactive component follows the standard
$^{56}$Ni--$^{56}$Co decay heating rate scaled by $M_{\rm Ni}$. For the CSM
interaction component, we used a simple steady-wind motivated form,

\begin{equation}
L_{\rm CSM}(t) = L_{\rm CSM,1}\,t^{-3/5},
\end{equation}

where the $-3/5$ scaling corresponds to the steady-wind case for an outer ejecta density profile $\rho_{\rm ej}\propto r^{-7}$ and a constant-density inner ejecta structure \citep{Moriya2013}. We do not fit the full physical CSM interaction model parameters, and instead treat $L_{\rm CSM,1}$ as a free normalization. Following the Arnett model \citep{1982ApJ...253..785A}, the magnetar and radioactive-decay inputs were converted to luminosities using a single effective diffusion timescale
$t_{\rm diff}$. This accounts for photon diffusion through the ejecta but does not model the ejecta structure or radiation transport.

Finally, we also consider fallback accretion, which we parameterize as \citep{Dexter2013_FallbackAccretion}:

\begin{equation}
    L_{\rm fallback}(t) = L_{\rm fallback,1}
    \left(\frac{t}{1~{\rm s}}\right)^{-5/3},
\end{equation}

where $L_{\rm fallback,1}$ is a normalization constant.

We fit nine models: magnetar, fallback, CSM interaction, nickel decay, magnetar with CSM interaction, magnetar with nickel decay, fallback with CSM interaction, fallback with nickel decay, and CSM with nickel decay. In addition to the parameters for each power source, all models included the diffusion timescale, $t_{\rm diff}$, the rise time from explosion to bolometric maximum, $t_{\rm rise}$, and an extra scatter term, $\sigma_{\rm frac}$ which helps account for uncertainty that may have not been considered when deriving the pseudo-bolometric light curve. 

\begin{figure}
    \centering
    \includegraphics[width=\linewidth]{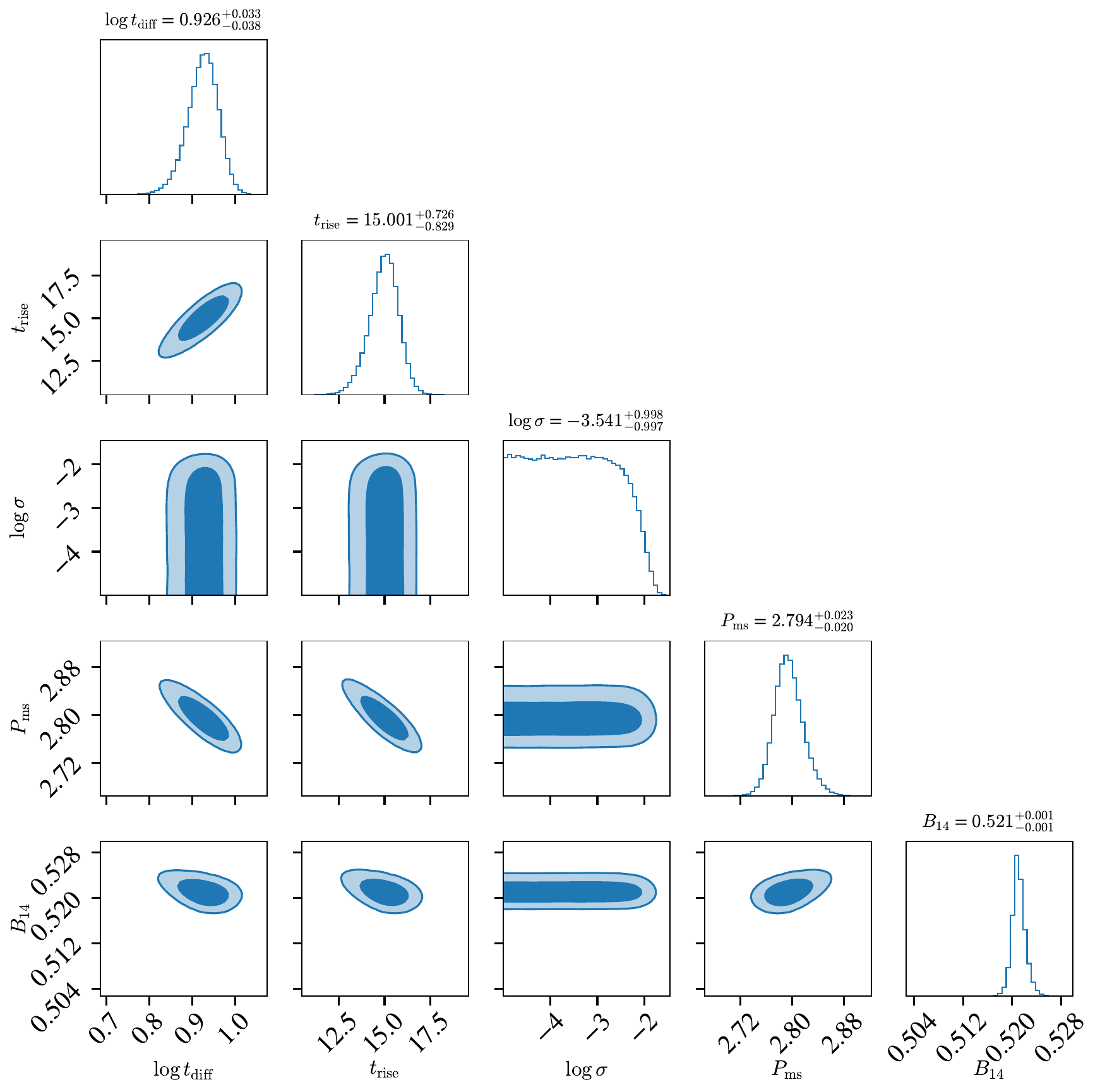}
    \caption{Corner plot of the magnetar-only pseudo-bolometric fit. Parameter values are given as the 16th-50th-84th percentiles. Plotted are the 68\% and 95\% levels.}
    \label{fig:magnetar_corner}
\end{figure}

We performed MCMC fitting with the uniform priors listed in Table~\ref{tab:bolometric-fit-priors}. For each model, we used 64 walkers, 10,000 burn-in steps, and 20,000 iterations. We show the resulting fits in Figure~\ref{fig:bolometric_model_fits}. The best fits generally require a magnetar component, either as the sole power source or in combination with CSM-interaction or radioactive $^{56}$Ni decay. For clarity, we show the magnetar-only corner plot in Figure~\ref{fig:magnetar_corner} and report the 68\% credible intervals for the inferred magnetar parameters. The fit yields plausible values of $P \approx 2.8$\,ms and $B\approx5\times10^{13}$\,G, consistent with the range typically inferred for SLSNe-I in \citet{Gomez2024_SLSNeI}.

These results provide a simple first-pass test of whether the luminosity evolution can be reproduced by a single power source or instead favors multiple contributing components. However, importantly, these fits are not self consistent; they performed on the empirical pseudo-bolometric light curve, which integrates only over 1500--4230\,{\AA} and therefore does not capture the full photometric evolution or color information of SN\,2025wny. Moreover, this simple semi-analytic prescription does not allow us to infer ejecta or CSM properties. We therefore also model the data with the \textsc{Redback} code \citep{Sarin2024a}, which supports multi-band fitting and more detailed models capable of constraining more physical parameters.

\subsection{Modeling with redback}

\begin{figure}
    \centering
    \includegraphics[width=\linewidth]{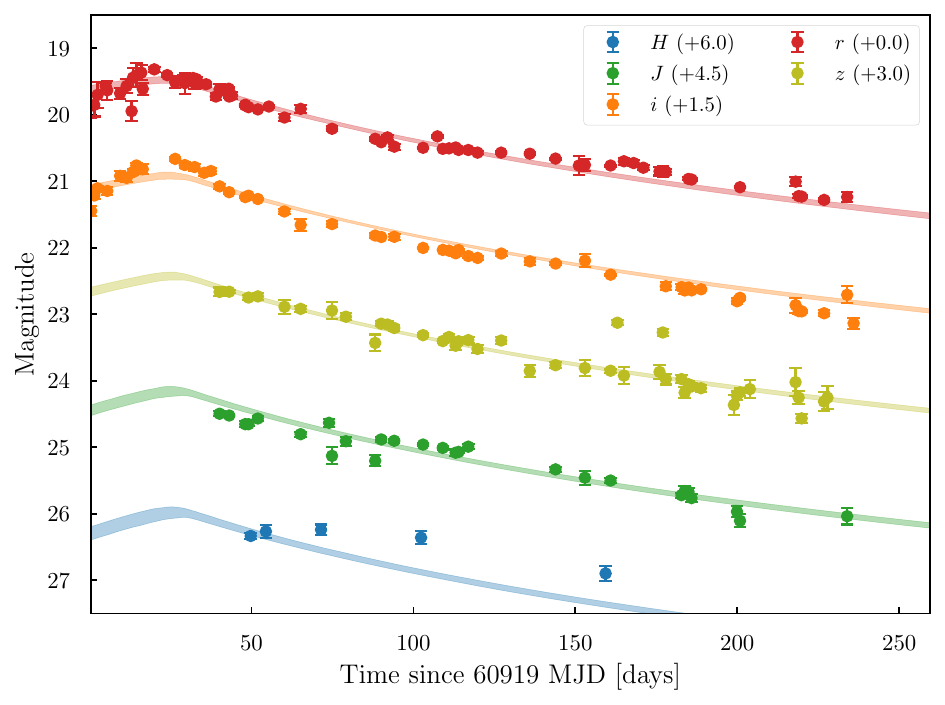}
    \caption{The $grizJH$ light curve of SN\,2025wny fit to a two-component model combining fallback accretion and radioactive decay of $^{56}$Ni, implemented using \textsc{Redback}. The magnification factor is set free. The shaded band indicates the 90\% credible interval from our posteriors for each band. The best-fit parameters include a magnification of 11, ejecta mass of $6$\,M$_\odot$, and a nickel mass of $\sim0.12$\,M$_\odot$. The model provides a reasonable match to the observed data, but the fit should be interpreted with caution due to the limited wavelength coverage and the unusual properties of SN\,2025wny.}
    \label{fig:redback-lightcurve-result}
\end{figure}

We fit the $grizJH$ data using the \textsc{Redback} package \citep{Sarin2024a} (version 1.17.0), which includes implementations of magnetar~\citep{Sarin2022, Omand2024}, ejecta interactoin with massive C/O-rich CSM \citep{Chatzopoulos2012}, radioactive nickel decay~\citep{1982ApJ...253..785A}, and fallback accretion powering mechanisms~\citep{Moriya2018_Models}. We assume a simple demagnification correction (division by $|\mu|$) and use a prior on $\mu$ informed by the lens modeling and parameter ranges informed by low-$z$ studies (default wide priors in \textsc{Redback}), sampling the posterior with the \texttt{pymultinest} sampler~\citep{Feroz2009} implemented in \textsc{Bilby}~\citep{Ashton2019}. We find that none of the individual models provides an acceptable fit. Magnetar and CSM models can reproduce the data only if the magnification is increased to $\sim100$, which is implausibly large and inconsistent with the posterior constraints reported by \citet{Mortsell2026}. We therefore rule out single-mechanism explanations under realistic magnification (at least with the same set of SED assumptions as is typical for nearby SLSNe).

\begin{figure}
    \centering
    \includegraphics[width=\linewidth]{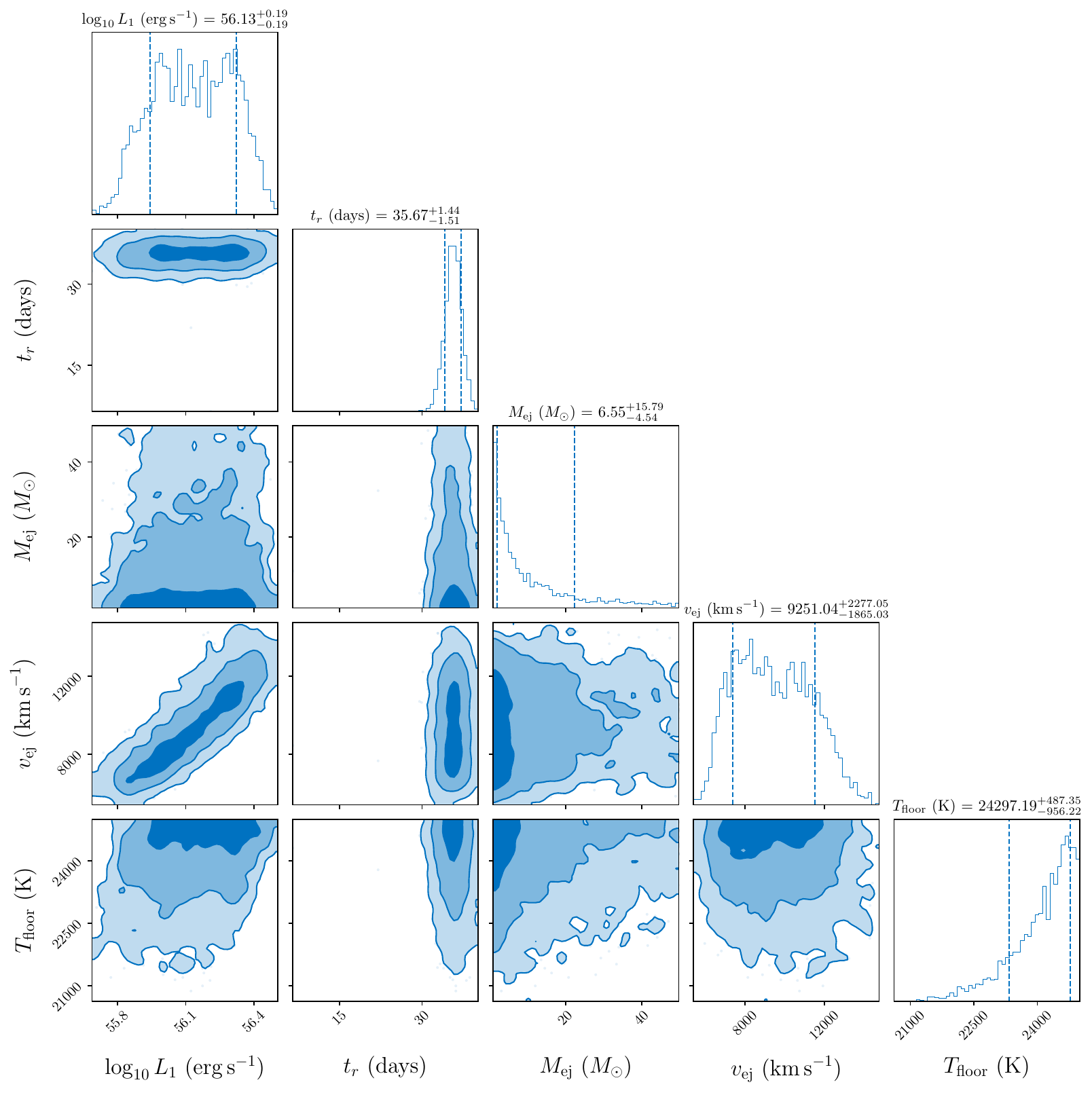}
    \caption{Corner plots (68\% credible interval) of the engine parameters of the two-component fallback accretion and radioactive nickel decay fit, implemented with \textsc{Redback}.}
    \label{fig:redback-corner1}
\end{figure}


This inconsistency is a by-product of a few competing effects. First, the implied temperature and spectral evolution of SN\,2025wny strongly suggest no UV suppression and a blackbody temperature above $\sim20{,}000$\,K for a significant period. Second, despite this high-temperature plateau, the light curve declines rapidly and, in the source frame, faster than usually predicted by these models. To compensate between these two effects, the sampler prefers a solution where the SN is intrinsically fainter and magnified more intensely to the observed magnitudes. These competing effects help explain why the bolometric light curve is described adequately by various models, but the same models fail to reproduce the SN with a consistent magnification. More flexible CSM interaction models with eruptive mass-loss histories~\citep{Sarin2026_csm} do not alleviate this tension as they are not hot enough over all epochs implied by the photometry and spectra. This preliminary analysis suggests the SED evolution of SN\,2025wny is significantly different from SLSNe in the local universe, where these models are better calibrated.

\begin{figure}
    \centering
    \includegraphics[width=\linewidth]{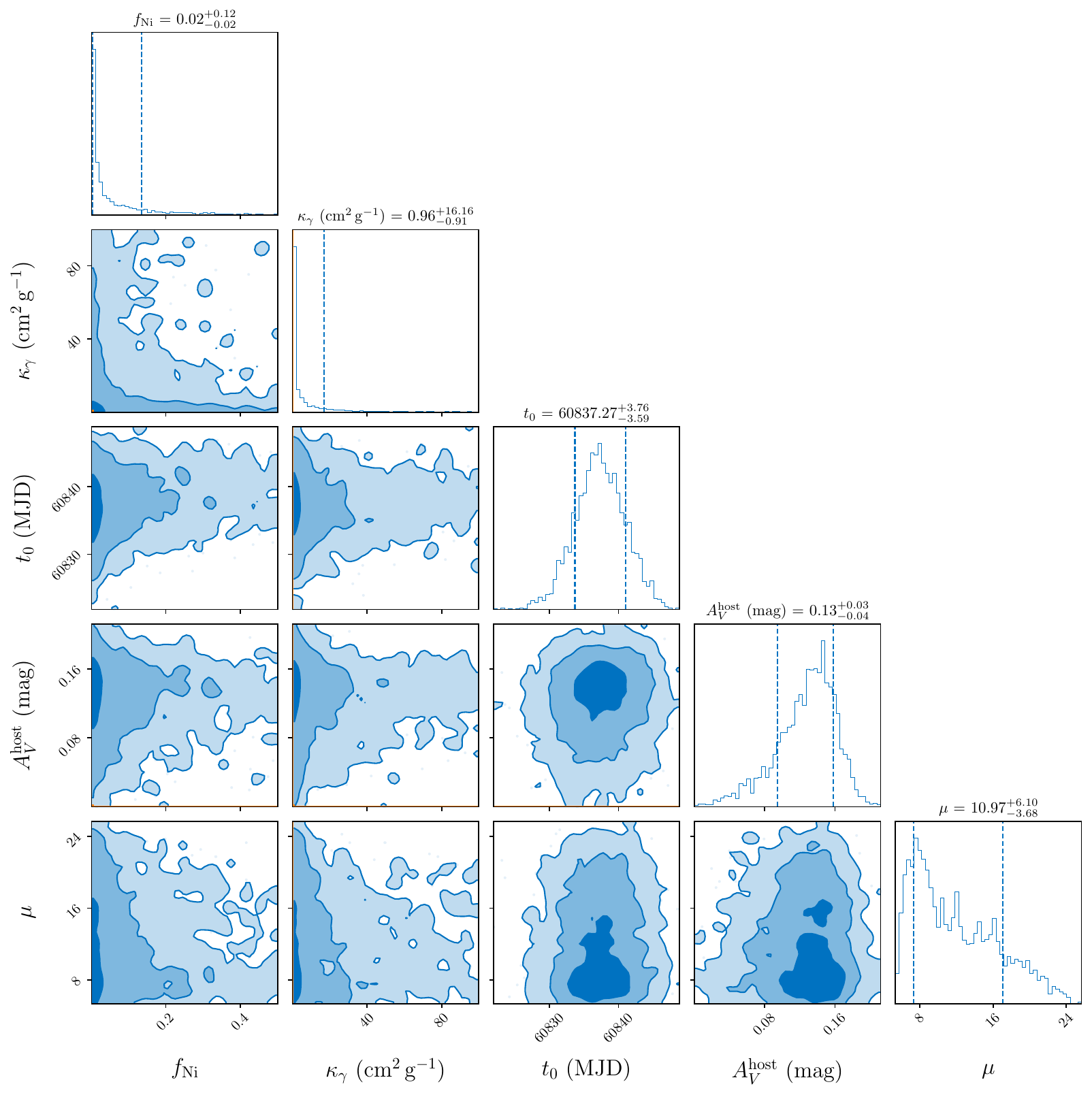}
    \caption{Corner plots (68\% credible interval) of the nuisance parameters of the two-component fallback accretion and radioactive nickel decay fit, implemented with \textsc{Redback}.}
    \label{fig:redback-corner2}
\end{figure}

Since the spectra instead point toward multiple emission components and, potentially, multiple powering mechanisms, we also explored two-component scenarios. However, we did not attempt a combined magnetar plus CSM-interaction fit in \textsc{Redback}. A self-consistent magnetar and CSM model requires co-evolving the CSM-shock solution alongside the evolving energy-injection from the magnetar. This is non-trivial and not feasible to implement in the scope of this work.

One physically motivated case for a two-component power source involves a very massive progenitor that forms a black hole with incomplete fallback of synthesized nickel. In this scenario, the emission would arise from a combination of fallback accretion and the decay of $^{56}$Ni to $^{56}$Co and $^{56}$Co to $^{56}$Fe. Figure~\ref{fig:redback-lightcurve-result} shows the fit combining the fallback model of \citet{Moriya2018_Models} with a $^{56}$Ni decay component \citep{1982ApJ...253..785A}, with Figures~\ref{fig:redback-corner1} and \ref{fig:redback-corner2} displaying the posterior distributions of the fit. The fit yields an ejecta mass of $6^{+15}_{-4}$\,M$_\odot$, a nickel fraction of $f_{\rm Ni} = 0.02^{+0.11}_{-0.02}$ (corresponding to a nickel mass of $\sim0.12$\,M$_\odot$), a photospheric velocity of $9100^{+2300}_{-1900}$\,km\,s$^{-1}$, and a lensing magnification of $\mu = -10.9^{+3.6}_{-6.1}$, consistent with the independent constraint from the lens model of \citet{Mortsell2026}.

These results should be interpreted with caution. First, we stress that the observed $griz$ photometry probes the rest-frame UV ($\sim1500$--2990\,{\AA}), while the $J$ and $H$ bands approximately sample the rest-frame $g$ and $V$ bands. As such, the data span only a narrow rest-frame wavelength interval around the peak of the blackbody emission. Second, the apparent success of this multi-component model relies on exploiting the model flexibility to simultaneously reproduce the unusual properties of SN\,2025wny, including the relatively modest ($\sim10\times$) magnification, the apparent lack of UV suppression, the overall light-curve morphology, and the limited wavelength coverage. In particular, the best-fit transition timescale ($t_r \sim 36$ days) is significantly longer than expected for physical fallback accretion, where debris circularization occurs on timescales of days. Furthermore, the inferred fallback luminosity scale, $\log_{10} L_1 = 56.1^{+0.18}{-0.21}$\,erg\,s$^{-1}$, exceeds the range found by \citet{Moriya2018_Models} for a sample of 36 low-redshift SLSNe ($\log_{10} L_1 = 54.4$-$55.8$\,erg\,s$^{-1}$), further suggesting that the model is being driven to unphysical regions of parameter space to compensate for the unusual SED and light-curve properties of SN\,2025wny. Similarly, the photosphere temperature plateaus at an incredibly hot, $T \sim {25,000}$K.


The light-curve fitting presented here should not be interpreted as a definitive determination of SN\,2025wny's powering mechanism or progenitor. Instead, it serves to explore the parameter space of the powering mechanisms commonly invoked for well-observed, low-redshift SLSNe, and highlight that under a motivated prior on magnification, these same models struggle to explain the properties of SN\,2025wny.

\begin{figure*}
    \centering
    \includegraphics[width=\linewidth]{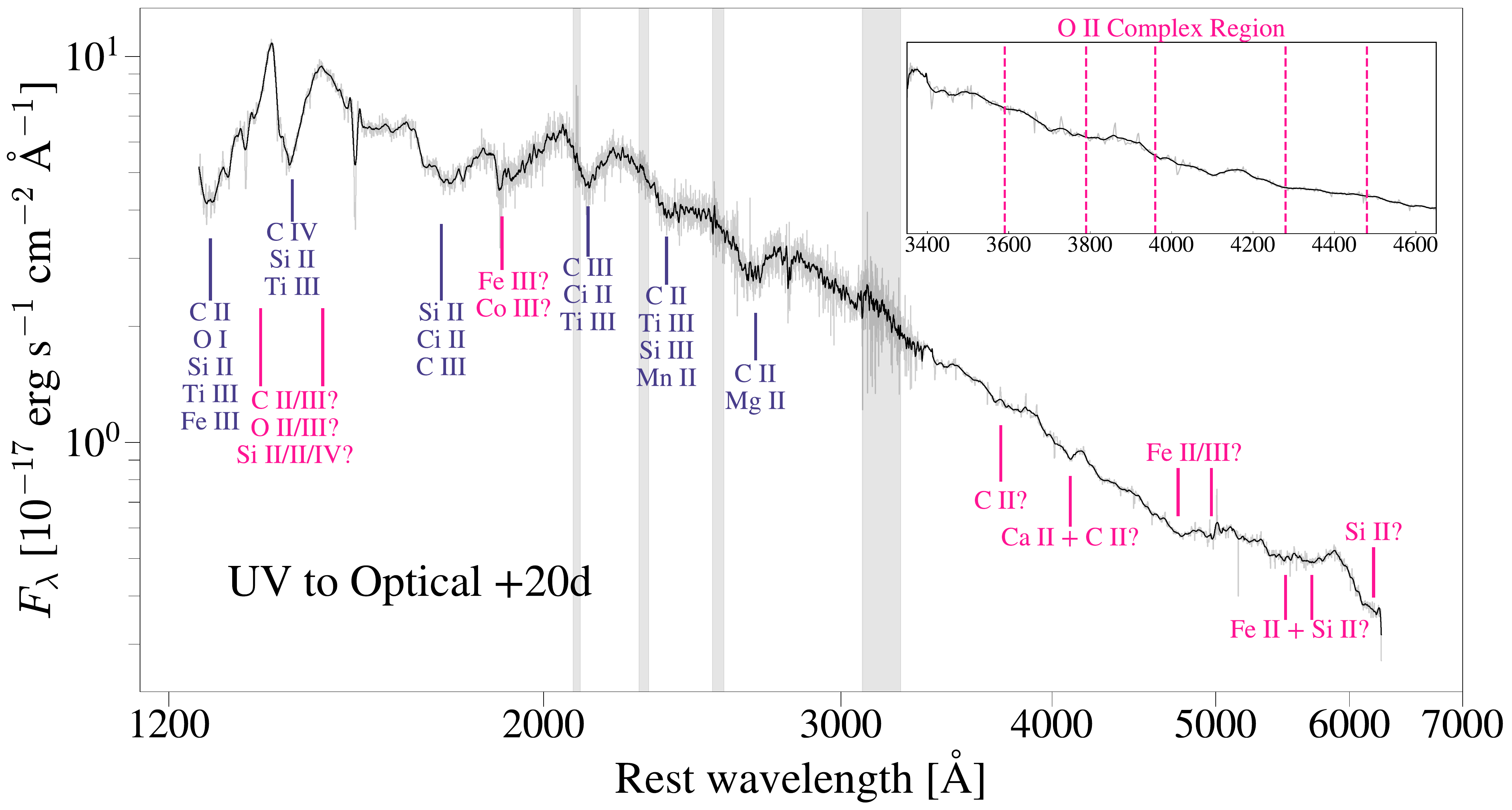}
    \caption{Rest-frame UV-to-optical spectrum of SN\,2025wny at $+20$\,d, constructed with the $+29$\,d LRIS and $+20$\,d \textit{JWST} spectra. Line identifications are labeled, with uncertain features in pink; grey shaded regions are telluric absorption bands. The inset shows the candidate \ion{O}{2} region. We correct for Milky Way extinction. No magnification correction is applied.}
    \label{fig:uv_to_opt_20d}
\end{figure*}

\begin{figure*}
    \centering
    \includegraphics[width=\linewidth]{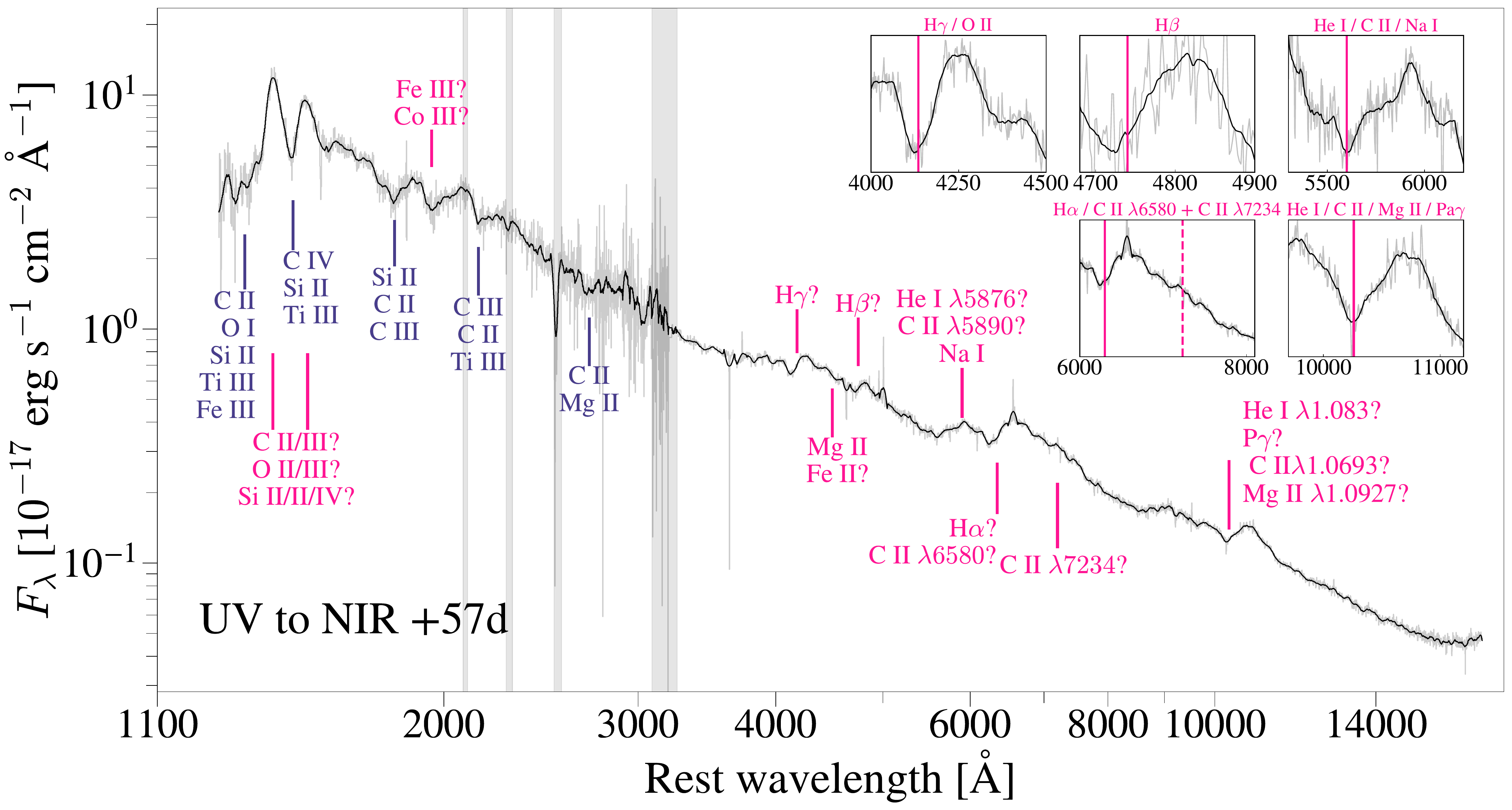}
    \caption{Rest-frame UV-to-NIR spectrum of SN\,2025wny at $+57$\,d, constructed using the $+57$\,d FORS2 and \textit{JWST} spectra. Line identifications are labeled, with uncertain features in pink; grey shaded regions indicate telluric absorption bands. Insets show regions with candidate features. We correct for Milky Way extinction. No magnification correction is applied.}
    \label{fig:uv_to_nir_58d}
\end{figure*}

\section{Spectral analysis}
\label{sec:spectra}


\subsection{Composite spectra}
We present the first contemporaneous rest-frame UV-to-optical and UV-to-NIR spectral templates of a SLSN-I at $z\approx2$ by constructing stitched spectra at $+20$\,d and $+57$\,d, shown in Figures~\ref{fig:uv_to_opt_20d} and \ref{fig:uv_to_nir_58d}. The spectra are combined by scaling overlapping wavelength regions and calibrating the final spectra onto an absolute flux scale using the FTW/WWFI photometry. For the $+20$\,d composite, we combine the \textit{JWST} spectrum with the +29\,d Keck/LRIS spectrum, since no closer UV spectrum with overlapping wavelength coverage is available; this phase offset is shorter than the rest-frame UV $e$-folding times discussed in Section~\ref{subsec:lc_evolution}. We also construct a composite between the $+20$\,d \textit{JWST}/NIRSpec spectrum and the $+10$\,d Keck/LRIS spectrum for continuum analysis, shown in Section~\ref{subsubsec:uv-excess}. For the $+57$\,d spectrum, we combine the contemporaneous \textit{JWST}/NIRSpec and VLT/FORS2 spectra.

\begin{figure}
    \centering
    \includegraphics[width=\linewidth]{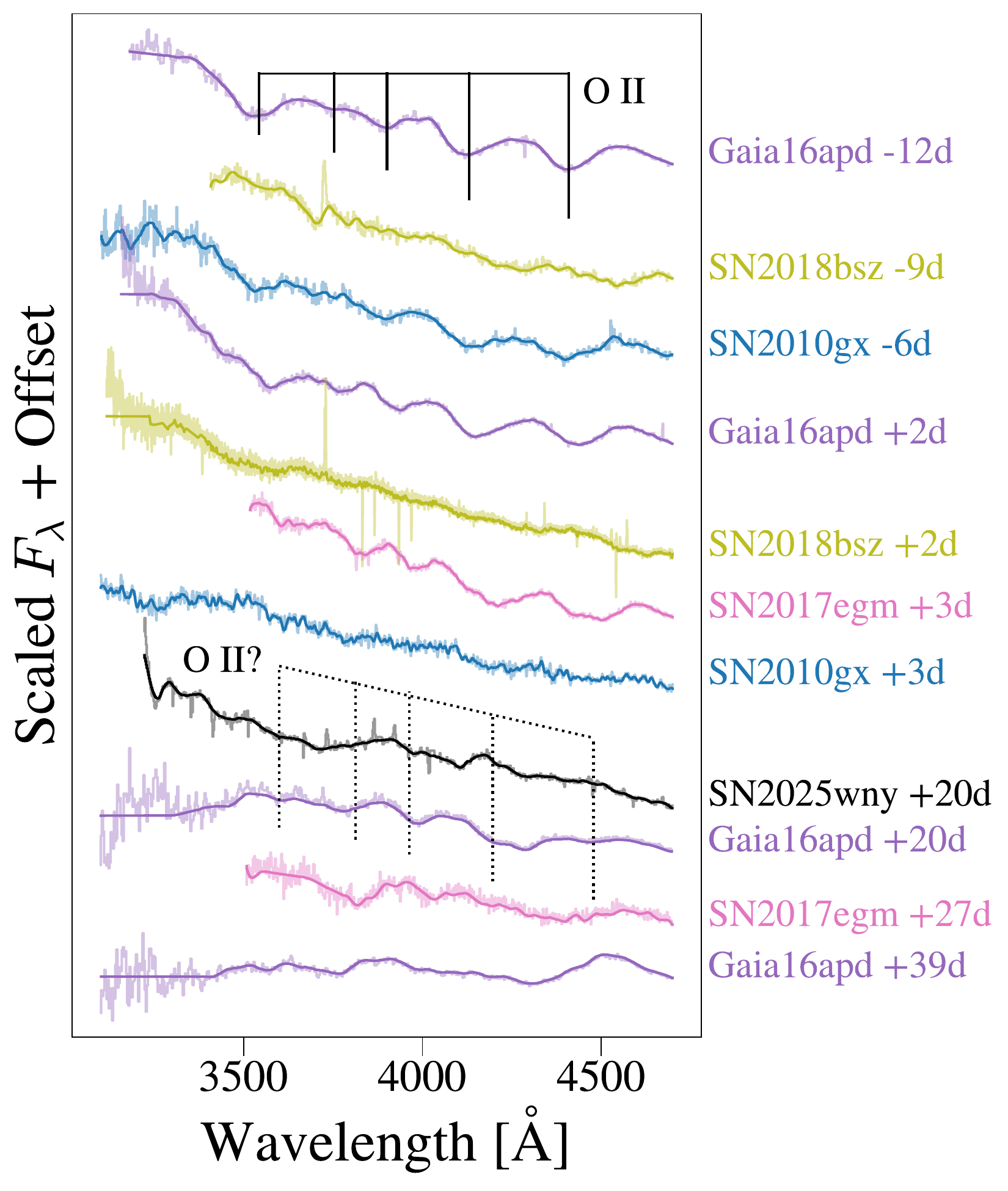}
    \caption{Optical spectral evolution of several SLSNe-I, including Gaia16apd (purple; \citealt{2017ApJ...840...57Y, 2017ApJ...840...57Y}) and SN\,2017egm (pink; \citealt{2017ApJ...845L...8N, 2018ApJ...858...91Y, Zhu2023_SN2017egm}), the fast-evolving SLSN-I SN\,2010gx (blue; \citealt{2010ApJ...724L..16P}), and the carbon-rich SLSN-I SN\,2018bsz (green; \citealt{2022AnA...666A..30P}). SN\,2025wny is shown in black. Labeled features indicate possible line identifications, including the potentially ambiguous \ion{O}{2} absorption complex.}
    \label{fig:20d_compare}
\end{figure}

In the following analysis, blueshift velocities are defined relative to absorption minima unless otherwise specified, and all line widths (FWHMs) are obtained from Gaussian or P Cygni profile fits. We present major line identifications and velocity measurements of the $+20$\,d and $+57$\,d stitched spectra in Appendix~\ref{app:line-profiles}.

\subsection{Ambiguous \ion{O}{2} in the $+20$\,d optical spectrum}

The $+20$\,d optical spectrum (Figure~\ref{fig:uv_to_opt_20d}) covers $\approx$3000--6000\,{\AA} in the rest frame.
As shown in Figure~\ref{fig:20d_compare}, the spectrum resembles those of early photospheric SLSNe-I, particularly Gaia16apd at a similar phase \citep{2017ApJ...840...57Y,2017MNRAS.469.1246K}. We therefore adopt the line identifications proposed by \citet{2017MNRAS.469.1246K} and \citet{2016ApJ...826...39N}, including \ion{Ca}{2} H\&K, Fe II, and possibly Fe III and Si II, as labeled in Figure~\ref{fig:uv_to_opt_20d}. We fit velocities to the tentative \ion{Fe}{2}/\ion{Fe}{3} feature as it is the most clearly identifiable absorption trough despite likely blending. If the redward component at 4900--5100\,{\AA} is associated with \ion{Fe}{2} $\lambda$5169, the absorption minimum implies a photospheric velocity of $\approx$12{,}700\,km\,s$^{-1}$.

We also investigate the possible presence of \ion{O}{2} in the $+20$\,d spectrum. In typical SLSNe-I, a blend of \ion{O}{2} lines produces five strong absorption troughs at effective wavelengths 3737.59\,{\AA}, 3959.83\,{\AA}, 4115.17\,{\AA}, 4357.97\,{\AA}, and 4650.71\,{\AA}, including a prominent W-shaped feature between 4300--4700\,{\AA}. These features are clearly visible in objects such as Gaia16apd \citep{2017ApJ...840...57Y} and SN\,2017egm \citep{2018ApJ...858...91Y} during the pre-peak to near-peak phases, and weaken as the ejecta cool and the ionization state decreases. As shown in Figure~\ref{fig:20d_compare}, slowly evolving events like Gaia16apd retain \ion{O}{2} absorption until $+20$\,d, whereas rapidly evolving events like SN\,2010gx lose their \ion{O}{2} features by $+3$\,d \citep[their Figure 16]{2015MNRAS.452.3869N}. 

\begin{figure*}
    \centering
    \includegraphics[width=\linewidth]{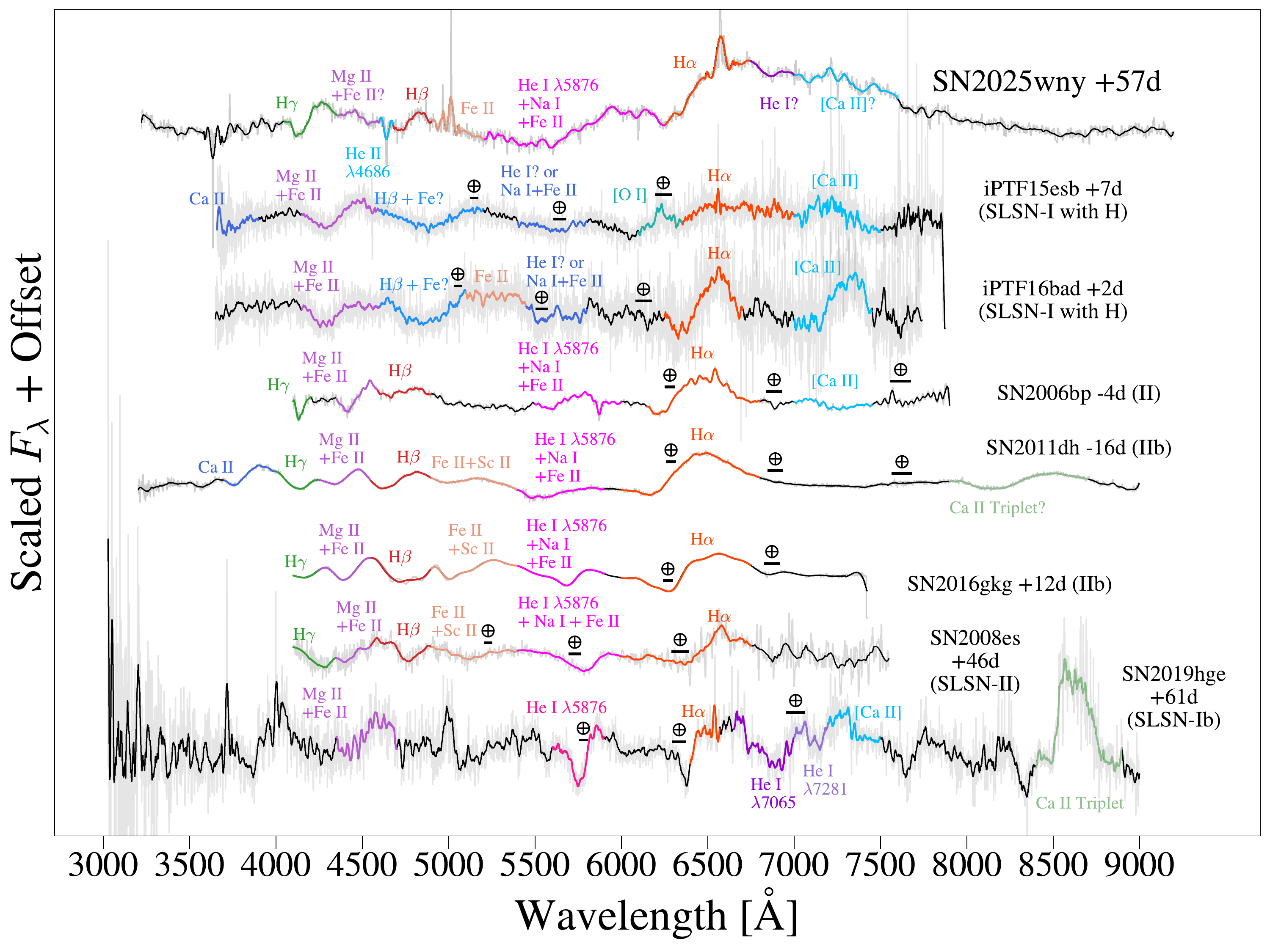}
    \caption{De-redshifted and continuum-normalized optical spectral comparison of SN\,2025wny at $+57$\,d (top) with hydrogen- and helium-rich SNe, including SLSNe-I with late-time hydrogen (iPTF15esb, iPTF16bad; \citealt{2017ApJ...848....6Y}), Type~II SN\,2006bp \citep{2007ApJ...666.1093Q}, Type~IIb SNe SN\,2011dh \citep{2011ApJ...742L..18A} and SN\,2016gkg \citep{2018Natur.554..497B}, SLSN-II SN\,2008es \citep{2009ApJ...690.1303M, 2009ApJ...690.1313G}, and SLSN-Ib SN\,2019hge \citep{Yan2020_HeI}. Smoothed versions are overlaid in black and prominent spectral features are highlighted in color. Telluric absorption bands are marked with $\oplus$ symbols.}
    \label{fig:jwst_58d_comparison}
\end{figure*}

\citet{2024ApJ...967...13S} show that photospheric temperatures of $\sim$14{,}000--16{,}000\,K are sufficient to produce \ion{O}{2} absorption complexes. From the bolometric and continuum modeling (Sections~\ref{subsec:bol-lc} and \ref{subsubsec:uv-excess}), SN\,2025wny at $+20$\,d should therefore be hot enough to sustain \ion{O}{2} line formation. However, unlike typical SLSNe-I at comparable temperatures, the spectrum does not exhibit a clear \ion{O}{2} complex. We discuss this further in Section~\ref{subsec:line-implications}.


If \ion{O}{2} is present, we consider the 3500--3900\,{\AA} absorption trough in SN\,2025wny to plausibly include contributions from \ion{O}{2} $\lambda$3737.59 and $\lambda$3959.83, together with \ion{C}{2} $\lambda$3921, which commonly contributes to this region. The resulting blueshifts are $\sim$14{,}200--17{,}400\,km\,s$^{-1}$ (see Appendix~\ref{app:line-profiles}), consistent with the typical \ion{O}{2} velocity range in SLSNe-I (6000--21{,}000\,km\,s$^{-1}$; \citealt[their Figure 1]{Chen2023_PopulationStudyII}). The absorption minimum associated with the \ion{O}{2} $\lambda$3959 feature has a blueshift of $\sim$17{,}400\,km\,s$^{-1}$. However, this feature is instead more plausibly attributed to C II $\lambda$3921 (see Appendix~\ref{app:line-profiles}), which has a lower blueshift of 15{,}800\,km\,s$^{-1}$. Therefore, the $\sim$17{,}400\,km\,s$^{-1}$ association with \ion{O}{2} $\lambda$3959 is treated instead as an upper limit on the velocity rather than a secure \ion{O}{2} measurement. The inferred \ion{O}{2} velocities are broadly comparable to those from \ion{Fe}{2} $\lambda$5169, although a direct comparison is limited by the weak and blended nature of both the \ion{O}{2} and \ion{Fe}{2} features. The C II $\lambda$3921 identification is also consistent with the velocity inferred from C II $\lambda$4267.

Overall, SN\,2025wny at $+20$\,d most closely resembles Gaia16apd at the same phase. The identification of \ion{O}{2} is ambiguous, but contributions from \ion{Ca}{2} H\&K, \ion{C}{2}, and \ion{Fe}{2} are likely present. All measurable line velocities are broadly consistent with $\sim$12{,}000--17{,}000\,km\,s$^{-1}$.

\begin{figure*}
    \centering
    \includegraphics[width=\linewidth]{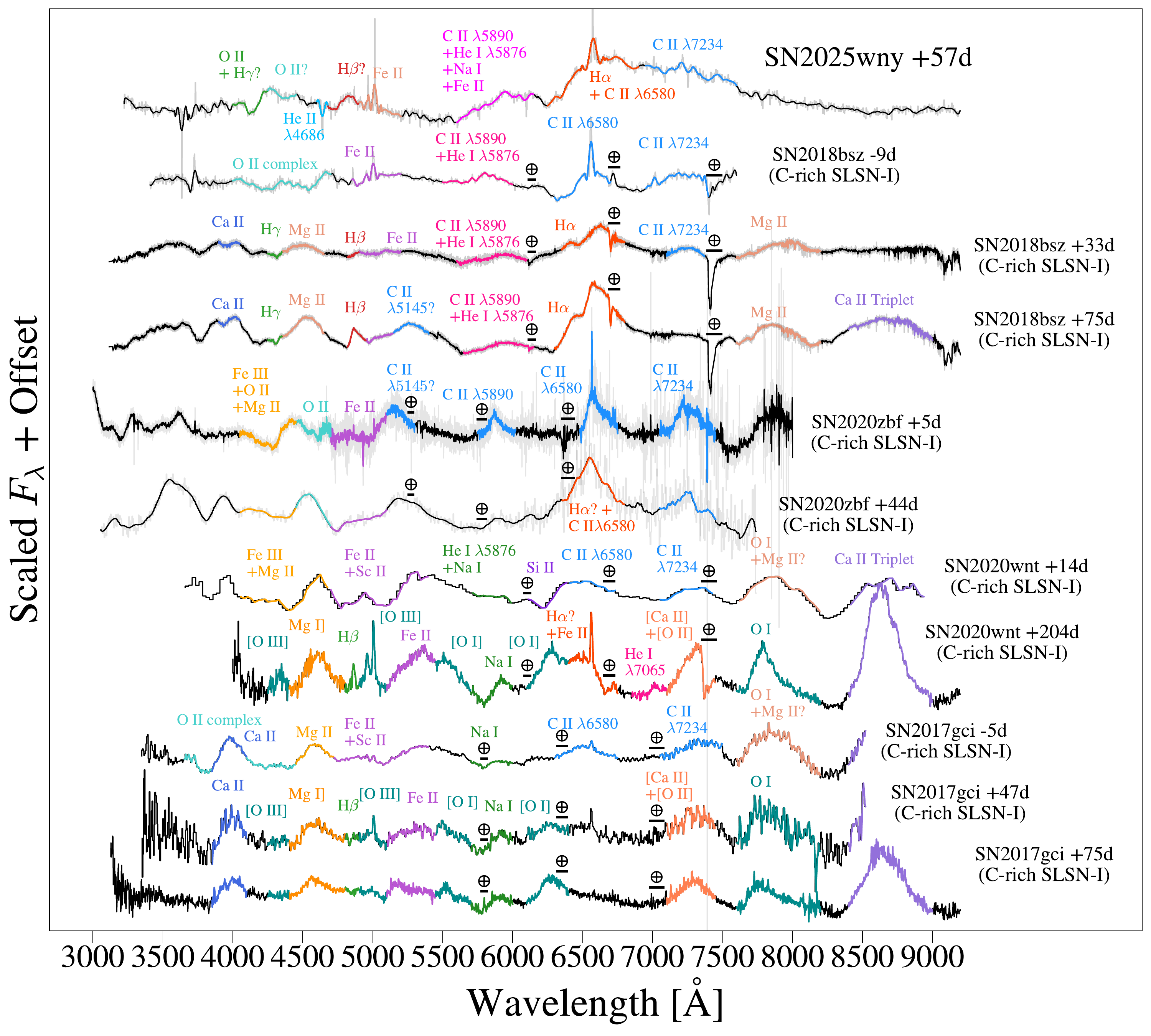}
    \caption{De-redshifted and continuum-normalized optical spectral comparison of SN\,2025wny at $+57$\,d (top) with carbon-rich SLSNe-I across a range of phases, including SN\,2018bsz \citep{2022AnA...666A..30P}, SN\,2020zbf \citep{2024AnA...685A..20G}, SN\,2020wnt \citep{2022MNRAS.517.2056G}, and SN\,2017gci \citep{2021MNRAS.502.2120F}. Smoothed versions are overlaid in black and prominent spectral features are highlighted in color. Telluric absorption bands are marked with $\oplus$ symbols.}
    \label{fig:jwst_58d_compare_carbon}
\end{figure*}

\subsection{The $+57$\,d optical to NIR spectrum}

\subsubsection{Carbon and hydrogen}

The $+57$\,d optical-to-NIR spectrum (Figure~\ref{fig:uv_to_nir_58d}) covers approximately 3200--17{,}500\,{\AA} in the rest frame and differs substantially from that of canonical SLSNe-I at comparable phases. There are five prominent features, including (1) a feature centered near 4130\,{\AA} and extending to 4700\,{\AA} (blend of \ion{Mg}{2}, \ion{Fe}{2}, and possibly H$\gamma$); (2) a broad bump near 4820\,{\AA} (possibly H$\beta$); (3) a feature spanning 5600--6300\,{\AA} (\ion{C}{2}, \ion{Na}{1}, and possibly \ion{He}{1}); (4) a complex feature extending from approximately 6300--7600\,{\AA} (H$\alpha$ and \ion{C}{2}); and (5) a prominent P~Cygni profile around 1\,$\mu$m (\ion{C}{2} and possibly \ion{He}{1}).

Although SN\,2025wny was initially classified as a SLSN-I based on template matching to its early rest-frame UV spectra, template matching of the $+57$\,d optical spectrum using both the Supernova Identification \citep[SNID]{1979AJ.....84.1511T, 2007ApJ...666.1024B} and Next Generation SuperFit \citep{2022TNSAN.191....1G} codes instead yielded best matches to hydrogen- and helium-rich Type~IIb and Type~IIP SNe near peak brightness. To investigate the presence of hydrogen and helium features in the optical spectrum of SN\,2025wny, we compare the $+57$\,d spectrum to those of Type~II SNe, Type IIb SNe, SLSNe-II, SLSNe-I exhibiting late-time hydrogen emission, and SLSNe-Ib. We present the comparison in Figure~\ref{fig:jwst_58d_comparison} along with ion identifications under a Type II interpretation. However, in hydrogen-poor stripped-envelope SNe, \ion{C}{2} commonly produces broad optical features that overlap with the Balmer transition wavelengths. We therefore also compare the spectrum to carbon-rich SLSNe-I, as shown in Figure~\ref{fig:jwst_58d_compare_carbon}, with identifications made under a carbon-rich interpretation. All spectra are shifted to the rest frame, continuum-normalized using a \textsc{scipy} \citep{2020SciPy-NMeth} univariate spline, and smoothed using the Savitzky--Golay filter implemented in \textsc{scipy}.

The broad P~Cygni-like profile between 6300--7600\,{\AA} is not consistent with a single line profile and instead appears to consist of two broad components. Based on the comparisons in Figures~\ref{fig:jwst_58d_comparison} and \ref{fig:jwst_58d_compare_carbon}, we interpret the left component as a blend of H$\alpha$ and \ion{C}{2} $\lambda$6580, while the redward component is plausibly dominated by \ion{C}{2} $\lambda$7234. Under this interpretation, we fit a P-Cygni and Gaussian profile to the feature with fit results (see Appendix~\ref{app:line-profiles}). The inferred blueshift velocities are $\approx$11{,}000--12{,}400\,km\,s$^{-1}$ for the H$\alpha$ and \ion{C}{2} $\lambda$6580 features. The \ion{C}{2} $\lambda$7234 component does not have a clear absorption profile, but its inferred emission hump is at a $\approx1400$\,km\,s$^{-1}$ blueshift.

The coexistence of carbon- and hydrogen-related features in SN\,2025wny is not unprecedented among spectroscopically unusual SLSNe-I. For example, the carbon-rich SLSN-I SN\,2018bsz evolved from exhibiting strong and relatively sharp \ion{C}{2} features at early phases to showing increasingly prominent H$\alpha$ emission after $\sim+30$\,d \citep{2022AnA...666A..30P}. SN\,2020zbf \citep{2024AnA...685A..20G} and SN\,2020wnt \citep{2022MNRAS.517.2056G} show potentially similar behavior. Moreover, as shown in Figure~\ref{fig:jwst_58d_compare_carbon}, SN\,2018bsz and SN\,2020zbf exhibit comparatively sharp \ion{C}{2} emission profiles at early times, whereas SN\,2020wnt \citep{2022MNRAS.517.2056G} and SN\,2017gci \citep{2021MNRAS.502.2120F} show broader but still separate \ion{C}{2} features. In particular, SN\,2020zbf has been identified as one of the clearest examples of a carbon-rich SLSN-I due to its prominent and well-defined \ion{C}{2} profiles. In contrast, the 6300--7600\,{\AA} feature in SN\,2025wny does not closely resemble any of these morphologies, instead appearing as a genuine blend of multiple components, consistent with simultaneous contributions from H$\alpha$ and \ion{C}{2} $\lambda6580$.

We also identify weaker candidate Balmer features at shorter wavelengths. The emission feature near $\sim$4800\,{\AA} resembles H$\beta$ seen in Type II SNe and the carbon-rich SLSN-I SN\,2025wny. Additionally, the absorption trough near $\sim$4130\,{\AA} may correspond to H$\gamma$, implying a blueshift of $\sim$14{,}000\,km\,s$^{-1}$; however, H$\gamma$ is expected to be weaker than H$\beta$, while in SN\,2025wny it appears stronger. A more likely interpretation is that the wavelength region spanning the potential H$\gamma$ to H$\beta$ features is significantly blended, plausibly with contributions from \ion{Mg}{2} and \ion{Fe}{2}.

The feature spanning 5600--6300\,{\AA} likely arises from a blend of multiple species, including \ion{C}{2} $\lambda$5890, \ion{Na}{1} $\lambda\lambda$5890,5896, and possibly \ion{He}{1} $\lambda$5876, as shown in several of the comparison spectra. The blueshifts are $\approx$14{,}000$-$15{,}000\,km\,s$^{-1}$ for these lines. The velocity for \ion{C}{2} $\lambda$5890 is also broadly consistent with the range inferred from \ion{C}{2} $\lambda$6580.


Overall, we interpret the $+57$\,d optical spectrum as having evidence for both hydrogen and carbon. The inferred blueshift velocities across the identified optical features are also mutually consistent at the $\sim10{,}000-15{,}000$\,km\,s$^{-1}$ level.

\subsubsection{The 1 micron feature}
\label{subsec:nir-pcygni}

\begin{figure}
    \centering
    \includegraphics[width=\linewidth]{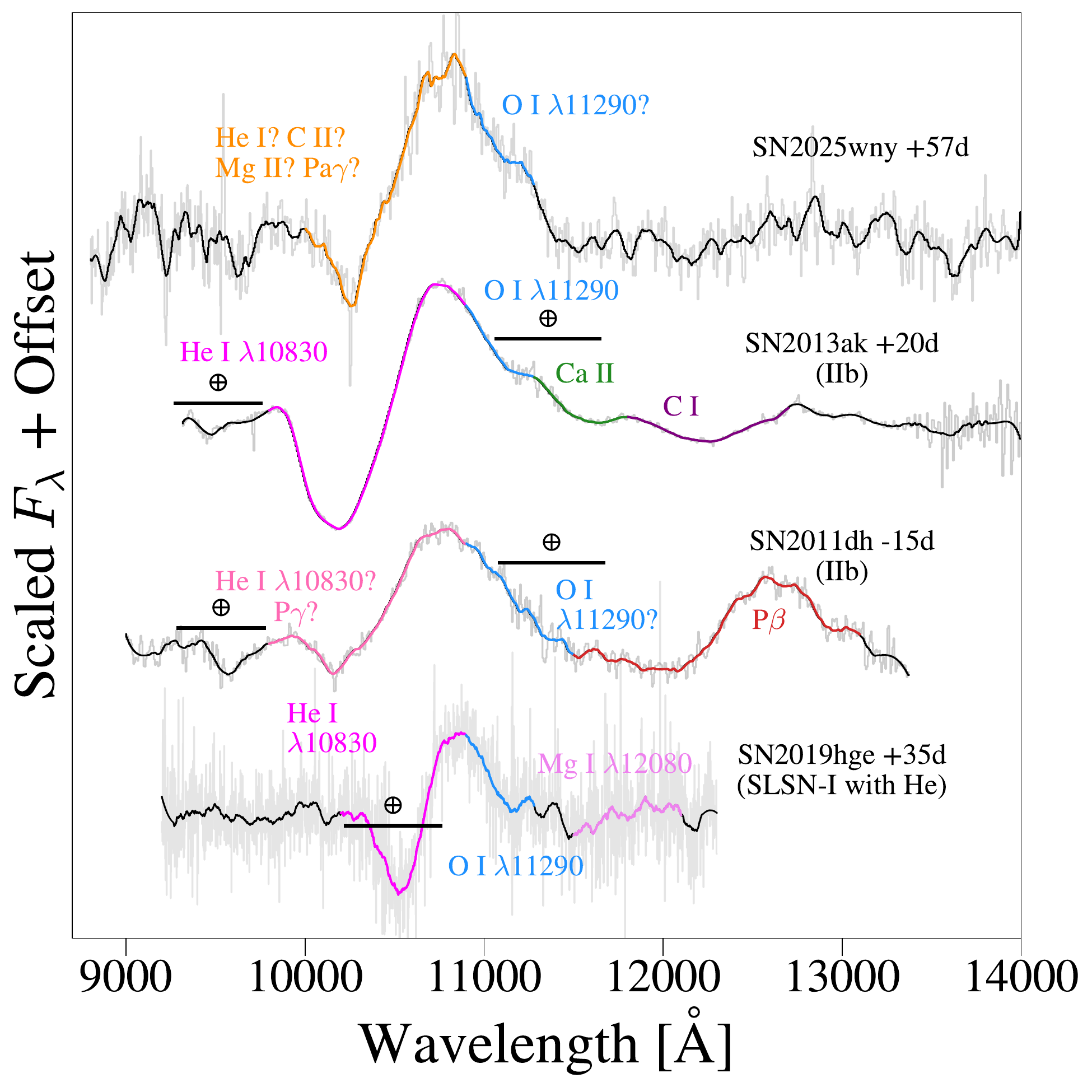}
    \caption{De-redshifted and continuum-normalized NIR spectral comparison of SN\,2025wny (top) with the Type~Ib SN\,2013ak \citep{2013CBET.3437....1C}, Type~IIb SN\,2011dh \citep{2011ApJ...742L..18A}, and SLSN-Ib SN\,2019hge \citep{Yan2020_HeI}. Smoothed versions are overlaid in black and prominent spectral features are highlighted in color. Telluric absorption bands are marked with $\oplus$ symbols.}
    \label{fig:jwst_58d_nir}
\end{figure}

The $+57$\,d NIR spectrum exhibits a prominent P-Cygni profile at $\approx$1.03\,$\mu$m. The emission component of this profile shows a subtle dip that we associate with \ion{O}{1} 1.1290\,$\mu$m, as suggested by comparison with the spectra presented in Figure~\ref{fig:jwst_58d_nir}. In stripped-envelope SNe, the prominent P~Cygni profile near 1\,$\mu$m is commonly attributed to a blend of \ion{He}{1} $\lambda$1.0830\,$\mu$m, \ion{C}{1} $\lambda$1.0693\,$\mu$m, Pa$\gamma$, and \ion{Mg}{2}$\lambda$1.0927\,$\mu$m \citep{2019PASP..131a4002H, 2022ApJ...925..175S}. We examine their respective contributions to this feature.

In SN\,2025wny, the observed velocity structure disfavors Pa$\gamma$ as a dominant contributor; the implied blueshift of $\approx$18{,}800\,km\,s$^{-1}$ is significantly higher than that inferred from the optical Balmer lines ($\sim$10{,}000\,km\,s$^{-1}$), and the non-detection of Pa$\beta$ further argues against a strong Pa$\gamma$ contribution. We also disfavor \ion{Mg}{2} $\lambda$1.0927\,$\mu$m as a primary contributor due to its high blueshift of $\approx$18{,}400\,km\,s$^{-1}$ and the lack of convincing \ion{Mg}{2} features in spectrum. 

\begin{figure*}
    \centering
    \includegraphics[width=\linewidth]{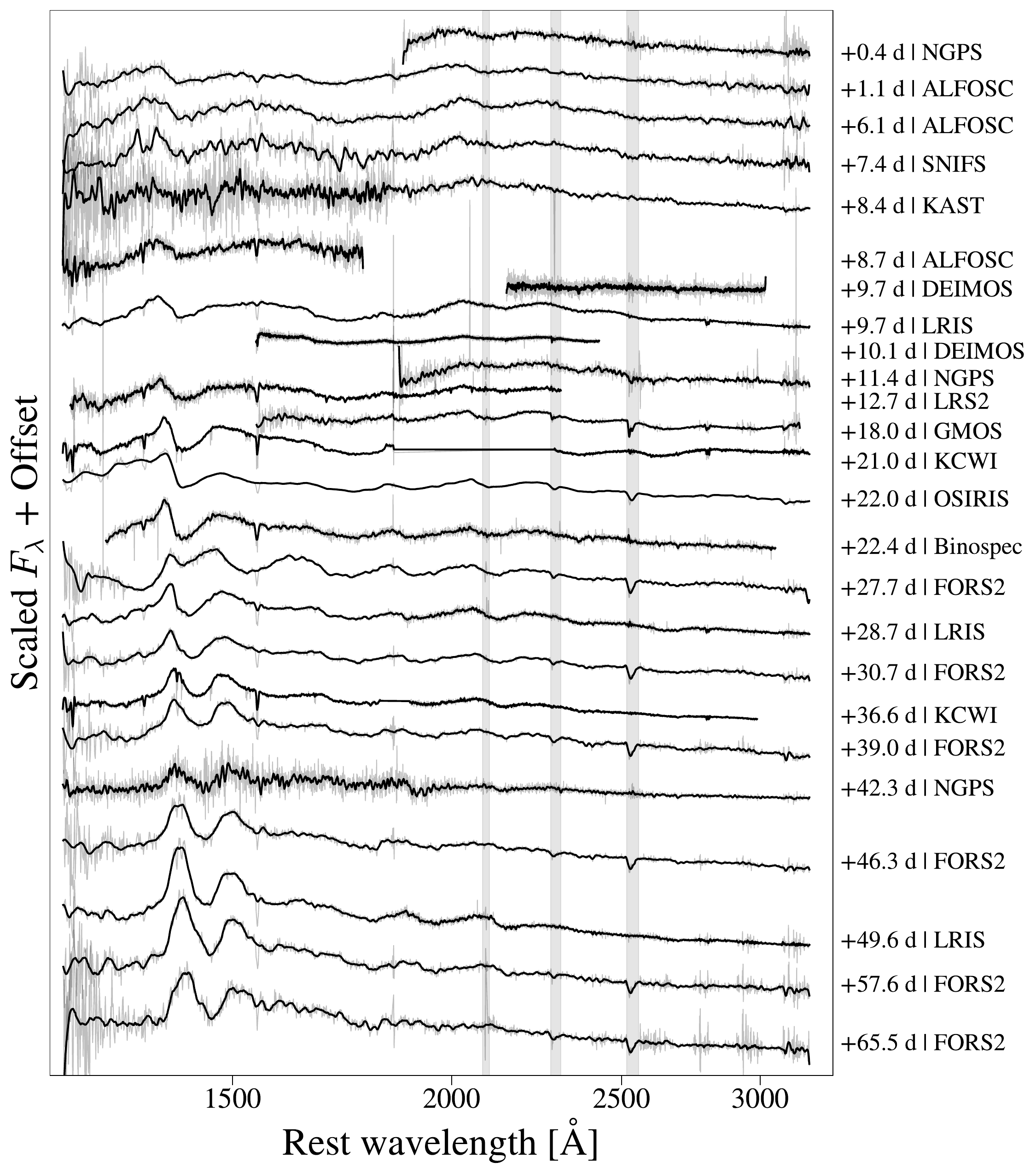}
    \caption{Rest-frame UV spectroscopic evolution of SN\,2025wny from $+1$\,d to $+66$\,d rest-frame days from bolometric peak (MJD\,=\,60940.7). Each spectrum is labeled by phase and instrument, scaled in $F_\lambda$, and offset vertically for clarity; smoothed versions produced with a Savitzky--Golay filter are overlaid. Grey shaded regions indicate telluric absorption bands.}
    \label{fig:uv-spec-evo}
\end{figure*}

\begin{figure}
    \centering
    \includegraphics[width=\linewidth]{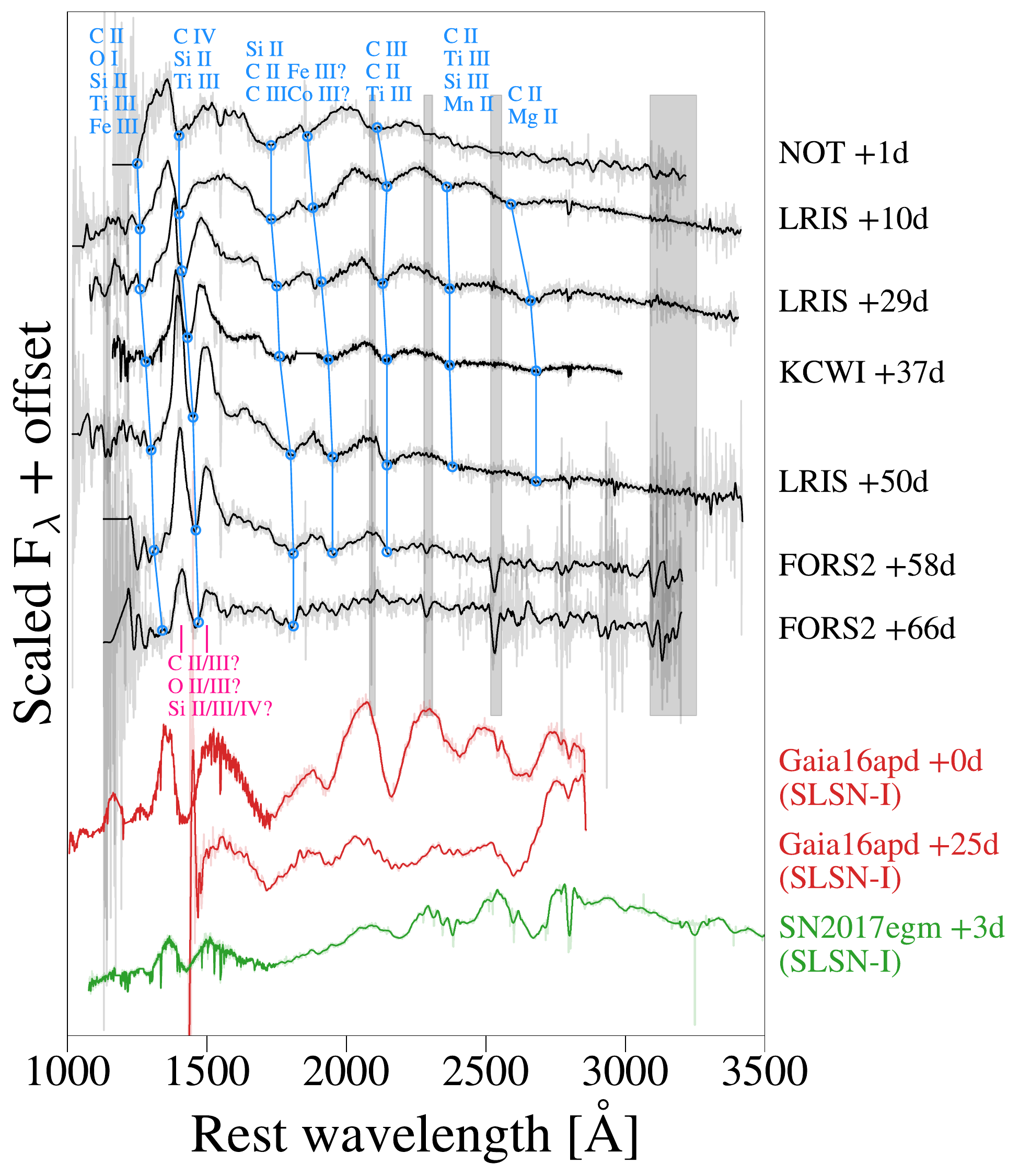}
    \caption{Rest-frame UV spectra of SN\,2025wny at select epochs chosen for high signal-to-noise. Spectra are shown in black with smoothed versions overlaid. Identified features are labeled at top. Blue circles connected by lines trace the velocity evolution of prominent features across epochs. Pink tick marks indicate unidentified features. Grey shaded regions denote telluric absorption bands. For comparison, rest-frame UV spectra of the canonical SLSNe-I Gaia16apd (red; \citealt{2017MNRAS.469.1246K, 2017ApJ...840...57Y}) and SN\,2017egm (green; \citealt{2017ApJ...845L...8N, 2018ApJ...858...91Y}) are shown at bottom; features are broadly consistent across the SLSNe-I.}
    \label{fig:uv-line-id}
\end{figure}

However, \ion{C}{1} $\lambda$1.0693\,$\mu$m yields a blueshift of $\approx$12{,}300\,km\,s$^{-1}$, roughly consistent with velocities inferred from optical \ion{C}{2} features. Interpreting the feature as \ion{He}{1} $\lambda$1.0830\,$\mu$m instead implies a velocity of $\approx$15{,}900\,km\,s$^{-1}$, also broadly consistent with the tentative optical \ion{He}{1} $\lambda$5876 identification. If the optical spectrum indeed contains an H$\alpha$+\ion{C}{2} blend, a corresponding \ion{He}{1}+\ion{C}{2} blend in the NIR is likewise plausible. Overall, the 1\,$\mu$m P~Cygni profile is most naturally interpreted as a multi-component blend dominated by \ion{C}{2} $\lambda$1.0693\,$\mu$m and possibly \ion{He}{1} $\lambda$1.0830\,$\mu$m, with \ion{Mg}{2} $\lambda$1.0927\,$\mu$m and Pa$\gamma$ contributing weakly if present.

\subsection{Rest-frame UV spectroscopic evolution}
\subsubsection{Evolution of UV features}

We present the full UV spectral sequence in Figure~\ref{fig:uv-spec-evo}. The most prominent features in the UV spectra at $\lesssim66$\,d are absorption troughs at roughly 1300\,{\AA}, 1400--1450\,{\AA}, 1700--1800\,{\AA}, 1900-2000\,{\AA}, 2100\,{\AA}, 2400\,{\AA}, and 2700\,{\AA}. 

We select spectra with high signal-to-noise ratios and label their ion features in Figure~\ref{fig:uv-line-id}. We adopt the same ion identifications as \citet{Johansson2025_SN2025wny}, which uses \citet{2018ApJ...858...91Y} as reference.

The early UV spectra of SN\,2025wny bear a close resemblance to those of well-studied SLSNe-I such as Gaia16apd \citep{2017ApJ...840...57Y} and SN\,2017egm \citep{2018ApJ...858...91Y}, showing \ion{C}{2}, \ion{O}{1}, \ion{Mg}{2}, \ion{Si}{2}, \ion{Ti}{2}/\ion{Ti}{3}, \ion{Fe}{3}, and high-ionization species including \ion{C}{4} \citep{Johansson2025_SN2025wny}. The majority of these features remain until the $+50$\,d LRIS spectrum. By $+50$\,d, the $\approx2400$\,{\AA} feature attributed to \ion{C}{2}, \ion{Ti}{3}, \ion{Si}{3}, and \ion{Mn}{2} has disappeared, while the $\approx2700$\,{\AA} feature associated with \ion{C}{2} and \ion{Mg}{2} weakens before fading at later epochs. The two sharp emission features near $\approx$1400\,{\AA} and $\approx$1500\,{\AA} appear between $+20$--60\,d before fading in subsequent spectra; we discuss these in Section~\ref{subsec:fuv-emission}. The absorption features at $\approx$1950 and $\approx2100$\,{\AA} disappear by $+66$\,d.


\subsubsection{Sharp FUV features}
\label{subsec:fuv-emission}

\begin{figure}
    \centering
    \includegraphics[width=\linewidth]{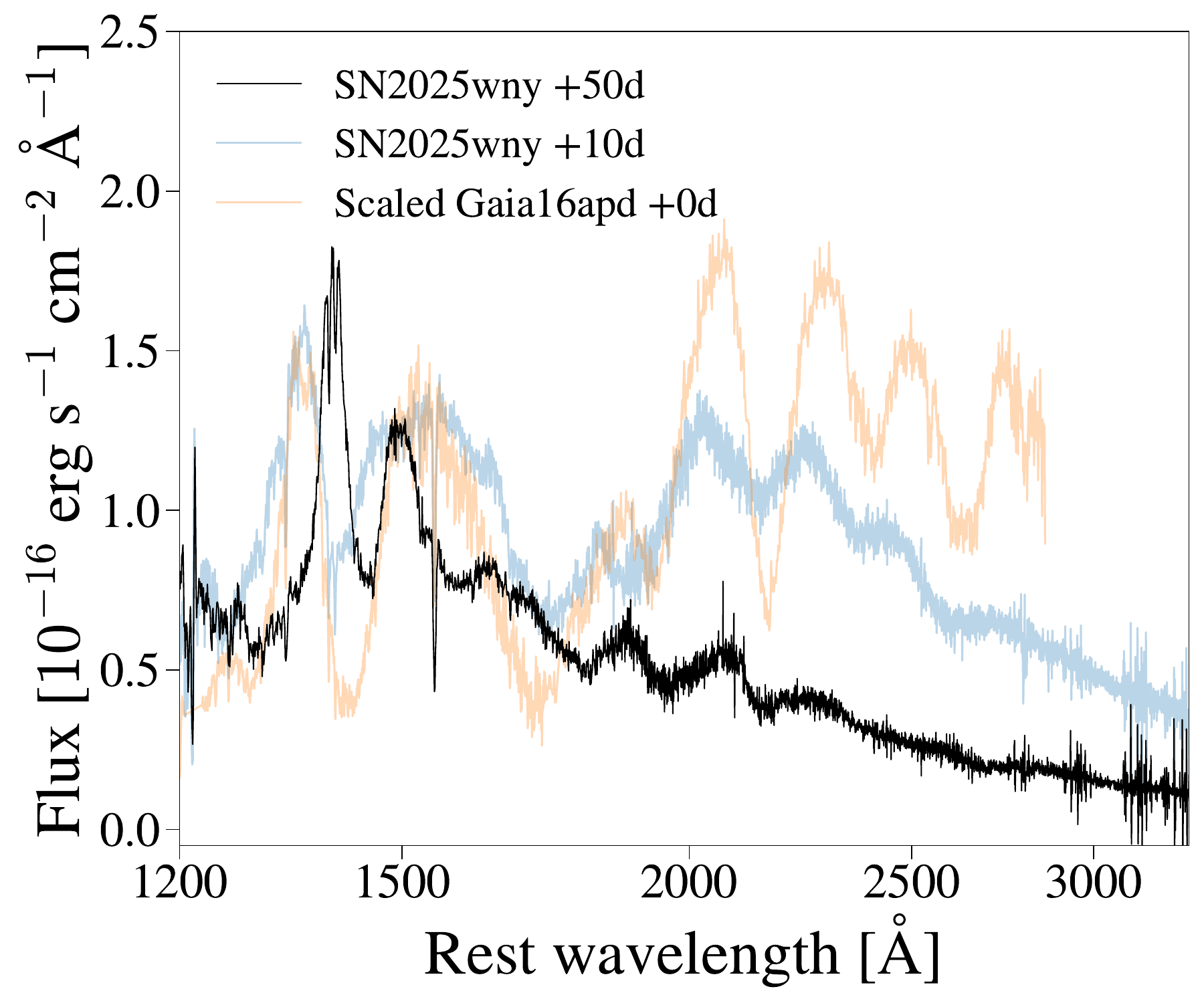}
    \caption{Evolution of the FUV region of SN\,2025wny from +10\,d (blue) to +50\,d (black), plotted on an absolute flux scale. No magnification correction is applied. We plot a scaled spectrum of Gaia16apd at +0\,d for reference.}
    \label{fig:fuv_emission_analyze}
\end{figure}

The rest-frame UV spectra show two sharp features centered near 1400\,{\AA} and 1500\,{\AA}. In Figure~\ref{fig:fuv_emission_analyze}, we highlight the emergence of these features. In the +10\,d spectrum, the 1400-1500\,{\AA} region can be interpreted as absorption features superposed on a hot UV continuum. However, by +50\,d, the surrounding continuum has faded while the features at 1400\,{\AA} and 1500\,{\AA} remain largely unchanged in flux.

This behavior admits two interpretations. One possibility is that the ejecta has a long-lived FUV continuum component, with the apparent 1400\,{\AA} and 1500\,{\AA} peaks tracing the continuum shape around an intervening absorption trough. In this scenario, the flux at $\lesssim1300$\,{\AA} also remains highly absorbed for the observed duration of SN\,2025wny. We find evidence of a possible multi-component emitting region in Section~\ref{subsubsec:uv-excess} to support this interpretation.

\begin{figure}
    \centering
    \includegraphics[width=\linewidth]{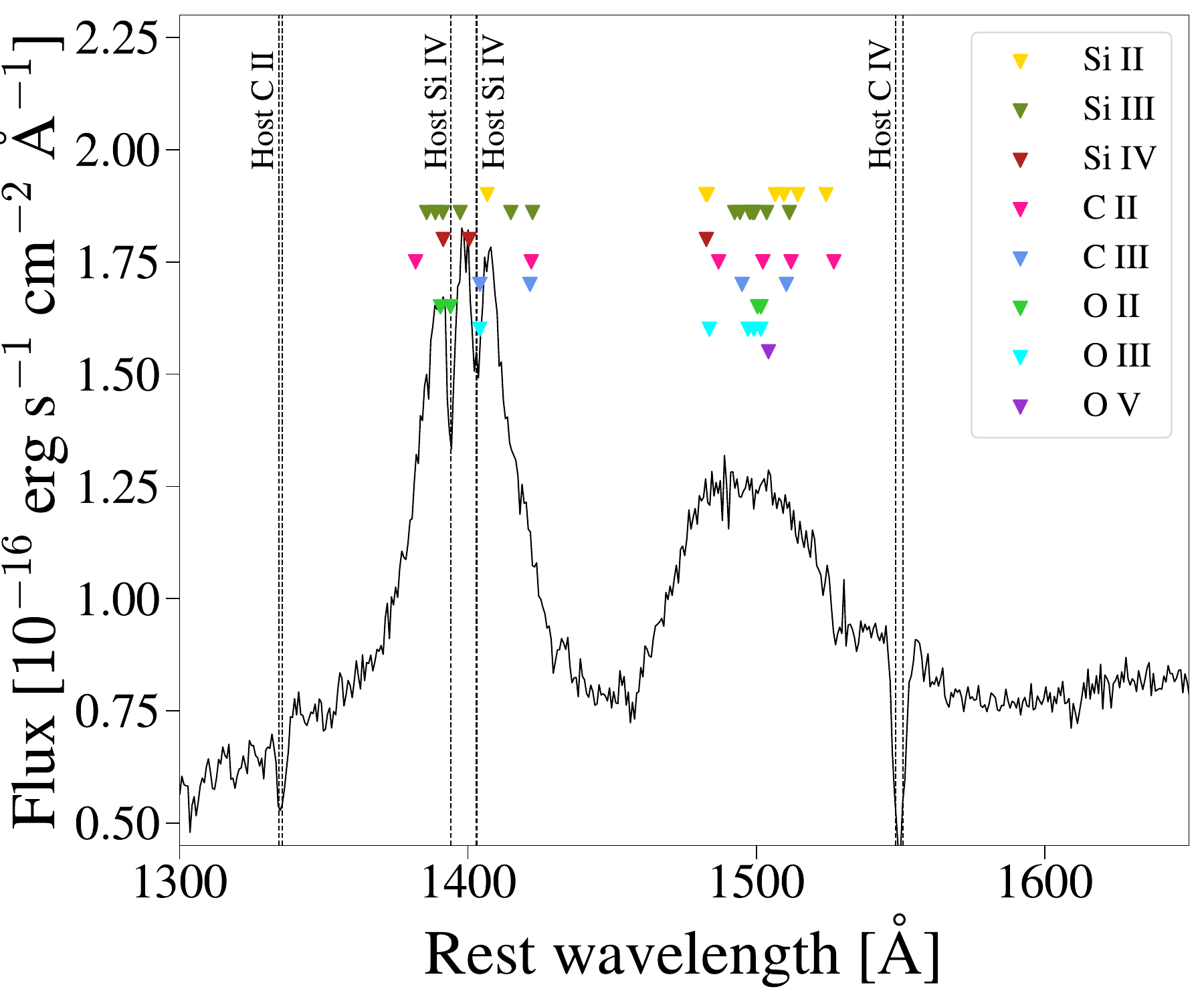}
    \caption{Zoom-in on the sharp rest-frame FUV features in the +50\,d LRIS spectrum. Assuming the features are emission, triangles mark candidate contributions from various ions at a blueshift of 500\,km\,s$^{-1}$. Dashed vertical lines indicate possible host-galaxy absorption features. No magnification correction is applied.}
    \label{fig:500kms_fuv_emission_lines}
\end{figure}

Alternatively, the 1400\,{\AA} and 1500\,{\AA} features may correspond to emission lines emerging above the fading UV continuum. We show a zoom-in of this region from the +50\,d spectrum in Figure~\ref{fig:500kms_fuv_emission_lines}. 
Gaussian fits give broad FWHMs of $\sim8{,}100$\,km\,s$^{-1}$ and 11{,}800\,km\,s$^{-1}$ for the 1400\,{\AA} and 1500\,{\AA} features, respectively. If these are emission features, candidate ion contributors include \ion{C}{2}, \ion{C}{3}, \ion{O}{2}, \ion{O}{3}, \ion{Si}{2}, \ion{Si}{3}, and \ion{Si}{4}. We consider carbon, since carbon is present in the optical spectrum at +57\,d; we consider oxygen, as it is often expected to accompany carbon-rich material; silicon is considered from previous absorption-line identifications in this wavelength region. 

We also identify several narrow absorption features likely associated with the host galaxy in this region, based on \citet{2017ApJ...845L...8N}. We identify \ion{Si}{4} $\lambda\lambda1393.7546,1402.7697$ on top of the 1400\,{\AA} emission complex. We also identify \ion{C}{2} $\lambda\lambda1334.5326,1335.7077$, \ion{C}{4} $\lambda\lambda1548.2043,1550.784$. We defer to \citet{Qin2026} for an in-depth host galaxy analysis.

If the 1400\,{\AA} and 1500\,{\AA} features are emission, their presence at +20--60\,d is unusual. Normally, SLSNe-I are slow evolving, and exhibit photospheric absorption at these phases. One possibility is that these emission features arise from ejecta--CSM interaction, where the expanding ejecta shocks and heats nearby CSM, producing transient line emission. However, this would differ from the interaction signatures commonly seen in interacting SLSNe-II and Type II CCSNe, which typically show narrow line cores of only a few hundred km\,s$^{-1}$ together with broader electron-scattering wings. We also do not find convincing evidence for interaction-powered \ion{Mg}{2} $\lambda\lambda2795.53,2802.70$ emission in any of the spectra, seen in strongly interacting Type II SNe \citep{2026ApJ..1004...23B}. Although weak \ion{Mg}{2} $\lambda\lambda2795.53,2802.70$ absorption is present, their narrow FWHM velocities suggest that these features are likely host-galaxy absorption lines rather than signatures of CSM interaction. Together, these observations disfavor a standard CSM-interaction scenario, but do not necessarily rule out its contribution. We discuss the possibility of a mixed power source for SN\,2025wny in Section~\ref{sec:discussion}. 

\begin{figure}
    \centering
    \includegraphics[width=\linewidth]{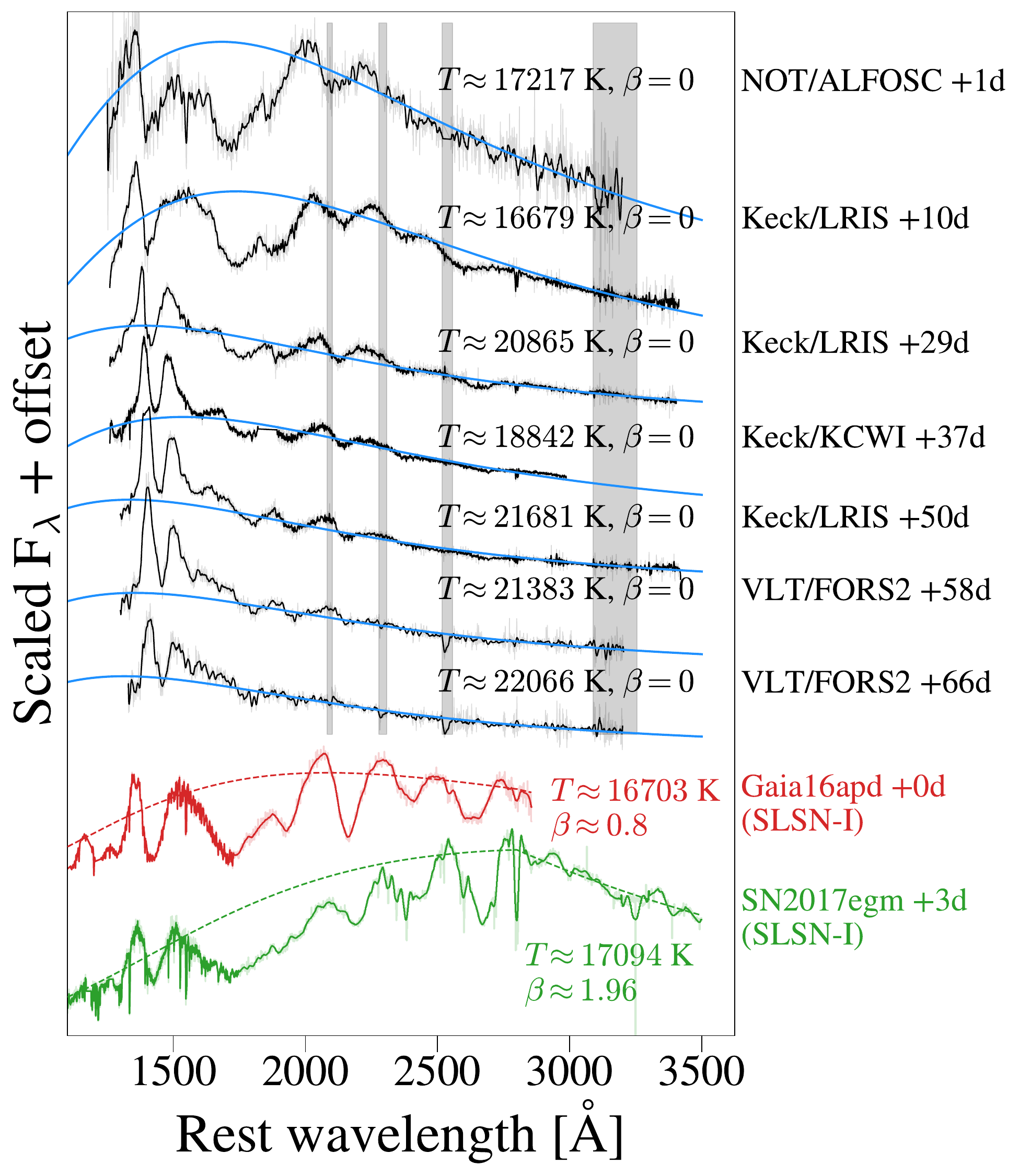}
    \caption{Blackbody fits to rest-frame UV spectra of SN\,2025wny at select epochs. The best-fit model (blue) is overlaid on each spectrum (black). For comparison, modified blackbody fits to Gaia16apd \citep{2017MNRAS.469.1246K, 2017ApJ...840...57Y} at $+0$\,d (red dashed) and SN\,2017egm \citep{2017ApJ...845L...8N, 2018ApJ...858...91Y} at $+3$\,d (green dashed) are shown at bottom, illustrating the weaker UV suppression in SN\,2025wny.}
    \label{fig:uv-bb-fit}
\end{figure}

\subsubsection{UV continuum excess and delayed line blanketing}
\label{subsubsec:uv-excess}

\begin{figure}
    \centering
    \includegraphics[width=\linewidth]{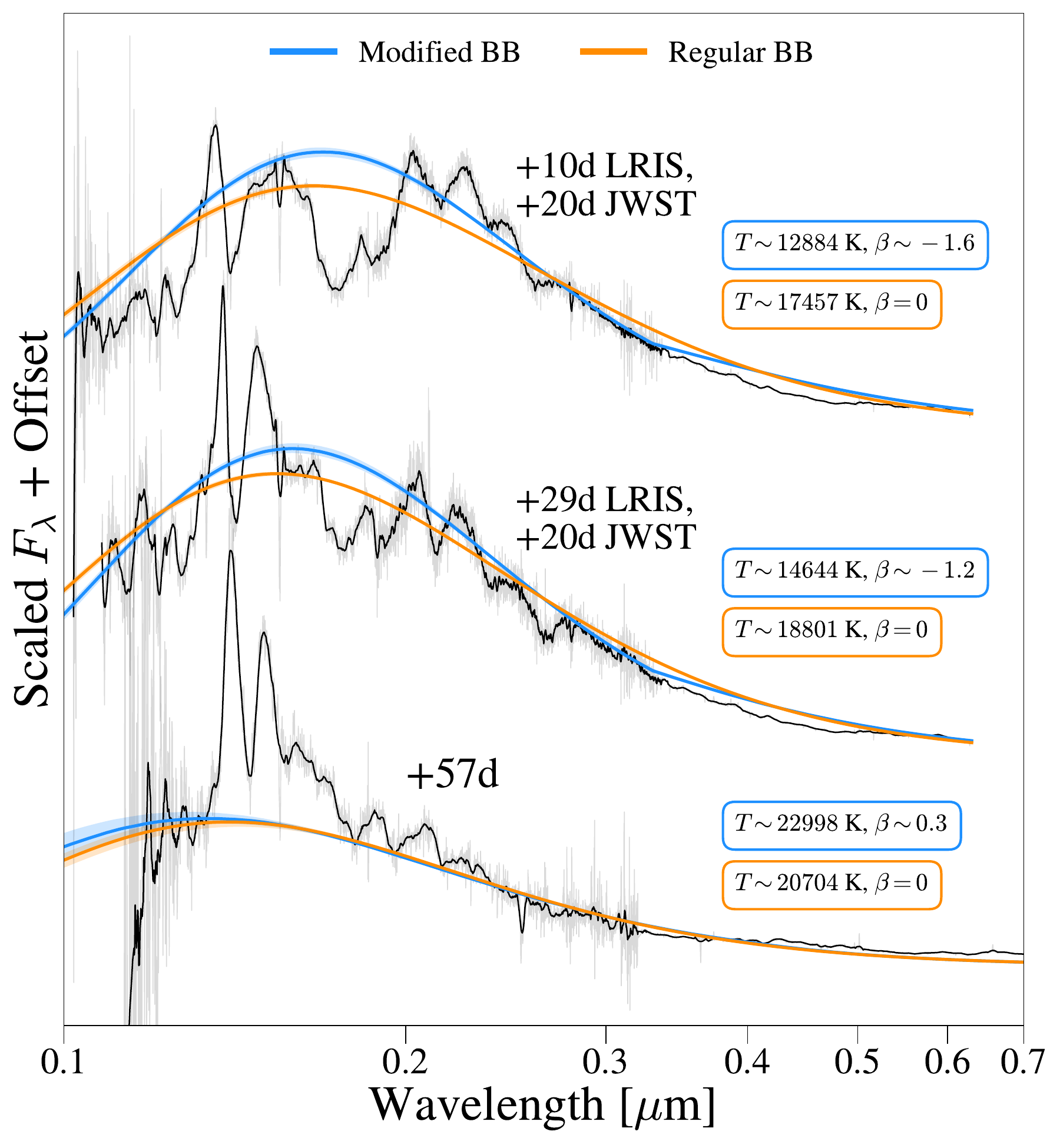}
    \caption{Blackbody fits to the stitched spectra of SN\,2025wny at $+20$\,d and $+57$\,d. For the $+20$\,d epoch, fits are shown for both the $+10$\,d LRIS/$+20$\,d \textit{JWST} stitch (top) and the $+29$\,d LRIS/$+20$\,d \textit{JWST} stitch (middle), with best-fit modified blackbody (blue) and standard Planck blackbody (orange) overlaid. 
    Fitted temperatures and $\beta$ values are labeled for each epoch.}
    \label{fig:stitched-bb-bit}
\end{figure}

To characterize the evolution of the rest-frame UV spectral continuum, we fit blackbody models to select high-quality spectra (Figure~\ref{fig:uv-line-id}) as well as to the stitched spectra (Figures~\ref{fig:uv_to_opt_20d} and \ref{fig:uv_to_nir_58d}). In the photospheric phase, SLSNe-I often display significant rest-frame UV line blanketing from \ion{Fe}{2} and \ion{Fe}{3}. We therefore consider a modified blackbody incorporating a UV suppression term parameterized by $\beta$, following \citet{2018ApJ...858...91Y}. The modified blackbody takes the form $F_\lambda = B_\lambda(T)(\lambda/\lambda_0)^\beta$ for $\lambda < \lambda_0$, and $F_\lambda = B_\lambda(T)$ for $\lambda \geq \lambda_0$, where $\lambda_0$ is the onset wavelength of suppression, and a larger $\beta$ represents more severe UV suppression. $\beta=0$ represents the standard Planck function. We perform the fitting with MCMC. We present results of the UV-only spectra in Figure~\ref{fig:uv-bb-fit} and the stitched spectra in Figure~\ref{fig:stitched-bb-bit}.

The UV-only fits indicate that SN\,2025wny exhibits unusually weak FUV suppression through at least $+66$\,d. The best-fit modified blackbody models generally favor $\beta \approx 0$, with only the $+10$\,d spectrum exhibiting $\beta\approx0.5$. Fits to the stitched spectra also support this interpretation, as they are well reproduced by modified blackbodies with $\beta \approx 0$ or $\beta < 0$. Overall, there is little evidence for the strong UV line blanketing commonly observed in SLSNe-I. This behavior contrasts with canonical SLSNe-I near peak, such as SN\,2017egm ($\beta\approx1.95$) and PTF12dam ($\beta\approx1.78$) \citep{2018ApJ...858...91Y}. Among well-studied SLSNe-I, Gaia16apd exhibits the most similar behavior to SN\,2025wny, with relatively weak UV line blanketing ($\beta\approx0.8$; \citealt{2017ApJ...840...57Y, 2018ApJ...858...91Y}), which has been interpreted as evidence for either inefficient mixing of iron-group elements into the outer ejecta or a low-metallicity progenitor. However, the effect may also be from degeneracies between temperature and radius in the blackbody fits.

\begin{figure}
    \centering
    \includegraphics[width=\linewidth]{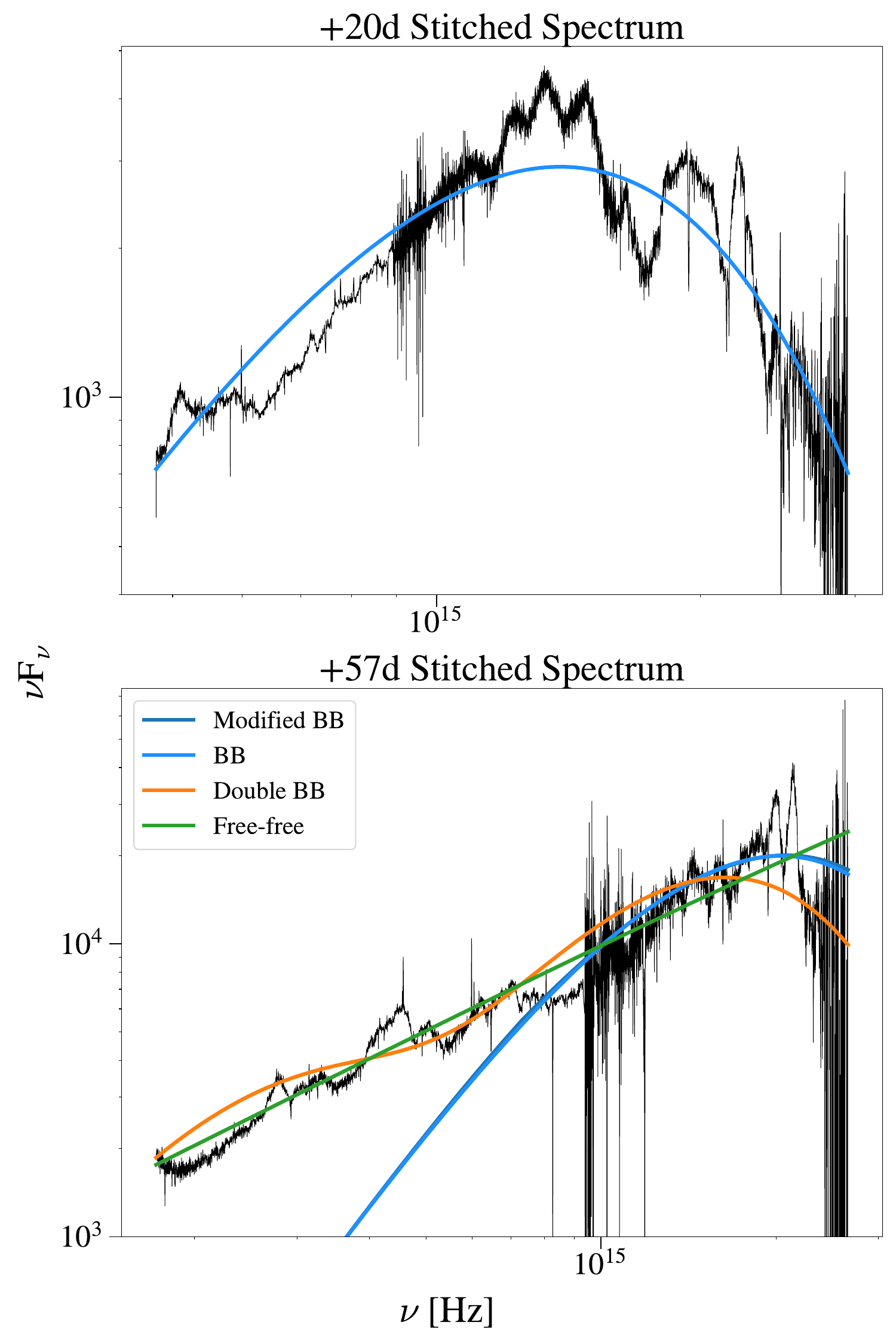}
    \caption{Continuum fits to the +20\,d and +57\,d stitched spectra in $\nu F_\nu$ space. The optically thin free-free model is from \citet[Chapter 5]{1979rpa..book.....R}.}
    \label{fig:nufnu-stitched-spectra}
\end{figure}

Evidence suggests that a single-blackbody interpretation is inadequate at $+57$\,d. First, Figure~\ref{fig:stitched-bb-bit} reveals a distinct FUV excess above the fitted blackbody continuum. Second, a blackbody continuum likely does not explain the entire spectrum well, even past the rest-frame UV. We illustrate this more clearly in $\nu F_\nu$--$\nu$ space in Figure~\ref{fig:nufnu-stitched-spectra}. While the $+20$\,d spectrum is roughly consistent with a blackbody, the $+57$\,d spectrum shows a clear excess at lower frequencies, corresponding to the rest-frame optical and NIR. We note that this is not apparent when only considering the rest-frame UV; i.e., in $\nu F_\nu$--$\nu$ space, blackbody fits to the rest-frame UV spectra are adequate. This may indicate the presence of an additional thermal component, as in a two-blackbody scenario, or that the continuum is instead dominated by a different emission process such as free--free emission. We discuss the free--free interpretation in more detail in Section~\ref{subsec:csm}.


Finally, throughout the observed evolution, the blackbody models favor high temperatures of $T\gtrsim16{,}000$\,K. In typical SLSNe-I, temperatures decline to $\sim$5{,}000--10{,}000\,K by around $+40$\,d post-peak \citep{Chen2023_PopulationStudy, Aamer2025_SLSNeI}, suggesting that SN\,2025wny remains unusually hot for an extended period. This behavior is inconsistent with simple adiabatic cooling from ejecta expansion alone.

\subsection{Comparisons to model spectra}

We compare the $+20$\,d and $+57$\,d stitched spectra to synthetic spectra of magnetar-powered Type Ic SLSNe presented in \citet{2019AnA...621A.141D}. These spectra are based on \textsc{CMFGEN} \citep{Hillier2012} non-local thermodynamic equilibrium time-dependent radiative transfer simulations and follow the spectral evolution of SLSNe from day one to $\sim$1--2 years after explosion \citep{Dessart2015,Dessart2016,Dessart2017a}. The main progenitor models for these spectra are 5p11 and r0 \citep{Yoon2010}, which are solar-metallicity carbon-rich Wolf-Rayet stars that explode with final masses of 5.11\,M$_\odot$ (rather than the 4.95\,M$_\odot$ quoted in \citealt{Yoon2010}; see \citealt{Dessart2015}) and 11.4\,M$_\odot$.

\begin{figure}
    \centering
    \includegraphics[width=\linewidth]{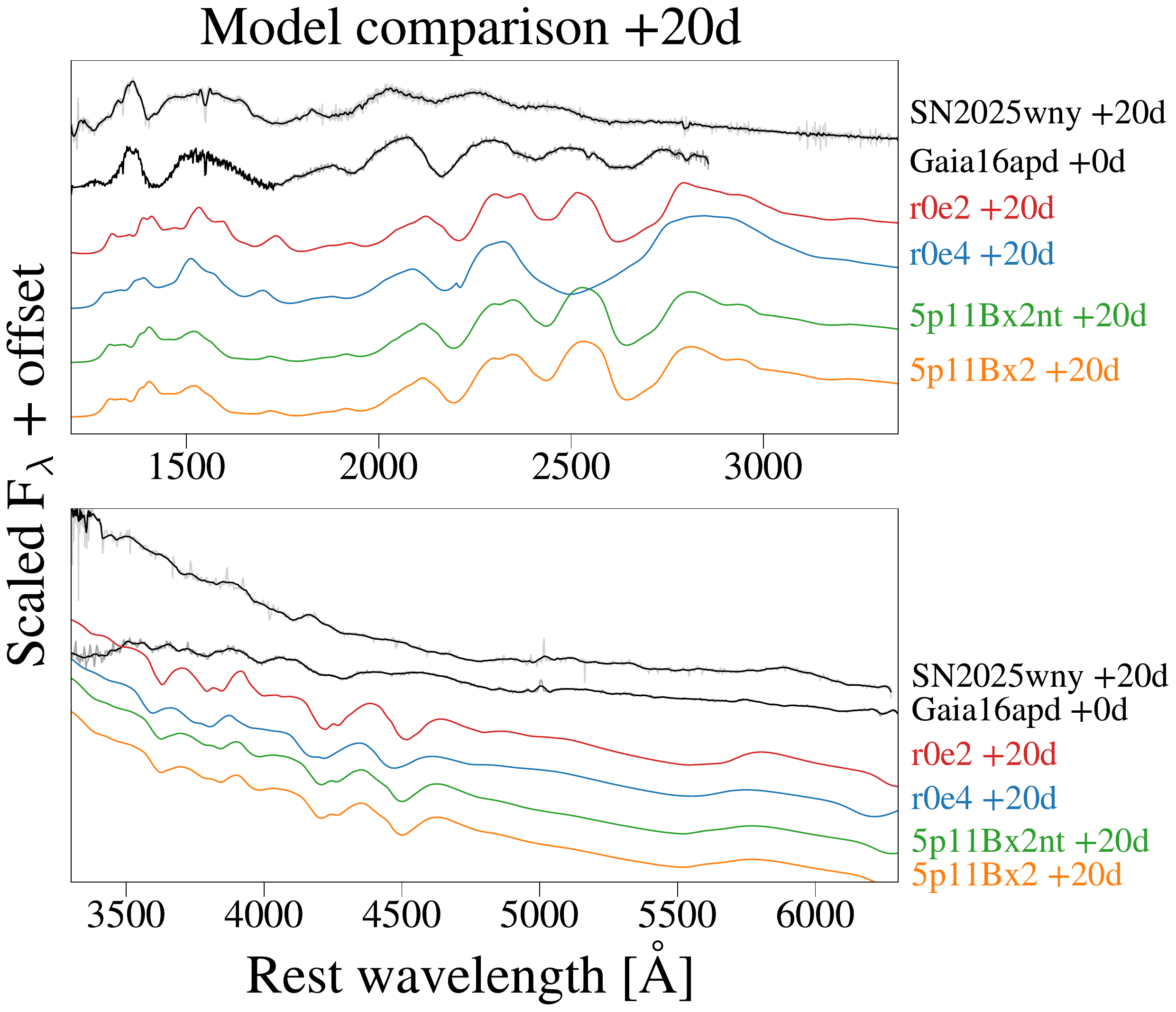}
    \includegraphics[width=\linewidth]{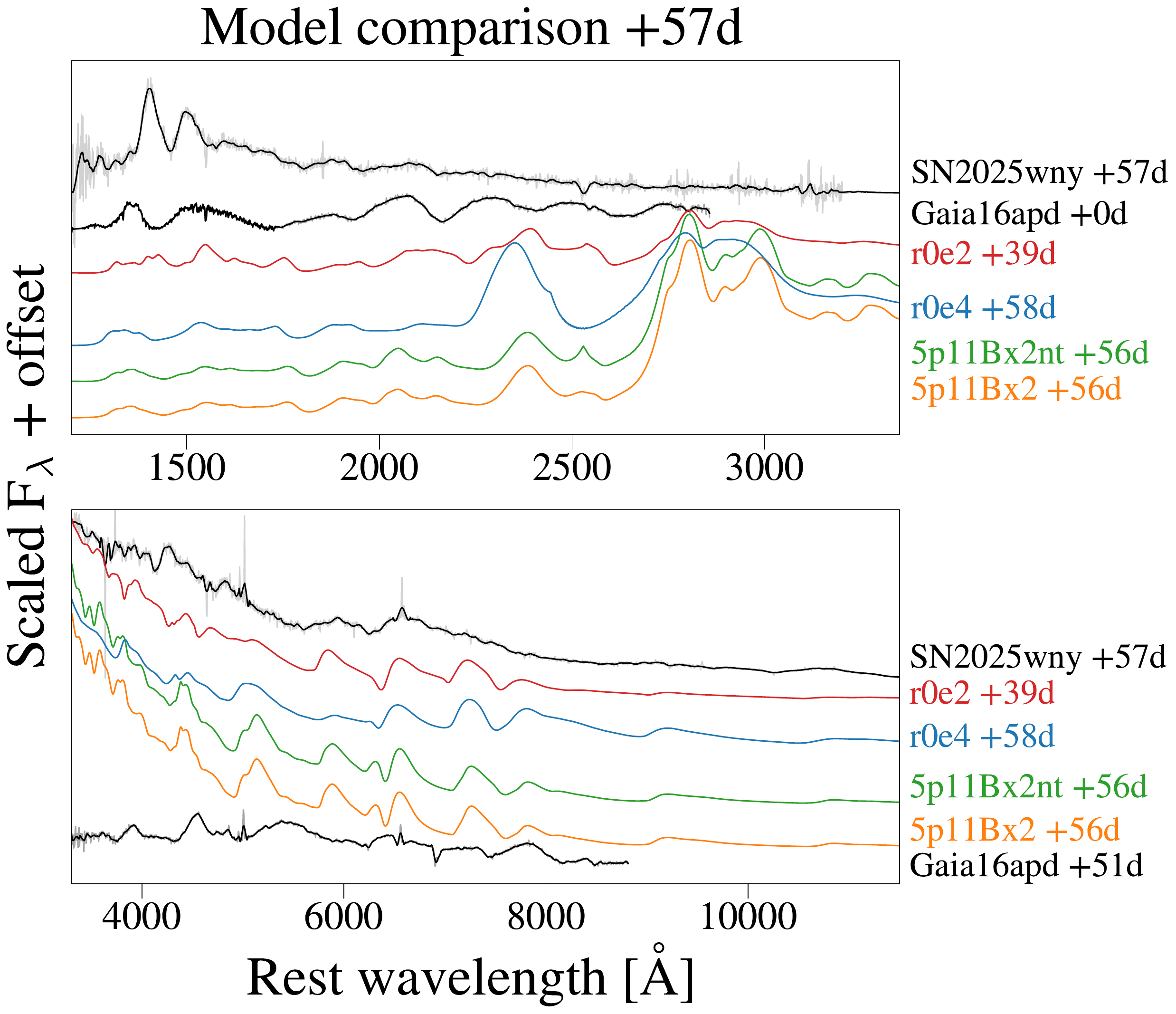}
    \caption{Comparisons between SN\,2025wny at $+20$\,d (top) and $+57$\,d (bottom) to Gaia16apd \citep{2017MNRAS.469.1246K, 2017ApJ...840...57Y} and select model spectra from \citet{2019AnA...621A.141D} at similar phases.}
    \label{fig:model_compare}
\end{figure}

No model provides a fully satisfactory match to the observed spectra, with the primary discrepancies arising from continuum mismatches in the rest-frame UV and differences in feature velocities and strengths. Nevertheless, as shown in Figure~\ref{fig:model_compare}, 5p11Bx2, 5p11Bx2nt, r0e2, and r0e4, reproduce subsets of the observed features reasonably well, including the 5500--7500\,{\AA} complex in the $+57$\,d spectrum and
the broad NIR excess near 1\,$\mu$m. Notably, the models attribute the 5500--7500\,{\AA} region to sharp carbon features, while the corresponding structure in SN\,2025wny is broader, consistent with contributions from additional species such as \ion{He}{1}, \ion{Na}{1}, and H$\alpha$ as previously discussed. Agreement in the 3000--4500\,{\AA} range is generally poor at both epochs, although the $+20$\,d spectrum may show some corresponding features.

Although the preference is not strong, the r0e2 model provides a somewhat better match than the other models, particularly in the continuum shape and overall broad feature morphology.  \citet[their Figure 29]{2019AnA...621A.141D} found that the early UV spectrum of Gaia16apd is also well reproduced by r0e2. Given the similarities between the $+20$\,d spectra of SN\,2025wny and those of Gaia16apd, the preference for r0 over 5p11 may be physically meaningful and could tentatively favor a higher-mass progenitor for SN\,2025wny.


Overall, additional model work is needed to achieve a better match. Still, the carbon-rich SLSN-I model spectra from \citet{2019AnA...621A.141D}, particularly r0e2, are able to reproduce some features in the $+20$\,d and especially the atypical $+57$\,d spectrum of SN\,2025wny, including the broad carbon-dominated structure in the 5500--7500\,{\AA} region. 

\section{Discussion}
\label{sec:discussion}

SN\,2025wny exhibits several unusual properties that distinguish it from typical SLSNe-I:
\begin{itemize}
    \item There is a rest-frame UV excess in the continuum and potential FUV emission features from +20--60\,d. This manifests in a plateau in the observed $g$-band light curve at the same phases, which plateaus while the $rizJ$ light curves steadily decline.
    \item We see little to no evidence of line blanketing throughout the observed evolution (Figure~\ref{fig:uv-bb-fit}).
    \item The late-time spectra show evidence for either carbon, hydrogen, or a blend of both.
    \item The spectra show no clear evidence for the characteristic \ion{O}{2} absorption features commonly seen in SLSNe-I that reach sufficiently high temperatures ($T\sim14{,}000-16{,}000$\,K) \citep{2024ApJ...967...13S}.
\end{itemize}

We explore possible interpretations of these properties in the following subsections.

\subsection{Can the host metallicity explain the observed properties?}
\label{subsec:host-metallicity}

Lower-metallicity environments may favor progenitor channels that produce hotter and more luminous explosions. Weaker metal-line driven winds allow massive stars to retain more mass and angular momentum before explosion, potentially yielding larger ejecta masses and higher luminosities. Lower metal abundances also reduce UV line blanketing, allowing more flux to emerge in the rest-frame UV. Thus, a low-metallicity environment could in principle help explain several properties of SN\,2025wny, including its FUV excess and weak FUV suppression compared to many lower-redshift SLSNe-I.

\begin{figure}
    \centering
    \includegraphics[width=\linewidth]{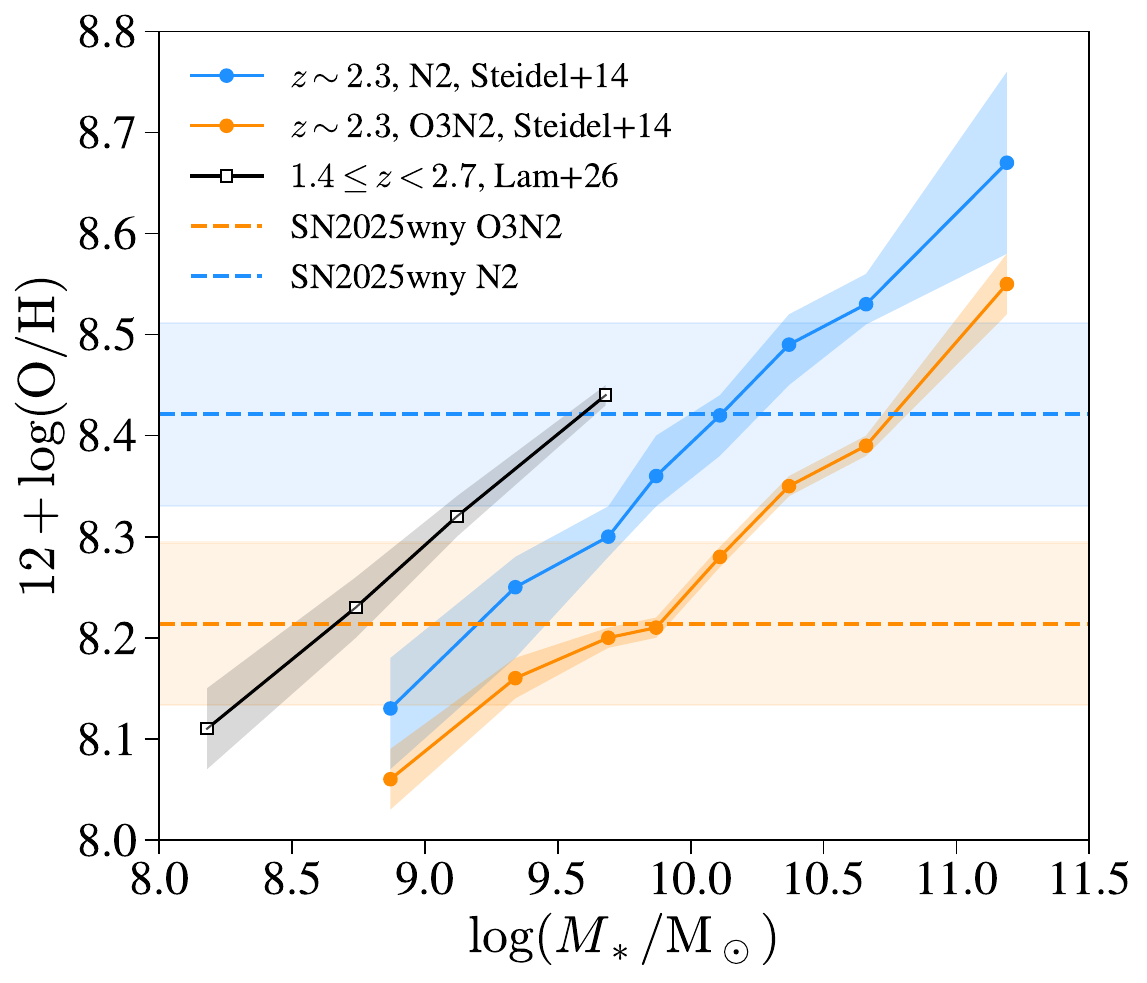}
    \caption{Mass--metallicity relation for star-forming galaxies at $z\sim2$ from \citet{2014ApJ...795..165S} and \citet{2026arXiv260530513L}. The inferred host-galaxy metallicities of SN\,2025wny, measured using the N2 (blue) and O2N3 (orange) diagnostics, are shown as horizontal lines. Both estimates fall within the range spanned by typical $z\sim2$ galaxies.}
    \label{fig:mzr}
\end{figure}

However, as shown in Figure~\ref{fig:mzr}, the host-galaxy metallicity inferred for SN\,2025wny does not appear to be unusually low. \citet{Qin2026}
measure $12+\log({\rm O/H})=8.21\pm0.08$ using the O2N3 diagnostic and $12+\log({\rm O/H})=8.42 \pm 0.09$ using the N2 diagnostic.
These values are typical for star-forming galaxies at $z\sim2$ \citep{2006ApJ...644..813E, 2014ApJ...795..165S, 2021ApJ...914...19S, 2026arXiv260530513L}. They are also broadly consistent with host-galaxy metallicities estimated for local SLSNe-I ($\sim8.3$--8.4; \citealt{2017MNRAS.470.3566C, 2018MNRAS.473.1258S}), including luminous events ($M\lesssim-22$ in any band; Table~\ref{tab:slsni-metallicities}). Thus, the host metallicity of SN\,2025wny does not offer an obvious explanation for its UV behavior, though it is consistent with SN\,2025wny having an otherwise typical luminosity among SLSNe-I. Lens modeling in \citet{Mortsell2026} also places SN\,2025wny close to its host center ($\sim$85\,pc), suggesting that the host metallicity may trace the progenitor environment.

This leaves two scenarios. First, SN\,2025wny may not be unusually metal poor relative to local SLSN-I hosts, in which case metallicity alone cannot explain the UV excess, and some other mechanism must be invoked. Second, SN\,2025wny may still have formed in a locally metal-poor region. Ongoing accretion of pristine or low-metallicity gas could produce localized metal-poor pockets even near the galaxy center. This process occurs in local star-forming dwarfs \citep{2014ApJ...783...45S} and in star-forming galaxies at $z\gtrsim2$ \citep{2010Natur.467..811C, 2016MNRAS.457.2605C}, similar to the host environments of SLSN-I. Thus, the measured host metallicity does not rule out a metal-poor SN progenitor. 

\input{tables/metallicities.tex}

\subsection{Implications of \ion{C}{2}, H$\alpha$, and \ion{He}{1} signatures}
\label{subsec:line-implications}
The +57\,d spectrum of SN\,2025wny shows possible signatures of \ion{C}{2}, H$\alpha$, and \ion{He}{1}, while lacking the clear \ion{O}{2} absorption complex commonly see in SLSNe-I that reach sufficiently hot temperatures \citep[$T\sim14{,}000$--16{,}000\,K]{2024ApJ...967...13S}.

One possible explanation is that the progenitor was only partially stripped prior to explosion, retaining an envelope containing some hydrogen and helium rich material. This interpretation is qualitatively similar to SN\,2019hge, whose presence of \ion{He}{2} have been interpreted as evidence for a partially stripped progenitor rather than a fully bare carbon-oxygen core \citep{Yan2020_HeI}. In this picture, oxygen may be hidden deeper in the ejecta, while the observed photosphere samples the H/He/C-rich layers. Another possible explanation is that the explosion was asymmetrical or ejecta was clumped, as invoked for SN\,2020wnt \citep{2022MNRAS.517.2056G}. Finally, it is also possible that the ejecta temperature is to hot for O II line formation, as the inferred blackbody temperatures range from $\sim17{,}000$--$22{,}000$\,K. However, Gaia16apd and SN\,2017egm show O II line formation $\sim17{,}000$\,K, inferred from modified blackbody fits to their rest-frame UV spectra \citep{2018ApJ...858...91Y}.

It is unclear whether these interpretations can explain the rest-frame FUV continuum excess or possible FUV emission features, especially with the high inferred photospheric temperatures ($T\gtrsim16{,}000$\,K; Figure~\ref{fig:temperature-and-radius}).

\subsection{A CSM-interaction interpretation}
\label{subsec:csm}

Interaction between SN ejecta and CSM has been proposed as a power source for some SLSNe, either alone or in combination with a central engine. As the ejecta collide with the CSM, kinetic energy is converted into radiation, producing additional luminosity. Observational signatures of CSM interaction can include enhanced UV emission and narrow emission lines from slow moving, photoionized CSM.

Evidence for CSM interaction is as follows. 
1) Between +20--60\,d the observed $g$-band light curve flattens and the spectra have an FUV excess, consistent with additional heating from ejecta--CSM interaction. 2) At +57\,d, the spectrum exhibits \ion{C}{2} $\lambda6580$, \ion{C}{2} $\lambda7234$, H$\alpha$, and possibly \ion{He}{1} $1.0830\,\mu$m. The broad emission near H$\alpha$/\ion{C}{2} $\lambda6580$ may also include a weak narrow component, although its significance remains uncertain. The emergence of these features may indicate that the expanding ejecta are encountering carbon, hydrogen, and helium-rich CSM. 3) We see no \ion{O}{2} absorption complex despite the high inferred temperatures, suggesting that these temperatures may not trace the oxygen-rich ejecta. Instead, they may trace the shocked CSM, while the oxygen-rich material remains deeper in the ejecta.

However, as discussed in Section~\ref{subsec:fuv-emission}, the possible sharp 1400\,{\AA} and 1500\,{\AA} features in SN\,2025wny do not resemble classical CSM-interaction lines, and we do not find convincing Mg\,II $\lambda\lambda2796,2803$ emission.
Together with the poor performance of the pure CSM-interaction \textsc{Redback} models, these results disfavor a standard CSM-interaction scenario.



Still, these results do not rule out CSM-interaction altogether as a contributing power source. For example, the observed interaction signatures may reflect the ejecta passing through a relatively thin CSM layer, while another mechanism dominates the luminosity and the underlying spectral evolution is set primarily by a partially stripped progenitor. Moreover, \textsc{Redback} fitting results suggest that a single power source is insufficient. A hybrid central-engine plus CSM-interaction model is possible, although we do not attempt such a fit in \textsc{Redback} because it would require coupling the dynamical evolution of the central-engine powered ejecta to the CSM shock evolution, which is not straightforward. 

Recent work also emphasizes that CSM interaction need not produce the canonical narrow emission lines associated with strongly interacting SNe. In particular, \citet{2026arXiv260118887G} show that for CSM interaction at moderate to high Thomson optical depth, lines may not be narrow and the continuum may show a roughly constant temperature of $\sim20{,}000$\,K, a non-blackbody spectrum, and an apparently decreasing blackbody radius. In this picture, X-ray photons produced by the interaction are reprocessed within the cold dense shell between the shocked ejecta and shocked CSM and re-emitted at UV and optical wavelengths. This interpretation was invoked by \citet{2026arXiv260118887G} for AT\,2018cow and warrants investigation for SN,2025wny, since its nearly constant $\sim20{,}000$\,K temperature after +20\,d post-peak, non-blackbody spectrum, apparently receding blackbody radius, and rest-frame FUV plateau (observed $g$ band) are all broadly consistent with this interaction scenario.

A CSM interaction interpretation for SN\,2025wny is also not unprecedented. Late-time hydrogen emission has been observed in other SLSNe. This includes nearly all carbon-rich SLSNe-I, as well as the SLSNe-Ib iPTF13ehe, iPTF15esb, and iPTF16bad, where broad H$\alpha$ features emerged weeks to months after peak and were attributed to ejecta interaction with hydrogen-rich CSM \citep{2017ApJ...848....6Y}. The emergence of carbon, hydrogen, and possible helium features observed in SN\,2025wny may reflect a similar interaction scenario.

As a follow-up to this work, we have scheduled two more epochs of \textit{JWST}/NIRSpec observations (PI Li; PID 12574) at roughly +130 and +170\,d post-peak. If these spectra develop strong and narrow hydrogen and helium lines, a CSM-interaction scenario may be supported.


\subsection{A magnetar-powered interpretation}

\begin{figure}
    \centering
    \includegraphics[width=\linewidth]{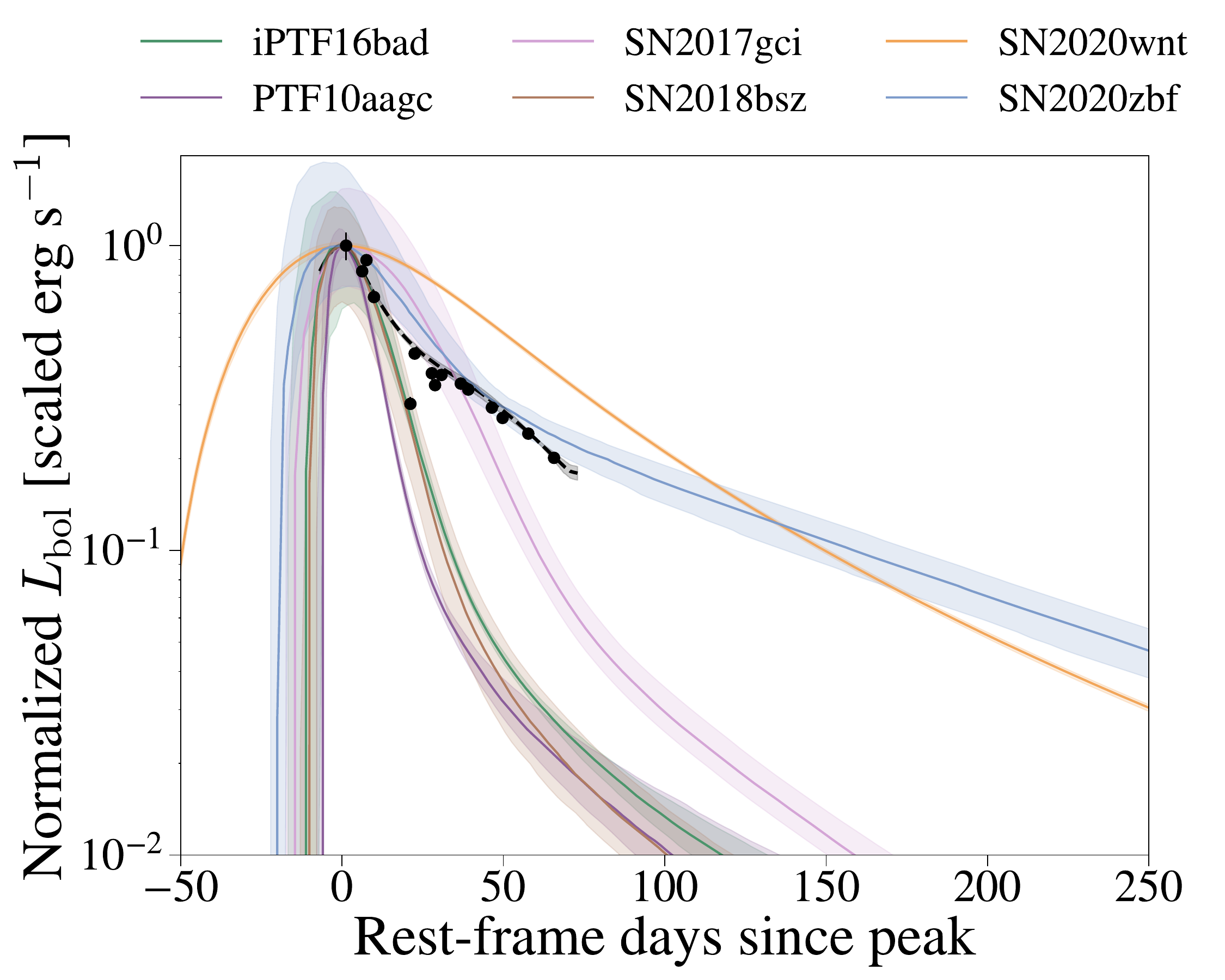}
    \caption{Comparison of the photometrically derived and spectroscopically derived pseudo-bolometric light curves of SN\,2025wny (black) with carbon-rich SLSNe-I in the \citet{Gomez2024_SLSNeI} sample.}
    \label{fig:bol-lc-compare-c-rich}
\end{figure}

A common interpretation for SLSNe-I is that their luminosity is powered, at least in part, by the spin-down of a newly formed magnetar \citep{1971ApJ...164L..95O, 1989ApJ...340..396A, Kasen2010_MagnetarModel, Chatzopoulos2012}. Rotational energy from a rapidly rotating, highly magnetized neutron star is injected into the expanding ejecta, increasing its internal energy on timescales of days to weeks. This additional heating can maintain hot temperatures and high ionization states after maximum light and temporarily shift the spectral energy distribution toward the FUV/NUV \citep{2018AnA...610L..10D}.

The evidence for a magnetar-powered interpretation is as follows. 1) The FUV excess observed in the light curves and spectra between +20--60\,d is similar to the behavior of the magnetar models of \citet{2019AnA...621A.141D}, in which ongoing magnetar energy injection can temporarily enhance FUV emission. 2) Comparisons with the \citet{2019AnA...621A.141D} spectral models also favor, albeit weakly, the r0e2 model, which is magnetar powered. 3) Simple semi-analytic fits to the pseudo-bolometric light curve favor a magnetar model, on its own or in combination with another power source.

Moreover, SN\,2025wny's pseudo-bolometric light curve most closely resembles that of the carbon-rich SLSN-I SN\,2020zbf (Figure~\ref{fig:bol-lc-compare-c-rich}), which has been interpreted as a likely magnetar-powered event via light curve modeling and matches to the C-rich magnetar model spectra in \citet{2019AnA...621A.141D}, although CSM interaction was not ruled out \citep{2024AnA...685A..20G}. Other carbon-rich SLSNe-I in the comparison sample have been linked to more interaction-dominated or otherwise complex scenarios. PTF10aagc and iPTF16bad developed late-time hydrogen emission attributed to interaction with H-rich CSM \citep{2015ApJ...814..108Y, 2017ApJ...848....6Y}; SN\,2018bsz has been argued to be interaction-powered by an aspherical CSM with a hidden central engine \citep{2018AnA...620A..67A, 2022AnA...666A..30P}; and SN\,2020wnt has been interpreted as either a pair-instability-like event or a magnetar-powered explosion with early CSM interaction \citep{2022MNRAS.517.2056G, 2023ApJ...951...34T}.

A SLSN that is centrally heated (e.g., by a magnetar) may also be able to explain the apparent photospheric reheating observed in Section~\ref{subsec:bb-lc} \citep{2019AnA...621A.141D}. Centrally heated SNe are generally hotter at lower velocities than at higher velocities because the heating source is located near the center. As the ejecta expand, they cool, and when the photosphere is close to the outer edge of the ejecta, this cooling dominates the observed temperature evolution. As the photosphere recedes into the ejecta, the ejecta may appear to reheat.

However, it is also unclear whether a simple magnetar model can account for the possible FUV emission features. These issues do not rule out magnetar spin-down as a contributing power source, but they suggest that a magnetar-dominated interpretation would require additional complexity, such as a non-uniform ejecta structure or an additional source of UV emission.

It is possible that SN\,2025wny is powered by both magnetar spin-down and ejecta--CSM interaction. Similar hybrid scenarios have been proposed for the carbon-rich SLSNe-I SN\,2020zbf \citep{2024AnA...685A..20G} and SN\,2020wnt \citep{2022MNRAS.517.2056G}. Together with the light-curve modeling results, which suggest that multiple power sources may contribute, a hybrid magnetar plus CSM-interaction scenario remains plausible for SN\,2025wny.



\subsection{SN\,2025wny is still a SLSN-I}
\label{subsec:slsn-ii}
SN\,2025wny was initially classified as a hydrogen-poor SLSN-I from template matching to its rest-frame UV spectrum \citep{Johansson2025_WinnyClassification}. However, the +57\,d rest-frame optical spectrum presented in this work shows possible signatures of hydrogen and helium. Moreover, despite high temperatures, we do not identify the five \ion{O}{2} absorption features that are characteristic of photospheric-phase SLSNe-I. These properties raise the possibility that SN\,2025wny is not a canonical SLSN-I, and may instead be related to the hydrogen-rich SLSN-II population. This classification could have been missed initially because published rest-frame UV spectra of SLSNe-II are sparse, and SN\,2025wny lacked rest-frame optical coverage at early times.

However, even if forthcoming \textit{JWST}/NIRSpec spectroscopy confirms the presence of hydrogen and helium, SN\,2025wny is still best be interpreted as a SLSN-I. As discussed, SLSNe-Ib and some carbon-rich SLSNe-I develop hydrogen and helium at later times. Moreover, most SLSNe-II exhibit SN\,IIn-like spectra, characterized by narrow H$\alpha$ emission from slowly moving hydrogen-rich CSM along with broader electron-scattering wings. The H$\alpha$ profile observed in SN\,2025wny does not show these characteristics. 

\section{Conclusions}
\label{sec:conclusions}

We present new observations and a detailed analysis of SN\,2025wny, a gravitationally lensed SLSN at $z=2.015$. Our main conclusions are as follows:

\begin{itemize}
\item SN\,2025wny differs from canonical SLSNe-I in several ways. It has (1) the possible emergence of FUV emission lines at +20--60\,d; (2) little evidence for UV line blanketing throughout most of the observed evolution, and instead an FUV continuum excess; (3) no clear \ion{O}{2} absorption complex despite sustained blackbody temperatures above $\sim16{,}000$\,K; and (4) evidence for carbon, hydrogen, and possibly helium in the +57\,d spectrum.

\item The host galaxy of SN\,2025wny is not unusually metal-poor compared to typical $z\sim2$ galaxies or local SLSN host galaxies. Host metallicity alone therefore cannot explain SN\,2025wny's persistent UV brightness or weak FUV line blanketing.

\item The observed properties of SN\,2025wny are not well explained by the standard-picture powering mechanisms invoked for SLSNe.



\end{itemize}

More detailed modeling is required to distinguish between a primarily interaction-powered scenario, a primarily central engine-powered scenario, or a hybrid interpretation. Forthcoming \textit{JWST}/NIRSpec and NIRCam observations (PI: Li, PID: 12574), scheduled roughly around +130\,d and +170\,d post-peak may help elucidate the dominant power mechanism.

Along with the SLSNe-I DES14C1fi and DES16C2nm \citep{Angus2019_DES_SLSNe}, SN\,2025wny is the third SLSN at $z\gtrsim1$ with rest-frame UV to NIR spectroscopic coverage to date. However, SN\,2025wny provides by far the most extensive dataset. 
This dataset provides the most complete empirical view of a high-$z$ SLSN to date and can serve as a valuable template for identifying and characterizing similar events in upcoming wide-field surveys like the Vera C. Rubin Legacy Survey of Space and Time and the Nancy Grace Roman Space Telescope's High Latitude Time Domain Survey, where large numbers ($\sim100$s) of high-$z$ SLSNe are expected to be discovered every year.

\section{Data Availability}
Data and modeling results are available upon request.

\section{Acknowledgements}

We thank William Januszewski, Joel Green, Peter Zeidler, and Elaine Frazer from the Space Telescope Science Institute for their help verifying and scheduling James Webb Space Telescope observations. We thank Christoph Ries, Michael Schmidt, and Silona Wilke for obtaining the HET/LRS observations. We also thank Yuxin Dong and Aswin Suresh for carrying out Keck/LRIS observations.

A.T., S.D., and H.C.T. acknowledge support from  UK Research and Innovation (UKRI) under the UK government’s Horizon Europe funding Guarantee EP/Z000475/1. N.S. is supported by the Kavli Foundation. E.M. acknowledges support from the Swedish Research Council under Dnr VR 2024-03927. R.L. and A.G. are funded by the European Union (ERC, project number 101042299,´ TransPIre). Views and opinions expressed are however those of the author(s) only and do not necessarily reflect those of the European Union or the European Research Council Executive Agency. Neither the European Union nor the granting authority can be held responsible for them. 
A.G.\ acknowledges financial support from the research project grant “Understanding the Dynamic Universe” funded by the Knut and Alice Wallenberg under Dnr KAW 2018.0067, {\em Vetenskapsr\aa det}, the Swedish Research Council through grants projects Dnr 2020-03444 and 2025-03692 the G.R.E.A.T research environment, Dnr 2016-06012, the EDUCATE excellence center funded by the Swedish Research Council through grant Dnr 2022-06627 and the Swedish National Space Agency, Dnr 2023-00226. A.C.G. and the Fong Group at Northwestern acknowledge support by the National Science Foundation under grant Nos. AST-1909358, AST-2206494, AST-2308182 and CAREER grant No. AST-2047919. W. M. Keck Observatory access was supported by Northwestern University and the Center for Interdisciplinary Exploration and Research in Astrophysics (CIERA). W.J.-G.\ is supported by NASA through Hubble Fellowship grant HSTHF2-51558.001-A awarded by the Space Telescope Science Institute, which is operated for NASA by the Association of Universities for Research in Astronomy, Inc., under contract NAS5-26555. K.H. would like to thank the Caltech Presidential Postdoctoral Fellowship program and President Rosenbaum. C.L. is supported by DoE award \#\,DE-SC0025599. A. Singh acknowledges support from the Knut and Alice Wallenberg Foundation through the ``Gravity Meets Light" project. N.R. is supported by a Northwestern University Presidential Fellowship Award. M.W.C. acknowledges support from the National Science Foundation with grant numbers PHY-2117997, PHY-2308862 and PHY-2409481. E.J.M., E.P., A.R., and C.V. acknowledge the INAF project Supporto Arizona \& Italia.
 
Based on observations obtained with the Samuel Oschin Telescope 48-inch and the 60-inch Telescope at the Palomar Observatory as part of the Zwicky Transient Facility project. ZTF is supported by the National Science Foundation under Award \#2407588 and a partnership including Caltech, USA; Caltech/IPAC, USA; University of Maryland, USA; University of California, Berkeley, USA; Cornell University, USA; Drexel University, USA; University of North Carolina at Chapel Hill, USA; Institute of Science and Technology, Austria; National Central University, Taiwan, and the German Center for Astrophysics (DZA), Germany. Operations are conducted by Caltech's Optical Observatory (COO), Caltech/IPAC, and the University of Washington at Seattle, USA.

The ZTF forced-photometry service was funded under the Heising-Simons Foundation grant No. 12540303 (PI: Graham).

SED Machine is based on work supported by the National Science Foundation under grant No. 1106171.

The GROWTH Marshal was supported by the GROWTH project funded by the National Science Foundation under grant No. 1545949.

The data presented here were obtained in part with ALFOSC, which is provided by the Instituto de Astrofisica de Andalucia (IAA) under a joint agreement with the University of Copenhagen and NOT.

This paper used data obtained with the Large Binocular Telescope (LBT). The LBT is an international collaboration among institutions in the United States, Germany and Italy.

Based on observations obtained at the international Gemini Observatory, a program of NSF NOIRLab, which is managed by the Association of Universities for Research in Astronomy (AURA) under a cooperative agreement with the U.S. National Science Foundation on behalf of the Gemini Observatory partnership: the U.S. National Science Foundation (United States), National Research Council (Canada), Agencia Nacional de Investigación y Desarrollo (Chile), Ministerio de Ciencia, Tecnología e Innovación (Argentina), Ministério da Ciência, Tecnologia, Inovações e Comunicações (Brazil), and Korea Astronomy and Space Science Institute (Republic of Korea).

The Liverpool Telescope is operated on the island of La Palma by Liverpool John Moores University in the Spanish Observatorio del Roque de los Muchachos of the Instituto de Astrofisica de Canarias with financial support from the UK Science and Technology Facilities Council. Based on observations made with the Italian Telescopio Nazionale Galileo (TNG) operated on the island of La Palma by the Fundación Galileo Galilei of the INAF (Istituto Nazionale di Astrofisica) at the Spanish Observatorio del Roque de los Muchachos of the Instituto de Astrofisica de Canarias.

The W. M. Keck Observatory is operated as a scientific partnership among the California Institute of Technology, the University of California, and the National Aeronautics and Space Administration. The Observatory was made possible by the generous financial support of the W. M. Keck Foundation. W. M. Keck Observatory and MMT Observatory access was supported by Northwestern University and the Center for Interdisciplinary Exploration and Research in Astrophysics (CIERA). The authors wish to recognize and acknowledge the very significant cultural role and reverence that the summit of Maunakea has always had within the indigenous Hawaiian community. We are most fortunate to have the opportunity to conduct observations from this mountain.

This paper contains data obtained with the 2.1 m Fraunhofer Telescope of the Wendelstein observatory of the Ludwig-Maximilians University Munich. We thank the staff of the Wendelstein observatory for technical help and strong support, including observing targets for us, during the data acquisition. 

This paper includes data taken at The McDonald Observatory of The University of Texas at Austin. The Hobby-Eberly Telescope (HET) is a joint project of the University of Texas at Austin, the Pennsylvania State University, Stanford University, Ludwig-Maximilians-Universität München, and Georg-August-Universität Göttingen

This paper contains data obtained at the Lick Observatory, owned and operated by the University of California. A major upgrade of the Kast spectrograph on the Shane 3 m telescope at Lick Observatory was made possible through
generous gifts from the Heising-Simons Foundation as well as William and Marina Kast. Research at Lick Observatory is partially supported by a generous gift from Google.

Based on observations made with the Gran Telescopio Canarias (GTC), installed at the Spanish Observatorio del Roque de los Muchachos of the Instituto de Astrofísica de Canarias, on the island of La Palma. This work partially based on data obtained with the instrument OSIRIS$+$, built by a Consortium led by the Instituto de Astrofísica de Canarias in collaboration with the Instituto de Astronomía of the Universidad Autónoma de México. OSIRIS was funded by GRANTECAN and the National Plan of Astronomy and Astrophysics of the Spanish Government.

\bibliography{main}{}
\bibliographystyle{aasjournalv7}

\appendix

\onecolumngrid
\section{Photometry}
\label{app:photometry-table}
\input{tables/photometry_table.tex}
\input{tables/ztf_photometry_table}

\onecolumngrid
\section{Spectroscopic observations}
\label{app:spectra-log}

\twocolumngrid
\input{tables/spectra_log.tex}

We used spectroscopy of image A of SN\,2025wny obtained with the instruments listed below:

\begin{itemize}
    \item The IFU mode on the Spectral Energy Distribution Machine (SEDM; $R\sim100$; \citealt{2018PASP..130c5003B, Rigault_2019}) on the Palomar 60-inch telescope (P60). We exclude these spectra from our analysis due to their low resolution, but plot them in Figure~\ref{fig:sedm}.

    \begin{figure}
        \centering
        \includegraphics[width=\linewidth]{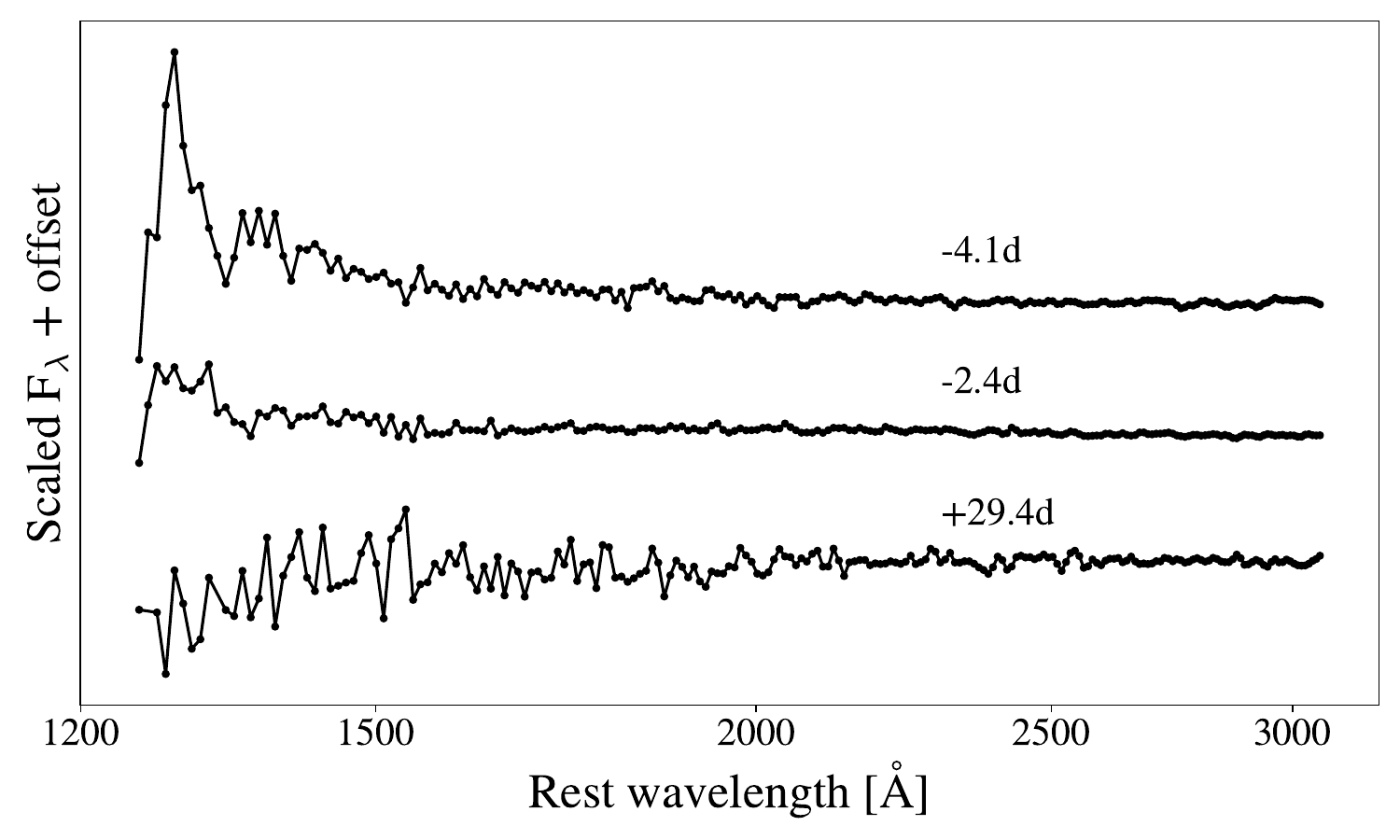}
        \caption{P60/SEDM IFU observations of SN\,2025wny.}
        \label{fig:sedm}
    \end{figure}

    \item The Next Generation Palomar Spectrograph (NGPS; $R\sim1000$; \citealt{2018SPIE10702E..2LJ}) on the
    Palomar 200-inch telescope (P200). For each NGPS epoch, sets of 900\,s exposures were taken using a 1\farcs5 slit and 2x3 spectral-to-spatial binning configuration. Data were processed using 
    \textsc{Quicklook-DRP} (Fremling et al. in prep.).

    \item The Andalucia Faint Object Spectrograph and Camera (ALFOSC; $R\sim350$) on the
    Nordic Optical Telescope (\textit{NOT}). See \citet{Taubenberger2025_SN2025wny} and \citet{Johansson2025_SN2025wny} for details.

    \item The Low Resolution Imaging Spectrometer (LRIS; $R\sim1000$; \citealt{Oke_1995}) on the Keck I 10-meter telescope. For each LRIS epoch, two 900\,s exposures were taken with a 1\farcs0 slit, B400/3400 grism, and R400/8500 grating. Data were processed using \textsc{LPipe} \citep{2019PASP..131h4503P}.

    \item The SuperNova Integral Field Spectrograph (SNIFS; $R\sim1000$; \citealt{2004SPIE.5249..146L}) on the University of
    Hawaii 88-inch telescope (\textit{UH88}). See \citet{Taubenberger2025_SN2025wny} for details.

    \item The Deep Imaging Multi-Object Spectrograph (DEIMOS; $R\sim2000$; \citealt{2003SPIE.4841.1657F}) on the Keck II 10-m telescope. See \citet{Johansson2026_inprep} for details.

    \item The Low Resolution Spectrograph 2 (LRS2; blue $R\sim2500$, red $R\sim1300$; \citealt{2014SPIE.9147E..0AC}) on the Hobby-Eberly Telescope
    (\textit{HET}). See \citet{Johansson2026_inprep} for details.

    \item The Gemini Multi-Object Spectrograph North (GMOS-N) on the Gemini North
    telescope. See \citet{Johansson2026_inprep} for details.

    \item The Keck Cosmic Web Imager (KCWI; using $R\sim3000$; \citealt{2018ApJ...864...93M}) on the Keck II 10-meter telescope. 
    See \citet{Qin2026} for details.

    \item The FOcal Reducer and low dispersion Spectrograph 2 (FORS2; \citealt{1998Msngr..94....1A}) on the
    Very Large Telescope (\textit{VLT}). See \citet{Johansson2026_inprep} for details.

    \item The Optical System for Imaging and low-Intermediate-Resolution
    Integrated Spectroscopy (OSIRIS$+$;  \citealt{1998Ap&SS.263..369C, 10.1117/12.395520}) on the Gran Telescopio Canarias (\textit{GTC}). See \citet{Johansson2026_inprep} for details.

    \item Binospec on the Multiple Mirror Telescope (\textit{MMT}; \citealt{2019PASP..131g5004F}). See \citet{Johansson2026_inprep} for details.

    \item The Near-Infrared Spectrograph (NIRSpec) on the James Webb Space
    Telescope (\textit{JWST}). See \citet{Goobar2026} for information about the first epoch of observations (MJD=61002). For the second epoch (MJD=61113; PID: 12510), we used the IFU mode and the G140M/F100LP, G234/F170LP, and G395M/F290LP grating/filter pairs ($R\sim1000$). Data were processed with the automatic \textit{JWST} Science Calibration Pipeline.
\end{itemize}

\onecolumngrid
\section{Gaussian process for intrinsic luminosity}
\label{app:gp-for-magnification}
\begin{figure*}[h]
    \centering
    \includegraphics[width=0.7\linewidth]{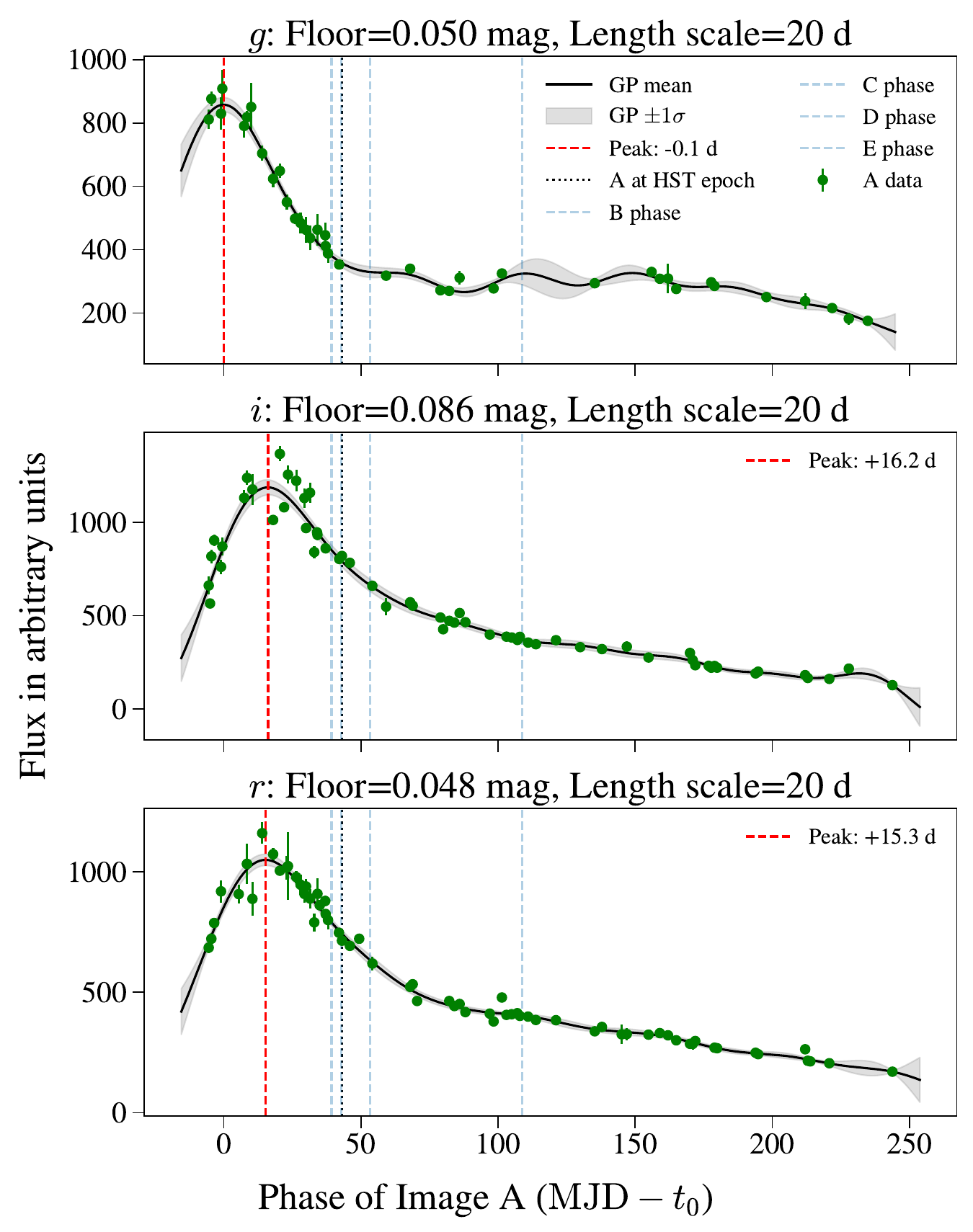}
    \caption{Gaussian-process fits to image A, used to estimate the change in source luminosity between the phases corresponding to Images A and D at the same HST observing epochs, after correcting for the observed time delays measured by \citet{Johansson2026_inprep}.}
    \label{fig:gp-for-mag-corr}
\end{figure*}

\onecolumngrid
\pagebreak
\section{SN\,2025wny light curve timescales}
\label{app:25wny_timescales}
\input{tables/efolding-times.tex}

\onecolumngrid
\section{UV SLSN-I light curve timescales}
\label{app:uv-slsn-timescales}
\input{tables/uv_efolding_comparison_table.tex}

\onecolumngrid
\pagebreak
\section{References for comparison light curves}
\label{app:comparison-lc-refs}
\input{tables/comparison_refs.tex}


\pagebreak
\onecolumngrid
\section{Line profiles}
\twocolumngrid
\label{app:line-profiles}


\begin{figure}[h]
    \centering
    \includegraphics[width=\linewidth]{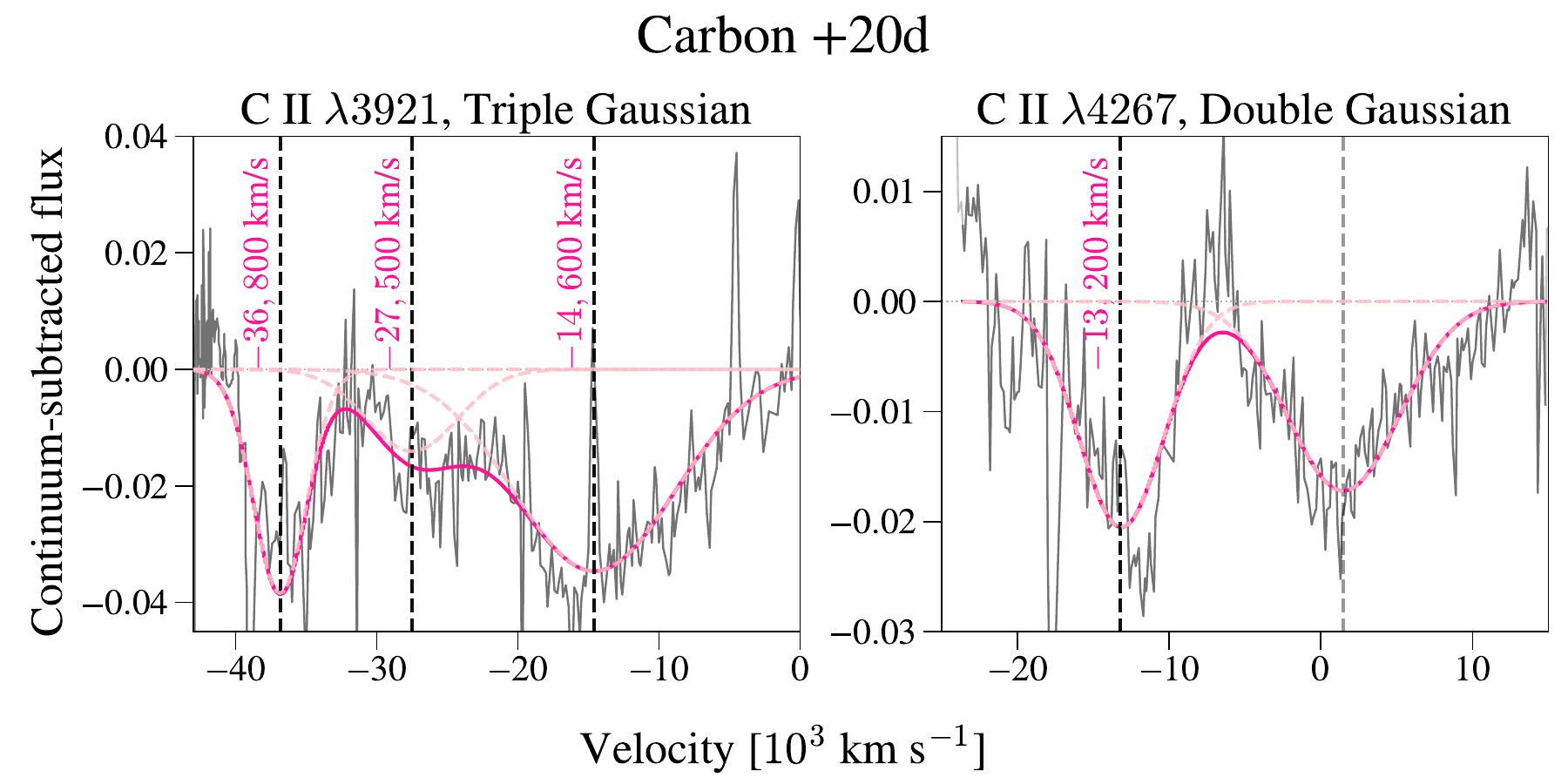}
    \includegraphics[width=\linewidth]{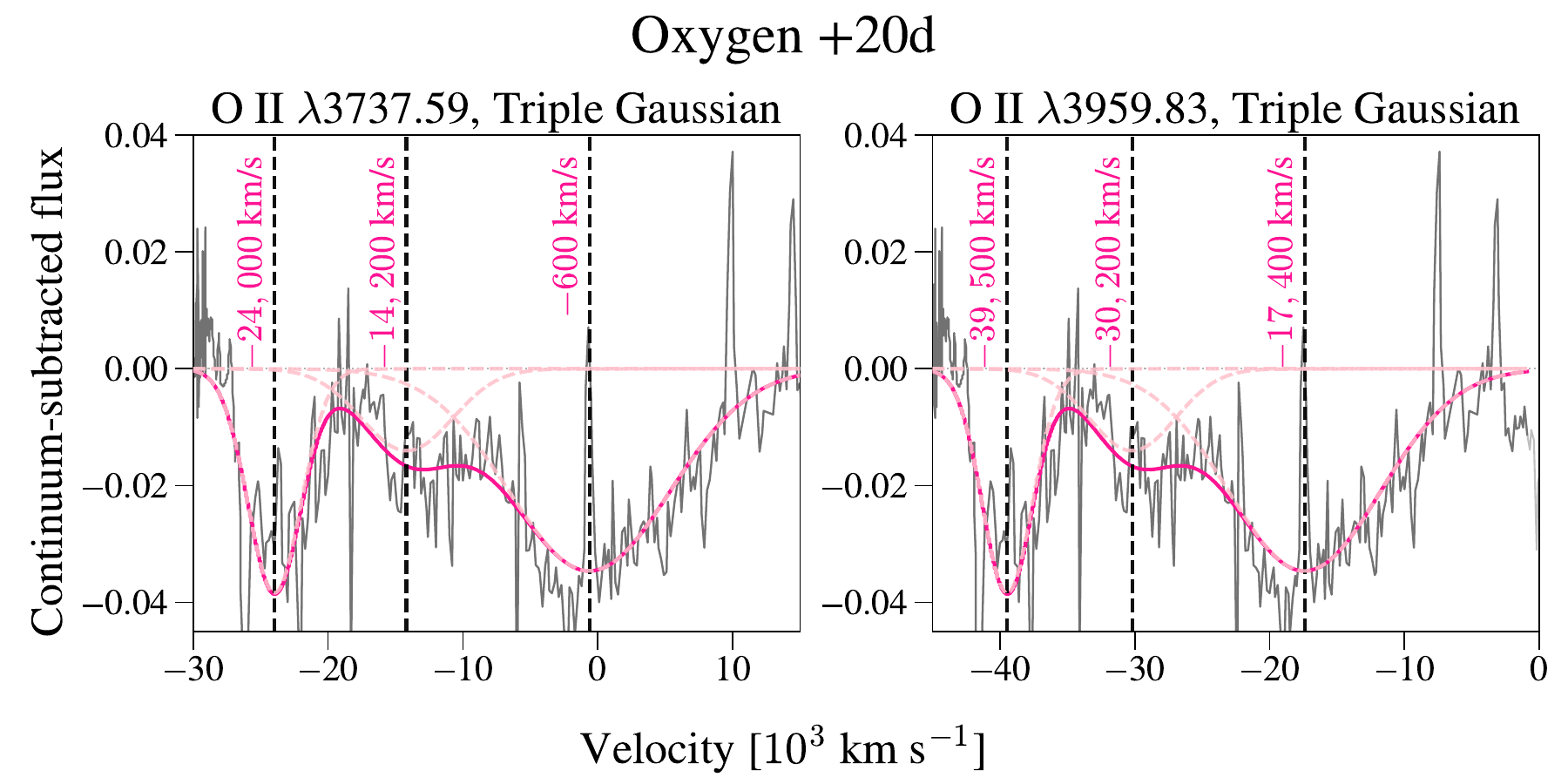}
    \caption{Gaussian fits to candidate \ion{C}{2} (top), and \ion{O}{2} (bottom) features in the $+20$\,d spectrum of SN\,2025wny, shown in velocity space relative to each transition. Continuum-subtracted flux is shown in black, with the total model in pink and individual components in light pink dashed lines. Dashed vertical lines indicate Gaussian line centers. No magnification correction is applied.}
    \label{fig:line_velocities_20d}
\end{figure}

\begin{figure}[h]
    \centering
    \includegraphics[width=\linewidth]{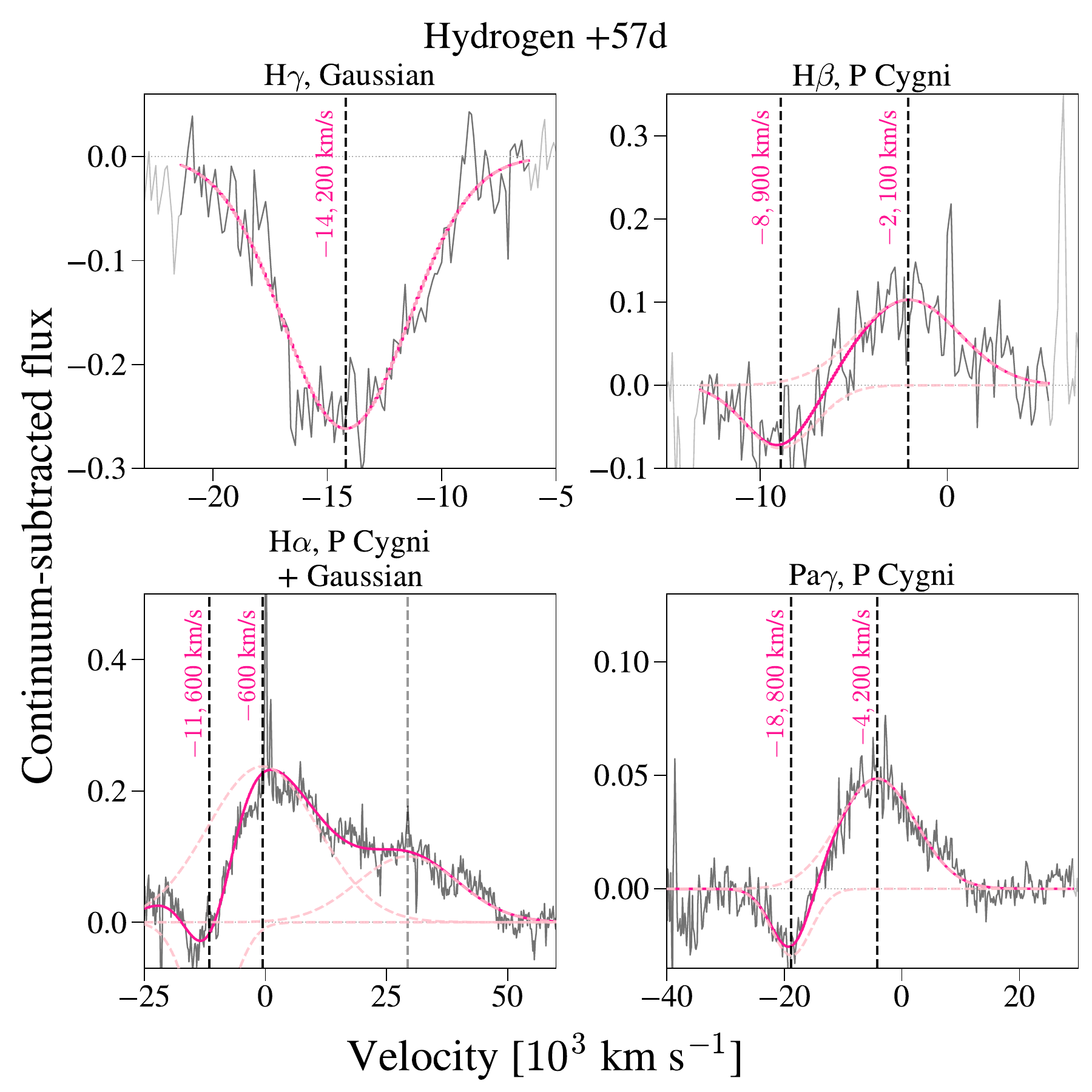}
    \caption{Profile fits to candidate \ion{H}{1} features in the $+57$\,d spectrum of SN\,2025wny, shown in velocity space relative to each transition. Continuum-subtracted flux is shown in black, the total model in pink, and components in light pink dashed lines. Dashed vertical lines mark the fitted absorption minima.}
    \label{fig:hydrogen-line-velocities-58d}
\end{figure}

\begin{figure}[h]
    \centering
    \includegraphics[width=\linewidth]{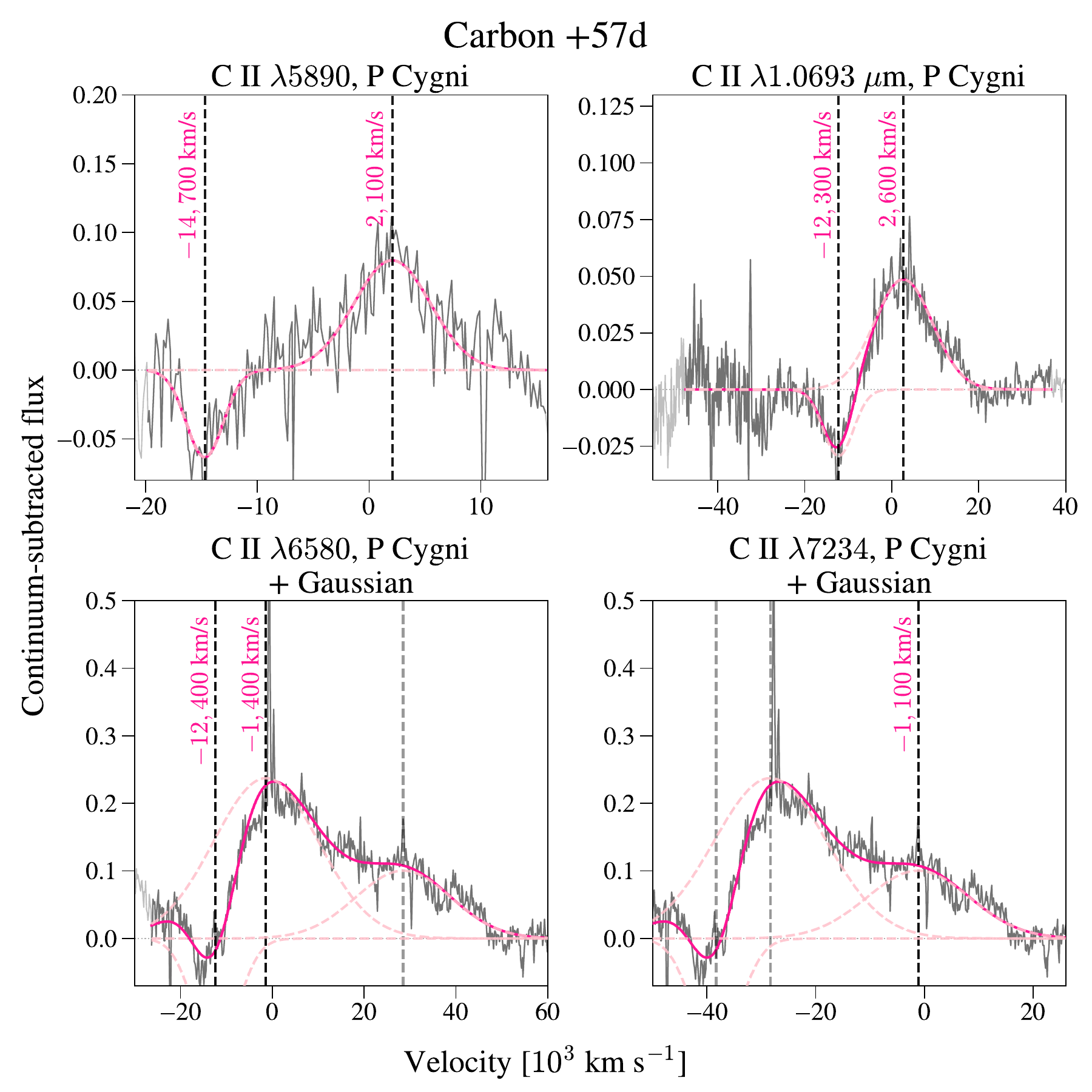}
    \caption{Profile fits to candidate \ion{C}{2} features in the $+57$\,d spectrum of SN\,2025wny, shown in velocity space relative to each transition. Continuum-subtracted flux is shown in black, the total model in pink, and components in light pink dashed lines. Dashed vertical lines mark the fitted absorption minima.}
    \label{fig:carbon-line-velocities-58d}
\end{figure}

\begin{figure}[h]
    \centering
    \includegraphics[width=\linewidth]{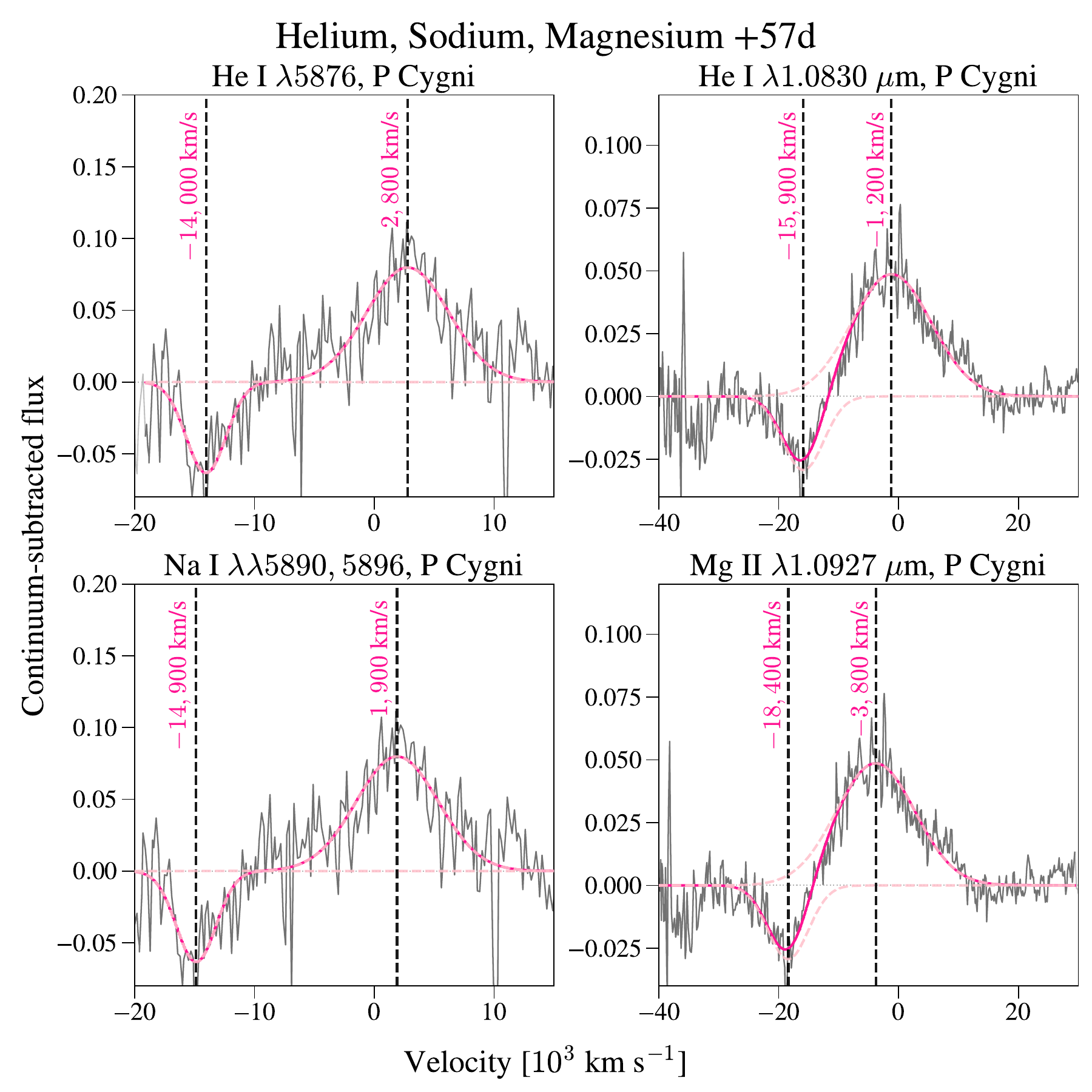}
    \caption{Profile fits to candidate \ion{He}{1} and blended features in the $+57$\,d spectrum of SN\,2025wny, shown in velocity space relative to each transition. Continuum-subtracted flux is shown in black, the total model in pink, and components in light pink dashed lines. Dashed vertical lines mark the fitted absorption minima.}
    \label{fig:other-line-velocities-58d}
\end{figure}

\end{document}

%% file: affiliations.tex
\newcommand{\Caltech}{\affiliation{Cahill Center for Astronomy and Astrophysics, California Institute of Technology, Mail Code 249-17, Pasadena, CA 91125, USA}}

\newcommand{\CaltechPhys}{\affiliation{Division of Physics, Mathematics and Astronomy, California Institute of Technology, Pasadena, CA 91125, USA}}

\newcommand{\TelAviv}{\affiliation{School of Physics and Astronomy, Tel Aviv University, Tel Aviv 6997801, Israel}}

\newcommand{\CaltechOO}{\affiliation{Caltech Optical Observatories, California Institute of Technology, Pasadena, CA 91125, USA}}

\newcommand{\OKCAstro}{\affiliation{Department of Astronomy, Oskar Klein Center, Stockholm University, SE-106 91 Stockholm, Sweden}}

\newcommand{\Birmingham}{\affiliation{School of Physics \& Astronomy and Institute for Gravitational Wave Astronomy, University of Birmingham, Birmingham B15 2TT, UK}}

\newcommand{\KIC}{\affiliation{Kavli Institute for Cosmology, University of Cambridge, Madingley Road, Cambridge, CB3 0HA, UK}}

\newcommand{\WIS}{\affiliation{Department of Particle Physics and Astrophysics, Weizmann Institute of Science, 76100 Rehovot, Israel}}

\newcommand{\IoA}{\affiliation{Institute of Astronomy, University of Cambridge, Madingley Road, Cambridge CB3 0HA, UK}}

\newcommand{\OKC}{\affiliation{Department of Physics, Oskar Klein Centre, Stockholm University, SE-106 91, Stockholm, Sweden}}

\newcommand{\Liverpool}{\affiliation{Astrophysics Research Institute, Liverpool John Moores University, Liverpool,  L3 5RF, UK}}

\newcommand{\LMU}{\affiliation{University Observatory, Faculty of Physics, Ludwig-Maximilians-Universität, Scheinerstr. 1, 81679 Munich, Germany}}

\newcommand{\ORIGINS}{\affiliation{Excellence Cluster ORIGINS, Boltzmannstr. 2, 85748 Garching, Germany}}

\newcommand{\CIERA}{\affiliation{Center for Interdisciplinary Exploration and Research in Astrophysics (CIERA), 1800 Sherman Ave., Evanston, IL 60201, USA}}

\newcommand{\Northwestern}{\affiliation{Department of Physics and Astronomy, Northwestern University, 2145 Sheridan Rd, Evanston, IL 60208, USA}}

\newcommand{\SkAI}{\affiliation{NSF-Simons AI Institute for the Sky (SkAI), 172 E. Chestnut St., Chicago, IL 60611, USA}}

\newcommand{\INAFMontePorzioCatone}{\affiliation{Osservatorio Astronomico di Roma, via Frascati 33, 00040 Monte Porzio Catone, Italy}}

\newcommand{\INAFBologna}{\affiliation{Osservatorio di Astrofisica e Scienza dello Spazio di Bologna, via Piero Gobetti 93/3, I-40129 Bologna, Italy}}

\newcommand{\CSIC}{\affiliation{Instituto de Astrof\'{\i}sica de Andaluc\'{\i}a CSIC, Granada, E-18008, Spain}}

\newcommand{\LMUPhysics}{\affiliation{Fakult\"at f\"ur Physik, LMU M\"unchen, Geschwister-Scholl-Platz 1, 80539 Munich, Germany}}

\newcommand{\Minnesota}{\affiliation{School of Physics and Astronomy, University of Minnesota, Minneapolis, Minnesota 55455, USA}}

\newcommand{\Drexel}{\affiliation{Department of Physics, Drexel University, Philadelphia, PA 19104, USA}}

\newcommand{\UCLA}{\affiliation{Department of Physics and Astronomy, UCLA PAB 430 Portola Plaza Los Angeles, CA 90095-1547}}

%% file: authors.tex
\author[0009-0001-6911-9144]{Maggie~L.~Li}
\email[show]{maggieli@caltech.edu}
\Caltech

\author[0000-0003-1710-9339]{Lin~Yan}
\email{lyan@caltech.edu}
\CaltechPhys
\CaltechOO

\author[0009-0000-9383-2305]{Anamaria~Gkini}
\email{anamaria.gkini@astro.su.se}
\OKCAstro

\author[0000-0001-6343-3362]{Alice~Townsend}
\email{a.townsend@bham.ac.uk}
\Birmingham

\author[0000-0003-3658-6026]{Yu-Jing~Qin}
\email{qinyj.astro@gmail.com}
\Caltech

\author[0000-0001-6797-1889]{Steve~Schulze}
\email{steve.schulze@weizmann.ac.il}
\WIS

\author[0000-0003-2700-1030]{Nikhil~Sarin}
\email{nsarin.astro@gmail.com}
\KIC
\IoA

\author[0000-0002-2376-6979]{Suhail~Dhawan}
\email{s.dhawan@bham.ac.uk}
\Birmingham

\author[0000-0001-5975-290X]{Joel~Johansson}
\email{joeljo@fysik.su.se}
\OKC

\author[0000-0002-4163-4996]{Ariel~Goobar}
\email{ariel@fysik.su.se}
\OKC

\author[0009-0009-6243-8300]{Jacob~Osman~Hjortlund}
\email{jacob.hjortlund@fysik.su.se}
\OKC

\author[0000-0002-8380-6143]{Edvard~Mörtsell}
\email{edvard@fysik.su.se}
\OKC

\author[0000-0003-1546-6615]{Jesper~Sollerman}
\email{jesper@astro.su.se}
\OKCAstro

\author[0000-0001-9454-4639]{Ragnhild~Lunnan}
\email{ragnhild.lunnan@astro.su.se}
\OKCAstro

\author[0000-0001-8472-1996]{Daniel~A.~Perley}
\email{D.A.Perley@ljmu.ac.uk}
\Liverpool

\author[0000-0002-4534-7089]{Ehud~Nakar}
\email{ehud.nakar@gmail.com}
\TelAviv

\author[0000-0003-2091-622X]{Avinash~Singh}
\email{avinash21292@gmail.com}
\OKCAstro

\author[0000-0002-3653-5598]{Avishay~Gal-Yam}
\email{avishay.gal-yam@weizmann.ac.il}
\WIS

\author[0000-0002-5619-4938]{Mansi~M.~Kasliwal}
\email{mansi@astro.caltech.edu}
\Caltech

\author[0000-0002-9646-8710]{Conor~M.~B.~Omand}
\email{C.M.Omand@ljmu.ac.uk}
\Liverpool


\author[0009-0008-2714-2507]{Aleksandra~Bochenek}
\email{A.M.Bochenek@2023.ljmu.ac.uk}
\Liverpool

\author[0009-0001-0574-2332]{Malte~Busmann}
\email{m.busmann@physik.lmu.de}
\LMU
\ORIGINS

\author[0000-0001-8372-997X]{Kaustav~K.~Das}
\email{kdas@caltech.edu}
\Caltech



\author[0000-0002-4223-103X]{Christoffer~Fremling}
\email{fremling@caltech.edu}
\CaltechPhys
\CaltechOO

\author[0000-0002-5025-4645]{Alexa~C.~Gordon}
\email{alexagordon2026@u.northwestern.edu}
\CIERA

\author[0000-0003-3270-7644]{Daniel~Gruen}
\email{daniel.gruen@lmu.de}
\LMU
\ORIGINS

\author[0000-0002-3934-2644]{Wynn~Jacobson-Gal\'{a}n}
\email{wynnjg@caltech.edu}
\Caltech

\author[0000-0002-0129-806X]{K-Ryan~Hinds}
\email{khinds@caltech.edu}
\CaltechPhys

\author[0000-0002-7866-4531]{Chang~Liu}
\email{ptg.cliu@u.northwestern.edu}
\Northwestern
\CIERA
\SkAI

\author[0009-0006-0726-1328]{Zoë~McGrath}
\email{Z.McGrath@2024.ljmu.ac.uk}
\Liverpool

\author[0000-0001-8368-8565]{Ezequiel~J.~Marchesini}
\email{ezequiel.marchesini@inaf.it}
\INAFBologna

\author[0000-0002-8650-1644]{Christopher~Martin}
\email{martinc@caltech.edu}
\Caltech

\author[0009-0005-8228-0329]{Peter~Massey}
\Birmingham
\email{pxm588@student.bham.ac.uk}

\author[0000-0002-0610-2644]{Martijn~S.~S.~L.~Oei}
\email{oei@caltech.edu}
\Caltech

\author[0000-0002-8691-7666]{Eliana~Palazzi}
\email{eliana.palazzi@inaf.it}
\INAFBologna

\author[0000-0001-7145-8674]{Francisco~Prada}
\email{f.prada@csic.es}
\CSIC

\author[0000-0001-5847-7934]{Nikolaus~Z.~Prusinski}
\email{nik@astro.caltech.edu}
\Caltech

\author[0000-0002-5683-2389]{Nabeel~Rehemtulla}
\email{nabeelr@u.northwestern.edu}
\Northwestern
\CIERA
\SkAI

\author[0000-0002-8860-6538]{Andrea~Rossi}
\email{andrea.rossi@inaf.it}
\INAFBologna

\author[0000-0003-0427-8387]{R.~Michael~Rich}
\email{rmrastro@gmail.com}
\UCLA

\author[0000-0003-4725-4481]{Sam~Rose}
\email{srose@caltech.edu}
\Caltech

\author[0009-0000-7976-1416]{Killa~Santer}
\email{killasanter0@gmail.com}
\LMUPhysics

\author[0009-0006-1510-4648]{Surya~Shivaprasad}
\email{surya.shivaprasad@physik.lmu.de}
\LMU
\ORIGINS

\author[0009-0006-7102-3674]{Hannah~C.~Turner}
\Birmingham
\email{h.c.turner@bham.ac.uk}

\author[0009-0009-9751-9215]{Chiara~Ventura}
\email{chiara.ventura@inaf.it}
\INAFMontePorzioCatone

\author[0000-0003-0733-2916]{Jacob~L.~Wise}
\email{J.L.Wise@2022.ljmu.ac.uk}
\Liverpool


\author[0000-0002-8262-2924]{Michael~Coughlin}
\email{cough052@umn.edu}
\Minnesota

\author[0000-0001-7357-0889]{Argyro~Sasli}
\email{asasli@umn.edu}
\Minnesota

\author[]{Niharika~Sravan}
\email{niharika.sravan@gmail.com}
\Drexel

%% file: tables/bolometric_lc_table.tex
\begin{deluxetable}{lllll}[h]
\setlength{\tabcolsep}{3pt}
\tablecaption{Peak values as derived in Section~\ref{subsec:bol-lc}.}
\label{tab:bol-lc-results}
\tablehead{
\colhead{Method} &
\colhead{$t_\mathrm{peak}$} &
\colhead{$L_\mathrm{peak}$} &
\colhead{$T(L_\mathrm{peak})$} &
\colhead{$E_\mathrm{rad}$} \\
\colhead{} &
\colhead{(MJD)} &
\colhead{($10^{44}$\,erg\,s$^{-1}$)} &
\colhead{(K)} &
\colhead{($10^{51}$\,erg)}
}
\startdata
Photometry & 60940.7 & $\gtrsim3.9_{-1.0}^{+0.5}$ & $17900 \pm 720$ & $\gtrsim1.18_{-0.01}^{+0.01}$ \\
Spectra & \nodata & $4.4_{-0.8}^{+1.2}$ & $17200^{+720}_{-460}$ & \nodata
\enddata
\tablecomments{The values reported here are the 16th-50th-84th percentiles and are corrected for magnification (see Section~\ref{sec:magnification-correction}.}
\end{deluxetable}

%% file: tables/lbol-priors.tex
\begin{deluxetable}{lll}[t]
\setlength{\tabcolsep}{2pt}
\tablecaption{Priors used for the simple semi-analytic bolometric light-curve fits.
\label{tab:bolometric-fit-priors}}
\tablehead{
\colhead{Parameter} &
\colhead{Prior} &
\colhead{Description}
}
\startdata
$\log_{10}(t_{\rm diff}/{\rm d})$ & $\mathcal{U}(-0.5, 3.0)$ &
Diffusion timescale \\
$t_{\rm rise}$ & $\mathcal{U}(1, 300)$\,d &
Rest-frame days from \\
& & explosion to bol. peak \\
$\log_{10}\sigma_{\rm frac}$ & $\mathcal{U}(-5, 2)$ &
Additional noise term \\
$\log_{10}(E_{\rm p}/{\rm erg})$ & $\mathcal{U}(50, 55)$ &
Magnetar rotational energy \\
$\log_{10}(t_{\rm p}/{\rm d})$ & $\mathcal{U}(-2, 3)$ &
Magnetar spin-down timescale \\
$\log_{10}(M_{\rm Ni}/M_\odot)$ & $\mathcal{U}(-3, 2)$ &
Nickel mass \\
$\log_{10}L_{\rm CSM,1}$ & $\mathcal{U}(42, 53)$ &
CSM interaction normalization \\
$\log_{10}L_{\rm fallback,1}$ & $\mathcal{U}(42, 53)$ & Fallback normalization \\
\enddata
\tablecomments{
$\mathcal{U}(a,b)$ denotes a uniform prior between $a$ and $b$. The first three parameters are included in all models.
}
\end{deluxetable}

%% file: tables/metallicities.tex
\begin{deluxetable}{lccc}
\tablecaption{Host galaxy metallicities measured for various luminous ($M\lesssim-22$ in any band) SLSNe-I.\label{tab:slsni-metallicities}}
\tablehead{
\colhead{Name} &
\colhead{Peak SN abs. mag.} &
\colhead{$12 + \log_{10}({\rm O/H})$} & 
\colhead{Ref.}
}
\startdata
PTF09cnd & $M_V=-23$ & $8.21_{-0.71}^{+0.22}$ & P16 \\
PTF10uhf & $M_V=-22$ & $9.00_{-0.05}^{+0.06}$ & P16 \\
PTF10vqv & $M_V=-22.5$ & $8.38_{-0.10}^{+0.08}$ & P16 \\
PTF11dsf & $M_V=-22.1$ & $8.20_{-0.06}^{+0.05}$ & P16 \\
PTF12mxx & $M_V=-22.5$ & $<8.38$ & P16 \\
SN~2018hti & $M_g=-22.2$ & 8.13--8.21 & L20 \\
SN~2018lfe & $M_r=-22.1$ & $8.18\pm0.12$ & Y22 \\
SN~2020xga & $M_g=-22.3$ & $7.96_{-0.18}^{+0.22}$ & G25
\enddata
\tablecomments{P16: \citet{2016ApJ...830...13P}; L20: \citet{2020MNRAS.497..318L}; Y22: \citet{2022ApJ...931...32Y}; G25: \citet{2025AandA...694A.292G}.}
\end{deluxetable}

%% file: tables/photometry_table.tex
\begin{deluxetable*}{ccccccc}[h]
\tablecaption{Resolved and S-corrected photometry of SN~2025wny.\label{tab:photometry}.}
\tablehead{
\colhead{MJD} &
\colhead{Filter} &
\colhead{Mag\tablenotemark{a}} &
\colhead{$\sigma_{\rm mag}$} &
\colhead{Mag$_{\rm S-corr}$} &
\colhead{$\sigma_{\rm mag,S-corr}$} &
\colhead{Instrument}
}
\startdata
60919.51 & $g$ & 19.73 & 0.04 & 19.82 & 0.04 & P60/SEDM \\
60919.51 & $r$ & 19.91 & 0.04 & 19.90 & 0.04 & P60/SEDM \\
60919.51 & $i$ & 19.95 & 0.08 & 19.96 & 0.08 & P60/SEDM \\
60920.47 & $g$ & 19.64 & 0.03 & 19.73 & 0.03 & P60/SEDM \\
60920.48 & $r$ & 19.85 & 0.03 & 19.84 & 0.03 & P60/SEDM \\
60920.48 & $i$ & 19.72 & 0.05 & 19.73 & 0.05 & P60/SEDM \\
60921.50 & $r$ & 19.76 & 0.03 & 19.75 & 0.03 & P60/SEDM \\
60921.50 & $i$ & 19.61 & 0.04 & 19.62 & 0.04 & P60/SEDM \\
60924.47 & $g$ & 19.60 & 0.07 & 19.69 & 0.07 & P60/SEDM \\
\nodata
\enddata
\tablecomments{Only a portion of this table is shown here to demonstrate its form and content. A machine-readable version of the full table is available.}
\tablenotetext{a}{Magnitudes are reported in the AB system and are not corrected for Milky Way extinction.}
\end{deluxetable*}

%% file: tables/ztf_photometry_table.tex
\begin{deluxetable*}{ccccc}[h]
\tablecaption{ZTF unresolved photometry of SN~2025wny.}
\tablehead{
\colhead{MJD} & 
\colhead{Mag\tablenotemark{a}} & 
\colhead{Mag Err.} & 
\colhead{Filter} & 
\colhead{Quality}
}
\startdata
60913.50 & 20.63 & 0.20 & $ztfr$ & good \\
60914.50 & 20.46 & 0.20 & $ztfr$ & good \\
60914.51 & 20.40 & 0.20 & $ztfr$ & good \\
60916.51 & 20.41 & 0.18 & $ztfr$ & good \\
60917.50 & 20.22 & 0.13 & $ztfr$ & good \\
60917.51 & 20.24 & 0.15 & $ztfr$ & good \\
60922.50 & 19.68 & 0.09 & $ztfr$ & good \\
60922.51 & 19.70 & 0.09 & $ztfr$ & good \\
60922.51 & 19.61 & 0.08 & $ztfr$ & good \\
60923.51 & 19.45 & 0.07 & $ztfr$ & good \\
\nodata
\enddata
\tablecomments{Only a portion of this table is shown here to demonstrate its form and content. A machine-readable version of the full table is available.}
\tablenotetext{a}{Magnitudes are reported in the AB system and are not corrected for Milky Way extinction.}
\label{tab:ztf_photometry}
\end{deluxetable*}

%% file: tables/spectra_log.tex
\begin{deluxetable*}{lccccc}[h]

\tablecaption{Log of spectroscopic observations of SN2025wny.\label{tab:spectra-log}}
\tablehead{
MJD & 
Phase$^{a}$ & 
Rest Wave. & 
Instrument & 
PI &
Reference \\
&
(days) &
({\AA}) &
&
&
}
\startdata
60928 & -4 & 1200--3000 & P60/SEDM & Kulkarni & This work \\
60933 & -3 & 1200--3000 & P60/SEDM & Kulkarni & This work \\
60942 & 0 & 1880--3400 & P200/NGPS & Das & J25 \\
60944 & 1 & 1160--3220 & NOT/ALFOSC & Public & T25+J25 \\
60959 & 6 & 1160--3220 & NOT/ALFOSC & Public & T25+J25 \\
60963 & 7 & 1160--3220 & UH88/SNIFS & Public & T25 \\
60966 & 8 & 1170--3740 & Lick/KAST & Rich & This work \\
60967 & 9 & 1160--1780 & NOT/ALFOSC & Public & T25+J25 \\
60970 & 10 & 2150--3020 & Keck/DEIMOS & Rich & This work \\
60970 & 10 & 1020--3410 & Keck/LRIS & Qin & J25 \\
60971 & 10 & 1550--2430 & Keck/DEIMOS & Rich & This work \\
60975 & 11 & 1870--3400 & P200/NGPS & Fremling & This work \\
60979 & 13 & 1210--2310 & HET/LRS2 & Gruen & This work \\
60995 & 18 & 1550--3160 & Gemini/GMOS-N & Dhawan & This work \\
61004 & 21 & 1200--3380 & Keck/KCWI & Martin & {Q26} \\
61007 & 22 & 1000--3260 & GTC/OSIRIS & Prada & {J26}+This work \\
61008 & 22 & 1270--3060 & MMT/BINOSPEC & Rehemtulla & {J26}+This work\\
61024 & 28 & 1000--3320 & VLT/FORS2 & Dhawan & {J26}+This work \\
61027 & 29 & 1080--3400 & Keck/LRIS & Liu & {J26}+This work \\
61029 & 29 & 1200--3000 & P60/SEDM & Kulkarni & This work \\
61033 & 31 & 990--3210 & VLT/FORS2 & Dhawan & {J26}+This work \\
61051 & 37 & 1160--2990 & Keck/KCWI & Qin & {Q26} \\
61058 & 39 & 1130--3210 & VLT/FORS2 & Dhawan & {J26}+This work \\
61068 & 42 & 1010--3440 & P200/NGPS & Fremling &  {J26}+This work\\
61080 & 46 & 1130--3210 & VLT/FORS2 & Dhawan & {J26}+This work \\
61090 & 50 & 1020--3420 & Keck/LRIS & Kasliwal & {J26}+This work \\
61114 & 58 & 1130--3210 & VLT/FORS2 & Dhawan & {J26}+This work \\
61138 & 66 & 1130--3200 & VLT/FORS2 & Dhawan & {J26}+This work \\
\hline
61002 & 20 & 3220--6280 & JWST/NIRSpec & Goobar & {G26} \\
61113 & 57 & 3220--17500 & JWST/NIRSpec & Li & This work\\
\enddata
\tablecomments{T25: \citet{Taubenberger2025_SN2025wny}; J25: \citet{Johansson2025_SN2025wny}; G26: {Goobar et al. (2026)}; J26: {Johansson et al. (2026)}; Q26: {Qin et al. (2026)}.}
\tablenotetext{a}{Rest-frame phase relative to peak, with MJD$_{\rm peak}=60940.7$.}
\end{deluxetable*}

%% file: tables/efolding-times.tex
\begin{deluxetable*}{lccccccccc}[h]
\tablecaption{Timescales measured from the observed $grizJ$ and synthetic \textit{Swift}/UVOT photometry of SN\,2025wny.}
\label{tab:efolding-times}
\tablehead{
\colhead{Filter\tablenotemark{$\dagger$}} &
\colhead{$t_\mathrm{peak}$} &
\colhead{$\tau_\mathrm{rise}$} &
\colhead{Phase range\tablenotemark{$a$}} &
\colhead{$\tau_1$} &
\colhead{Phase range} &
\colhead{$\tau_2$} &
\colhead{Phase range} &
\colhead{$\tau_3$} &
\colhead{Phase range} \\
\colhead{} &
\colhead{(MJD)} &
\colhead{(d)} &
\colhead{(d)} &
\colhead{(d)} &
\colhead{(d)} &
\colhead{(d)} &
\colhead{(d)} &
\colhead{(d)} &
\colhead{(d)}
}
\startdata
1600\,{\AA} ($g$) & \nodata & \nodata & \nodata & $15 \pm 1$ & $0$--$12$ & $\gtrsim43$ & $12$--$55$ & $36 \pm 6$ & $55$--$73$ \\
2090\,{\AA} ($r$) & 60933.5 & $13 \pm 1$ & $-7$--0 & $20 \pm 1$ & $0$--$19$ & $60 \pm 3$ & $19$--$76$ & \nodata & \nodata \\
2580\,{\AA} ($i$) & 60945.5 & $11 \pm 2$ & $-7$--$-1$ & $18 \pm 2$ & $0$--$15$ & $44 \pm 1$ & $15$--$76$ & \nodata & \nodata \\
2990\,{\AA} ($z$) & \nodata & \nodata & \nodata & $35 \pm 1$ & $6$--$40$ & $40 \pm 5$ & $40$--$71$ & \nodata & \nodata \\
4100\,{\AA} ($J$) & \nodata & \nodata & \nodata & $44 \pm 2$ & $6$--$71$ & \nodata & \nodata & \nodata & \nodata \\
5400\,{\AA} ($H$) & \nodata & \nodata & \nodata & $57 \pm 14$ & $11$--$46$ & \nodata & \nodata & \nodata & \nodata \\
UVW2\tablenotemark{$\ddag$} & \nodata & \nodata & \nodata & $30 \pm 0$ & 0--25 & $70 \pm 0$ & 25--80 & \nodata & \nodata \\
UVM2\tablenotemark{$\ddag$} & \nodata & \nodata & \nodata & $25 \pm 0$ & 0--25 & $54 \pm 0$ & 25--80 & \nodata & \nodata \\
UVW1\tablenotemark{$\ddag$} & \nodata & \nodata & \nodata & $29 \pm 0$ & 0--30 & $49 \pm 0$ & 25--80 & \nodata & \nodata \\
\enddata
\tablenotetext{a}{Rest-frame days since peak MJD=60940.7.}
\tablenotetext{\dagger}{Approximate rest-frame effective wavelengths are listed, with the corresponding observer-frame filter given in parentheses.}
\tablenotetext{$\ddag$}{Synthetic photometry derived from integrating the rest-frame UV spectra of SN\,2025wny over the \textit{Swift}/UVOT bandpasses.}
\tablecomments{The decline timescales were measured by fitting linear segments to the light curves in magnitude space. The corresponding flux $e$-folding times are computed as $\tau = 1.0857/s$, where $s$ is the fitted slope in mag\,d$^{-1}$. The reported uncertainties are fitting uncertainties and do not include systematic uncertainties.}
\end{deluxetable*}

%% file: tables/uv_efolding_comparison_table.tex
\begin{deluxetable*}{lcccccc}[h]
\tablecaption{Rest-frame UV rise and decline $e$-folding timescales.}
\label{tab:uv-slsn-timescales}
\tablehead{
\colhead{Name} &
\multicolumn{2}{c}{UVW2} &
\multicolumn{2}{c}{UVM2} &
\multicolumn{2}{c}{UVW1} \\
\cline{2-3}\cline{4-5}\cline{6-7}
&
\colhead{$\tau_{\rm rise}$} &
\colhead{$\tau_{\rm decline}$} &
\colhead{$\tau_{\rm rise}$} &
\colhead{$\tau_{\rm decline}$} &
\colhead{$\tau_{\rm rise}$} &
\colhead{$\tau_{\rm decline}$}
}
\startdata
SN2015bn & $175 \pm 95$ & $24 \pm 1$ & $116 \pm 36$ & $22 \pm 1$ & $112 \pm 47$ & $31 \pm 1$ \\
SN2016eay & \nodata & $23 \pm 1$ & \nodata & $23 \pm 1$ & $154 \pm 149$ & $15 \pm 0$ \\
SN2017egm & $32 \pm 2$ & $14 \pm 0$ & $26 \pm 1$ & $13 \pm 0$ & $20 \pm 1$ & $18 \pm 1$ \\
SN2018ibb & \nodata & $44 \pm 43$ & \nodata & $46 \pm 9$ & \nodata & $54 \pm 11$ \\
SN2020qlb & \nodata & $64 \pm 6$ & $43 \pm 40$ & $53 \pm 3$ & $61 \pm 52$ & $54 \pm 2$ \\
SN2020znr & $50 \pm 6$ & $66 \pm 3$ & $57 \pm 9$ & $62 \pm 2$ & $40 \pm 3$ & $63 \pm 2$ \\
SN2021ek & \nodata & $45 \pm 19$ & $76 \pm 98$ & $57 \pm 50$ & \nodata & $32 \pm 8$ \\
SN2021lwz & $49 \pm 78$ & $5 \pm 1$ & $6 \pm 2$ & $8 \pm 1$ & $9 \pm 2$ & $7 \pm 0$ \\
SN2024ahr & $61 \pm 22$ & $92 \pm 18$ & $50 \pm 21$ & $54 \pm 10$ & $52 \pm 15$ & $54 \pm 6$ \\
SN2024rmj & $46 \pm 4$ & $72 \pm 9$ & $37 \pm 3$ & $61 \pm 5$ & $21 \pm 2$ & $78 \pm 5$ \\
SN2025afcr & $59 \pm 26$ & $17 \pm 1$ & $21 \pm 5$ & $19 \pm 1$ & $32 \pm 5$ & $16 \pm 1$ \\
\enddata
\end{deluxetable*}

%% file: tables/comparison_refs.tex
\begin{deluxetable*}{lc}[h]
\tablecaption{Sources of the photometric data used for individual objects in the light curve comparisons shown in Figures~\ref{fig:uv-lc-comparison} and \ref{fig:gband-compare}. \label{tab:comparison-refs}}
\tablehead{
    \colhead{Name} &
    \colhead{Ref.}
}
\startdata
ASASSN15no & \citet{2018MNRAS.476..261B} \\
iPTF15eov & \citet{2019AnA...621A..71T} \\
SN~2015bn & \citet{2016ApJ...826...39N, 2016ApJ...828L..18N, 2018ApJ...866L..24N} \\ 
SN~2016eay (Gaia16apd) & \citet{2017ApJ...840...57Y, 2017ApJ...835L...8N, Gomez2024_SLSNeI} \\
SN~2017egm & \citet{2017ApJ...845L...8N, 2018ApJ...858...91Y} \\
SN~2017gci & \citet{2021MNRAS.502.2120F} \\
SN~2018beh & \citet{2022ApJ...941..107G, Gomez2024_SLSNeI} \\
SN~2018bgv & \citet{2020ApJ...901...61L, Gomez2024_SLSNeI} \\
SN~2018bsz & \citet{2018AnA...620A..67A, Gomez2024_SLSNeI} \\
SN~2018don & \citet{2020ApJ...901...61L, Chen2023_PopulationStudy, Gomez2024_SLSNeI} \\
SN~2018ibb & \citet{Schulze2024_18ibb, Gomez2024_SLSNeI} \\
SN~2019dwa & \citet{2021MNRAS.508.4342P, Gomez2024_SLSNeI} \\
SN~2019hge & \citet{Yan2020_HeI, 2021MNRAS.508.4342P, 2022ApJ...941..107G}; \\
& \citet{Chen2023_PopulationStudy, Gomez2024_SLSNeI} \\
SN~2019ieh & \citet{Gomez2024_SLSNeI} \\
SN~2019unb & \citet{Yan2020_HeI, 2021MNRAS.508.4342P, 2022ApJ...941..107G}; \\
& \citet{Chen2023_PopulationStudy, Gomez2024_SLSNeI}\\
SN~2020qlb & \citet{2023AnA...670A...7W, Gomez2024_SLSNeI} \\
SN~2020tcw & \citet{Gomez2024_SLSNeI} \\
SN~2020wnt & \citet{2022MNRAS.517.2056G, Gomez2024_SLSNeI} \\
SN~2020znr & \citet{Gomez2024_SLSNeI} \\
SN~2021bnw & \citet{Gomez2024_SLSNeI} \\
SN~2021ek & \citet{Gomez2024_SLSNeI} \\
SN~2021lwz & \citet{2022ApJ...941..107G, Gomez2024_SLSNeI} \\
SN~2021uvy & \citet{2022ApJ...941..107G, Gomez2024_SLSNeI} \\
SN~2024rmj & \citet{Kumar2025_HeI} \\
SN~2024ahr & \citet{2025ApJ...987..127K} \\
SN~2025afcr & \citet{2025TNSAN.333....1W} \\
\enddata
\end{deluxetable*}